# Experimental realization of BCS-BEC crossover physics with a Fermi gas of atoms

by

**Cindy Regal**[1]

A thesis directed by Dr. Deborah Jin[1] and submitted to the Faculty of the Graduate School of the University of Colorado in partial fulfillment of the requirements for the degree of Doctor of Philosophy, Department of Physics

[1] *JILA, National Institute of Standards and Technology Quantum Physics Division and University of Colorado, and Department of Physics, University of Colorado, Boulder, CO 80309-0440, USA*

December 2005

## Abstract

This thesis presents experiments probing physics in the crossover between Bose-Einstein condensation (BEC) and BCS superconductivity using an ultracold gas of atomic fermions. Scattering resonances in these ultracold gases (known as Feshbach resonances) provide the unique ability to tune the fermion-fermion interactions. The work presented here pioneered the use of fermionic Feshbach resonances as a highly controllable and tunable system ideal for studying the cusp of the BCS-BEC crossover problem. Here pairs of fermionic atoms have some properties of diatomic molecules and some properties of Cooper pairs. I present studies of a normal Fermi gas at a Feshbach resonance and the work required to cool the gas to temperatures where superfluidity in the crossover is predicted. These studies culminated in our observation of a phase transition at the cusp of the BCS-BEC crossover through condensation of fermionic atom pairs. I also discuss subsequent work that confirmed the crossover nature of the pairs in these condensates.

# Acknowledgements

I could not have hoped to find a better place to pursue my graduate studies than JILA. I had the opportunity to work directly with and learn from an array of great scientists. Debbie provided me with an exceptional model of a researcher. Debbie is the ultimate debugger; from constructing a plot that captures the essence a technical problem to sorting out the important physics from a mess of effects, she always knows how to resolve a problem. As a graduate student you expect to be an underling, but as Debbie's student I always felt that I had a say in what I was working on and that there was always room for creativity. For this opportunity I am extremely grateful.

I also had the privilege to work with one of the new great minds of atomic physics, Markus Greiner. Markus brought an amazing store of technical knowledge to our lab. From Markus I learned innumerable facts about optics but also gained a very physical understanding of light. My successors, Jayson Stewart and John Gaebler, also contributed to the work in this thesis. Although my time working with them was short, I am sure they will do great work in the years ahead. I also thank Jon Goldwin and Shin Inouye for always pausing to answer my many questions and Michele Olsen for allowing me to help bolt together her UHV system. Jianying Lang, Travis Nicholson, Tyler Cumby, and Margot Phelps all deserve my respect for having the ability to discern whether my advice was useful or plain wrong. Everyone in the machine shop and in the electronics shop never left my questions unanswered in addition to building amazing components for our experiment.

As with every thesis from the JILA ultracold atom groups, Eric and Carl played a crucial role in my graduate work. I am indebted to Eric for letting me work in his lab in the summer of 2000, where I came to understand what a great place JILA is. Carl has been an amazing second advisor; I thank him for his honest advice and for the many bagels I have stolen from his group on Saturdays. As a graduate student, presenting work to Eric and Carl is always the ultimate challenge; after going through their interrogation the questions I received after an invited talk were never harder.

I have also had the unique opportunity to collaborate on a variety of problems with many theorists: John Bohn, Chris Ticknor, Chris Greene, Brett Esry, Murray Holland, Stefano Giorgini, and Marilu Chiofalo. The amazing aspect of JILA is that all these people are at or were visiting JILA during my graduate career, and they always had time to answer my peculiar breed of theory questions.

To Scott Papp I owe the happiness that I have enjoyed as a graduate student. I could not begin to convey how important Scott has been in my life over the past five years. He is my best friend, and he always makes me laugh. He taught me that my world would not fall apart if I did not quite finish Jackson E&M exercise 4(e) perfectly. He is an exceptional scientist, and I doubt there is anything he cannot build. It has been incredibly useful in my graduate work to walk home everyday with a talking electronics encyclopedia. Our many Sundays spent on the trails and roads of Colorado are times I will never forget.

JILA is full of so many other people that I have come to know well, I could not begin to name them all. However I do want to single out Josh Zirbel as a great friend. I thank him for the many hours he spent helping us with homework and his patience in tolerating Scott and me for so long. The members of the Longmont symphony have been invaluable for keeping the violin and music in my life. I also thank Colorado for being such a beautiful state; it is rare to see a sight more breathtaking than a Rocky Mountain alpine meadow.

The thanks for reaching this point is not relegated to those with a direct role in my years at JILA. My parents have always been there for me but still made me make all my own decisions, an invaluable gift. In my family a common motto was "Never say die, up man and try," and I always took it as given that the only honorable path was working towards a goal I felt was important to the world. Still coming from a Minnesota family whose favorite pastimes include felling dead trees and sawing them apart with handsaws, keeps one grounded, literally.

Most responsible for my entry to the world of physics were my professors at Lawrence University. John Brandenberger, David Cook, Matt Stoneking, and Jeff Collett filled their corner of the science hall with a genuine love of science and of teaching physics. Their hard work in building an accomplished small physics department attracted many students I looked up to and learned a great deal from. I must especially thank John Brandenberger for convincing me that I could become a physicist and that, above all, physics can be more interesting than biochemistry.

The generosity of the Fannie and John Hertz foundation has provided me with constant support throughout my graduate career. Thanks!

# Contents

# Chapter 1

# Introduction

## 1.1 Historical perspective

The phenomenon of superconductivity/superfluidity has fascinated and occupied physicists since the beginning of the 20th century. In 1911 superconductivity was discovered when the resistance of mercury was observed to go to zero below a critical temperature [1]. Although liquid $^4$He was actually used in this discovery, the superfluid phase of liquid $^4$He was not revealed until the 1930s when the viscosity of the liquid below the $\lambda$ point (2.17 K) was measured [2, 3]. Much later, $^3$He, the fermionic helium isotope, was also found to be superfluid at yet a much colder temperature than $^4$He [4]. Relatively recently in 1986, high-temperature superconductors in Copper-oxide materials further enlarged the list of superconducting materials [5].

These "super" systems, which I will refer to in general as superfluids, are listed in Fig. 1.1, but they are only classic examples. There are many other physical systems that display superfluid properties from astrophysical phenomena such as neutron stars, to excitons in semiconductors, to atomic nuclei [6]. Although the physical properties of these systems vary widely, they are all linked by their counterintuitive behaviors such as frictionless flow and quantized vorticity. The manifestation of these effects depends upon, for example, whether the system in question is electrically charged (superconductors) or neutral (superfluids). Besides these intriguing properties, there are many practical reasons for the intense research in this field; arguably the most useful super-systems are superconduc-tors, and if a robust room-temperature superconductor were created it would be an amazing discovery.

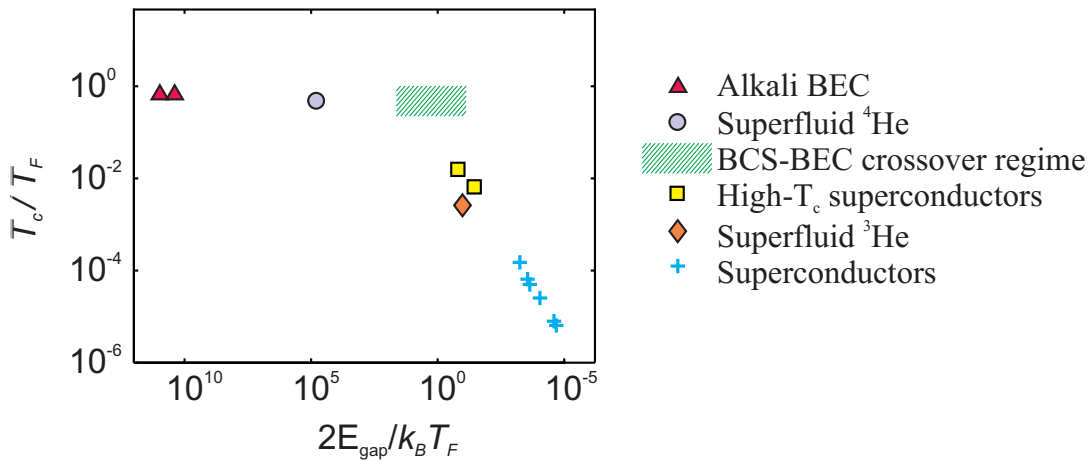

Figure 1.1: Classic experimental realizations of superfluidity/superconductivity arranged according to the binding energy (twice the excitation gap, $E_{gap}$) of the constituent fermions. The vertical axis shows the corresponding transition temperature, $T_c$, to a superfluid/superconducting state compared to the Fermi temperature, $T_F$. (Figure reproduced with permission from Ref. [7].)

Some of the first attempts to understand the phenomenon of superfluidity were in the context of Bose-Einstein condensation (BEC) of an ensemble of bosonic particles [8]. BEC is a consequence of the quantum statistics of bosons, which are particles with integer spin, and it results in a macroscopic occupation of a single quantum state (Fig. 1.2) [9, 10, 11]. Fritz London proposed in 1938 that superfluid $^4$He was a consequence of Bose-Einstein condensation of bosonic $^4$He [12]. ($^4$He is a boson because it is made up of an even number of 1/2 integer spin fermions - electrons, protons, and neutrons.) Physicists such as Blatt *et al.* pushed a similar idea in the context of superconductors in proposing that "at low temperature, charge carrying bosons occur, e.g., because of the interaction of electrons with lattice vibrations" [13]. For the case of tightly bound bosons, such as $^4$He, London's hypothesis turned out to be correct. However, the very strong interactions in $^4$He made it difficult to verify the existence of condensation [14], and for many years $^4$He studies did not mention BEC.

In the case of superconductors, discussion of BEC was overshadowed by the amazing success in 1957 of the Bardeen-Cooper-Schrieffer (BCS) theory of superconductivity [15, 16]. In 1956 Cooper found that a pair of fermions in the presence of a filled Fermi sea (Fig. 1.2) will form a bound pair with an arbitrarily small attractive interaction [17]. The BCS theory solved this problem in the case where many pairs can form in the Fermi sea. The result predicted (among other things) the formation of a minimum excitation energy, or energy gap, in the conductor below a critical temperature $T_c$. A great many properties of conventional superconductors can be understood as consequences of this energy gap.

Figure 1.1 sorts the classic superfluid systems according to the strength of the interaction between the fermions. A key aspect of the classic BCS theory is that it applies to the perturbative limit of weak attractive interactions and hence is only an exact theory for the far right side of Fig. 1.1. The theory perfectly described conventional superconductors for which the attraction between fermions is $\sim$10,000 times less than the Fermi energy, $E_F$. The BCS ground state was also able to accurately describe superfluid $^3$He and many (although arguably not all [18]) aspects of high-$T_c$ superconductors. The theory in its original form, however, did not at all apply to the case of the tightly bound boson, $^4$He.

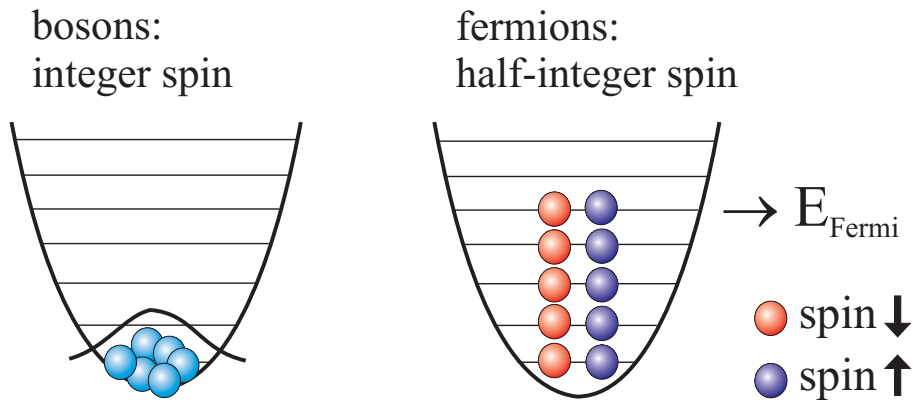

Figure 1.2: Quantum statistics: Bosons versus fermions with weak interactions at $T = 0$. Bosons form a BEC in which all of the bosons macroscopically occupy a single quantum state. Due to the Pauli exclusion principle fermions form a Fermi sea in which each energy state up to the Fermi energy is occupied.

In 1995 a completely new system joined $^4$He on the left side of Fig. 1.1.

Here the composite bosons were alkali atoms, such as $^{87}$Rb, that had been cooled as a gas down to nanoKelvin temperatures via laser cooling and evaporative cooling [19, 20, 21]. At these temperatures the thermal deBroglie wavelength ($\lambda_{deBroglie} = \sqrt{\frac{2\pi\hbar^2}{mk_bT}}$) of the particles becomes on order of the interparticle spacing in the gas and a Bose-Einstein condensate is formed (Fig. 1.3(a)) [22, 23]. In contrast to $^4$He the alkali BEC that was created was weakly interacting, making the condensation stunningly visible, as shown in Fig. 1.3(b). Experiments also observed that condensates behave as coherent matter waves [24] and verified the superfluid nature of a condensate [25, 26]. In this way both the BEC and superfluid properties could be clearly seen in one system and understood extremely well theoretically [27]. However, the fact that the dilute gas BEC was found to be a superfluid was not at all a surprise. Although a long and complicated history was required, physicists now understand the basic connection between BEC and superfluidity [28]. It is commonly accepted that superfluidity is always intimately connected to the macroscopic occupation of a quantum state.

Besides providing the first clear evidence for BEC, ultracold alkali gases opened a world of possibilities for studying superfluid systems. Many of the initial experiments with alkali BEC could be perfectly described by existing theories. However, recent work in the field of BEC has developed techniques to reach a regime that is more relevant for the outstanding theoretical questions in condensed matter physics, which are most commonly in strongly correlated systems. For example, experiments achieved BEC with much stronger interatomic interactions than typical alkali gases; furthermore, these interactions could even be controllably tuned [29, 30]. A phase transition to the highly-correlated Mott insulator state was observed through studies of quantum gases in optical lattice potentials [31]. These bosonic systems require theory that goes beyond mean-field interactions; yet they have a controllability rarely found in solid state materials.

At the same time as the creation of the first strongly interacting Bose gases, the techniques used to create alkali BECs were applied to the other class of quantum particles, fermions. To create a Fermi gas of atoms experimenters applied the same cooling techniques as those used to achieve BEC, simply replacing a bosonic atom, such as $^{87}$Rb or $^{23}$Na, with an alkali atom with an odd number of electrons, protons, and neutrons. (The two such stable alkali atoms are $^{40}$K and $^6$Li.) Still, evaporatively cooling fermions required ingenuity. Due to the quantum statistics of fermions the $s$-wave collisions required for evaporative cooling are not present at ultracold temperatures in a gas of spin-polarized, identical fermions. The solution to this problem was to introduce a second particle for the evaporative cooling, either another state of the fermionic atom or another species entirely. The first gas of fermionic atoms to enter the quantum degenerate regime was created at JILA in 1999 using $^{40}$K [32]. The observation in these experiments was not a phase transition, as in the Bose gas, but rather the presence of more and more energy than would be expected classically as the Fermi gas was cooled below the Fermi temperature. Many more Fermi gas experiments, using a variety of cooling techniques, followed [33, 34, 35, 36, 37, 38, 39, 40, 41].

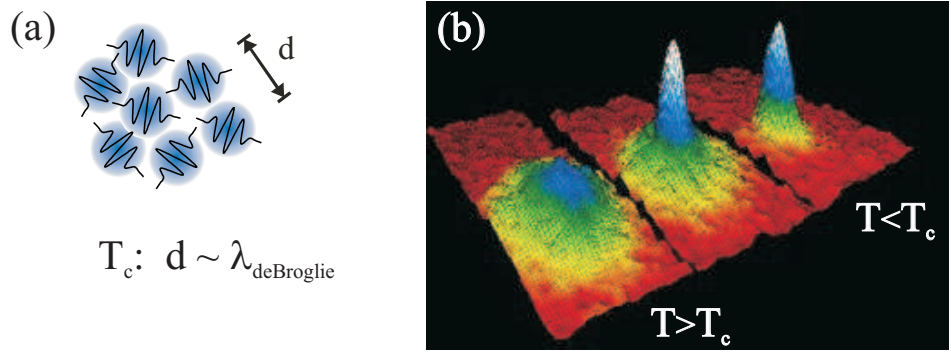

Figure 1.3: Bose-Einstein condensation in a dilute gas of $^{87}$Rb atoms. (a) Phase space density criterion for condensation. (b) Momentum distributions of $^{87}$Rb atoms at three values of the temperature compared to the critical temperature. (Figure adapted by M. R. Matthews from data in Ref. [22].)

The next goal after the creation of a normal Fermi gas of atoms was to form a superfluid in a paired Fermi gas. In conventional superconductors $s$-wave pairing occurs between spin-up and spin-down electrons. The hope was that $s$-wave pairing could similarly occur with the creation of a two-component atomic gas with an equal Fermi energy for each component. Such a two-component gas can be realized using an equal mixture of alkali atoms in two different hyperfine spin states. The simplistic view was that a BCS state would appear if the temperature of this two-component gas were cold enough and the interaction between fermions attractive and large enough. However, for typical interactions the temperatures required to reach a true BCS state were far too low compared to achievable temperatures (at that point) to imagine creating Cooper pairs. Stoof *et al.* noted that the interaction between $^6$Li atoms was large compared to typical values ($|a| \approx 2000$ $a_0$), as well as attractive, bringing the BCS transition temperature closer to realistic temperatures [42, 43]. It was then recognized that a type of scattering resonance, known as a Feshbach resonance, could allow arbitrary changes in the interaction strength. Theories were developed that explicitly treated the case where the interactions were enhanced by a Feshbach resonance [44, 7, 45].

In these proposals, however, increasing the interactions beyond the perturbative limit of BCS theory meant that the physical system would not be simply a BCS state, but rather something much more interesting. It was predicted that a Feshbach resonance would allow the realization of a system with an excitation gap on the order of the Fermi energy and provide the ability to continually tune within this region (green box in Fig. 1.1) [7]. This system, if achievable, would be the experimental realization of a theoretical topic that dated back to the late 1960s, the BCS-BEC crossover. In a theory originally put forth by Eagles and later by Leggett, it was proposed that the BCS wavefunction was more generally applicable than just to the weakly interacting limit [46, 47]. As long as the chemical potential is found self-consistently as the interaction is increased, the BCS ground state can (at least qualitatively) describe everything from Cooper pairing to BEC of composite bosons made up of two fermions, i.e., the fundamental physics behind all of the systems in Fig. 1.1 [46, 47, 48, 49, 50, 51]. After nearly a century of $^4$He and superconductors being considered separate entities, an experimental realization of a superfluid in the BCS-BEC crossover regime would provide a physical link between the two. More recent interest in crossover theories has also come in response to the possibility that they could apply to high-$T_c$ superconductors. These superconductors differ from normal superconductors both in their high transition temperature and the apparent presence of a pseudogap, which are both characteristics expected to be found in a Fermi gas in the crossover [8, 18].

Thus, starting in about 2001 (the year I started work on this thesis) a major goal in dilute Fermi gas experiments was to achieve a superfluid Fermi gas at a Feshbach resonance, often referred to as a "resonance superfluid" [7]. However, achievement of this experimental goal was a number of years and many steps away. The existence of Feshbach resonances had been predicted by atomic physicists both in the $^6$Li and in the $^{40}$K systems [52, 53], and the first step was to locate these resonances [54, 55, 56, 57]. Subsequent experimental studies appeared at an amazingly fast rate with contributions from a large number of groups, in particular those of R. Grimm (Innsbruck), R. Hulet (Rice), D. Jin (JILA), W. Ketterle (MIT), C. Salomon (ENS), and J. Thomas (Duke). Experimenters discovered interesting properties of the normal state of a strongly interacting Fermi gas [58, 59, 60, 61, 62]. Then Fermi gases were reversibly converted to gases of diatomic molecules using Feshbach resonances [63, 64, 65, 66]. The observation that these molecules were surprisingly long-lived created many opportunities for further study [64, 66, 65, 67]. Condensates of diatomic molecules in the BEC limit were achieved [68, 69, 70, 71, 72]; then these condensates were found to exist in the crossover regime [73, 74], signalling the existence of a phase transition in the BCS-BEC crossover regime. Collective excitations [75, 76, 77] and thermodynamic properties [39, 71, 78, 79] were also measured, and the nature of the pairs was probed in a variety of ways [80, 81, 72]. Most recently a vortex lattice was even created in the crossover [82].

Developing these techniques to access and probe the BCS-BEC crossover was a challenging adventure for the field. Experiments in the crossover are inherently difficult because the strong interactions make probing difficult. Some of the techniques used in the end were borrowed from those developed for alkali BEC, while others were taken from condensed matter physics. Some were new inventions that relied on the unique ability to tune the interaction at arbitrary rates using a Feshbach resonance.

So far the experiments that have been carried out with dilute Fermi gases near Feshbach resonances have been qualitatively consistent with classic BCS-BEC crossover theory. The excitation gap is on the order of the Fermi energy; the system crosses a phase transition to a superfluid state. Quantitatively though there is much work to be done. In tandem with these experiments, sophisticated theories of the crossover have been developed that are too numerous to list here, but are actively being pursued in groups such as those of A. Bulgac, J. Carlson, M. Chiofalo, S. Giorgini, A. Griffin, H. Heiselberg, T.-L. Ho, M. Holland, K. Levin, S. Strinati, B. Svistunov, and E. Timmermans. In time it is expected that the BCS-BEC crossover system provided by dilute Fermi gases should be able to rigorously test these theories.

The power to test these many-body theories comes from the ability to create a very clean strongly interacting Fermi system. In principle the density and two-fermion interaction in the sample can be known precisely and the $s$-wave pairing fully characterized as a function of the interaction strength. The end result could be a fully understood physical system that connects the spectrum of pairing from BCS to BEC, uniting the **basic physics** surrounding "super" systems. On the other hand, the complicated materials physics involved in, for example high-$T_c$ superconductors, cannot be elucidated in these clean crossover experiments. Still the hope is that an understanding of the basic physics will provide a solid foundation for studies of real materials.

## 1.2   Contents

In this thesis I will present some of the first experimental work studying fermions in the BCS-BEC crossover in a dilute atomic system. The contents come from experiments co-workers and I performed in the group of D. Jin at JILA between 2001 and 2005 using $^{40}$K atoms. As outlined above this was an exciting time in the field of Fermi gases. In 2001 in our lab at JILA the technology existed to create two-component Fermi gases at temperatures around 0.25 $T_F$ [83]. Predictions had been made for the existence of a superfluid state near a Feshbach resonance for a Fermi gas on the order of this temperature [44, 7, 45]. Yet in the early days of this work many physicists were skeptical about the feasibility of experimentally realizing such a state. The Feshbach resonances that had been observed in Bose gases were associated with extremely fast inelastic decay of the trapped gas [29, 84, 85]. These decay processes, which most often stem from three-body collisions, can quickly turn a hard-earned quantum gas into a classical gas of atoms [86, 87, 88, 89]. Carl Wieman's group at JILA produced the only experiments studying BEC near a Feshbach resonance over long time scales [30]. In this group $^{85}$Rb BECs were studied at very low densities, where two-body elastic collisions dominate over three-body collisions. For two-component Fermi gases three-body decay processes were expected to be suppressed compared to the Bose case [90, 91], due to the Pauli exclusion principle. Still there was a fair amount of contention about the degree of this suppression.

Further difficulties stemmed from the fact that not all researchers in the atomic physics community were familiar with theories from condensed matter physics, such as BCS-BEC crossover theory. Even when these theories were understood there was significant confusion about the relation between classic crossover theory and the physics at Feshbach resonances in $^6$Li and $^{40}$K. An additional challenge was creating a sufficiently cold Fermi gas near a Feshbach resonance. To be certain of achieving the predicted phase transition, temperatures well below the predicted $T_c$ would have to be achieved. A technical challenge was that the states required to access Feshbach resonances in the $^{40}$K and $^6$Li gases could not be confined in a magnetic trap, which was the most proven trap in studies of ultracold gases up to that point. Instead the experiments would have to achieve an ultracold Fermi gas in an alternative trap, such as an optical dipole trap.

Much of the work to create this thesis involved overcoming these difficulties and sources of confusion, and as we shall see there are now clear answers to many of these questions. These answers were found through careful studies of the normal state of a Fermi gas near a Feshbach resonance and significant work to cool a $^{40}$K gas to the coldest temperatures possible. In late 2003 we observed a phase transition in the crossover and since then have been able to study BCS-BEC crossover physics with a Fermi gas of atoms. In this thesis I will not include every aspect of the experiments we completed; instead I will concentrate on our key contributions to the understanding of BCS-BEC crossover physics in Fermi gases. The full range of experiments is outlined in Appendix A through a list of published articles.

In the first few chapters I discuss theory that is necessary to understand experiments presented in later chapters. The goal is not to rigorously derive modern theories, but rather to convey the mindset that many experimentalists currently use to think about the crossover problem with atomic Fermi gases. Chapter 2 presents the ideas of BCS-BEC crossover physics through simplified theory. Chapter 3 introduces Feshbach resonances and discusses how well the Feshbach resonance systems reproduce conditions for the classic "universal" BCS-BEC crossover problem.

The main body of the thesis presents experiments using an ultracold gas of $^{40}$K atoms. Chapter 4 describes cooling methods and temperature measurement techniques. Chapter 5 contains experiments that probe the presence of Feshbach resonances in the $^{40}$K system and study their ability to tune atomic scattering properties. Chapter 6 introduces the creation of molecules from a Fermi gas of atoms, which is the analog of the BCS-BEC crossover in the normal state. Chapter 7 concentrates on the stability of fermionic atoms and pairs against inelastic collisions in the presence of a Feshbach resonance; this is a subject of practical importance to the ability to study the BCS-BEC crossover with a Fermi gas of atoms. Chapter 8 describes the first experiments to observe a phase transition in a Fermi gas of atoms in the BCS-BEC crossover regime. Chapter 9 focuses on a measurement of the momentum distribution of the pairs in the crossover; this measurement is an important probe of the nature of the pairs in the crossover. Chapter 10 discusses some of the most relevant additions and changes to the apparatus of Ref. [32] that were required to access the BCS-BEC crossover physics described in this thesis.

# Chapter 2

# BCS-BEC crossover physics

To understand the experimental work presented in this thesis it is helpful to be familiar with BCS-BEC crossover physics. In this chapter I present the theory of the BCS-BEC crossover, first from a purely qualitative point of view and then from a slightly more quantitative perspective. In this quantitative perspective it is not my goal to present the most sophisticated theory, but rather a theory that introduces important parameters and illustrates the key differences between the crossover problem and the BCS limit. Lastly I discuss some of the unanswered questions related to the crossover problem; this conveys a sense of the importance of experimental studies of crossover physics.

## 2.1 Pairing in a Fermi gas of atoms

As we have seen, superfluids are fundamentally associated with the quantum properties of bosons. Since all visible matter is made up of fermions, creating a superfluid most often requires pairing of fermions. The simplest (although historically not the most famous) way to imagine pairing fermions is to create a **two-body** bound state of the two fermions. Two half-integer spin fermions when paired will produce an integer spin particle, which is a composite boson. In the case of experiments discussed in this thesis the fermionic particles are atoms; this makes such a two-body bound state a diatomic molecule. Below a critical temperature an ensemble of these diatomic molecules will form a BEC. The left side of Fig. 2.1 represents a superfluid containing these type of pairs. The two colors represent fermions in two different spin states; two states are required if the fermions are to pair via $s$-wave ($l = 0$) interactions.

**BEC** ⟷ **BCS**

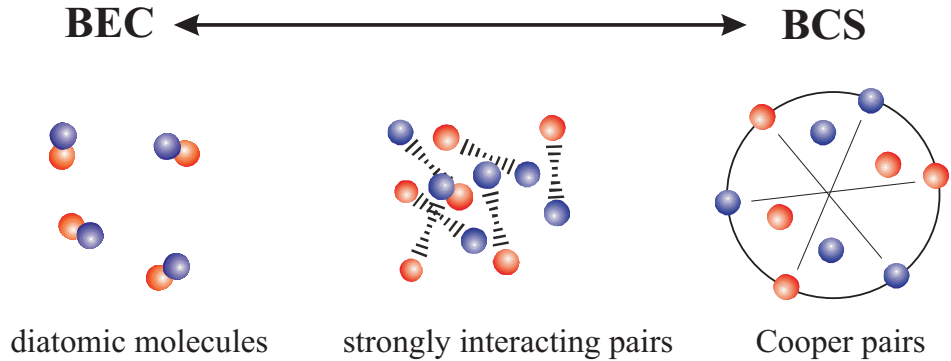

diatomic molecules    strongly interacting pairs    Cooper pairs

Figure 2.1: Cartoon illustration of the continuum of pairing in the BCS-BEC crossover.

In other pairing mechanisms, such as Cooper pairing, the underlying fermionic nature of the system is much more apparent. Cooper considered the problem of two fermions with equal and opposite momentum outside a perfect Fermi sea [17]. The energy of the two fermions turns out to be less than the expected value of $2\,E_F$ for arbitrarily weak attractive interactions. This result is in surprising contrast to the result of the problem of two fermions in vacuum; in this case there will not be a bound state until the interaction reaches a certain threshold. The key difference between the two situations arises from Pauli blocking, which in the Cooper pair case prevents the two fermions under consideration from occupying momentum states $k < k_F$, where $k_F$ is the Fermi wavevector [92].

Considering only one pair of electrons as free to pair on top of a static Fermi sea is not a sufficient solution to the pairing problem. All fermions should be allowed to participate in the pairing, and we expect that pairs should form until an equilibrium point is reached. At this equilibrium point the remaining ensemble of fermions is disturbed enough from a Fermi sea configuration to no longer lead to a bound state at the given interaction strength [93]. The BCS state is an approximate solution to this **many-body** problem. A description of the full BCS theory is beyond the scope of this current discussion, but is presented in the original papers [15, 16] and discussed in numerous books, for example Refs. [92, 93]. Qualitatively the BCS state consists of loose correlations between fermions across the Fermi surface in momentum space (Fig. 2.1 right side). Spatially the pairs are highly overlapping and cannot simply be considered to be composite bosons. In the BCS limit the momentum distribution only changes from the usual Fermi sea in an exponentially small region near the Fermi surface.

It is interesting to consider what happens if diatomic molecules become more and more weakly bound, to the point where the binding energy of the molecules, $E_b$, becomes less than the Fermi energy, $E_F$. One could also consider increasing the interaction energy of a Cooper paired state until it is close to $E_F$. The essence of the BCS-BEC crossover is that these two sentences describe the same physical state. As the interaction between fermions is increased there will be a continual change, or crossover, between a BCS state and a BEC of diatomic molecules. The point where two fermions in vacuum would have zero binding energy is considered the cusp of the crossover problem, and pairing in such a state is represented in the middle of Fig. 2.1. These pairs have some properties of diatomic molecules and some properties of Cooper pairs. Many-body effects are required for the pairing, as with the BCS state, but there is some amount of spatial correlation, as with diatomic molecules. The pair size is on the order of the spacing between fermions, and the system is strongly interacting.

## 2.2   Varying interactions

It is instructive to consider a physical situation that will allow the realization pairing throughout the crossover (Fig. 2.1) [47]. Suppose we start with an attractive potential between two atomic fermions in vacuum, such as the square well potential with characteristic range $r_0$ shown in Fig. 2.2. If this potential is very shallow there is a weak attractive interaction between the fermions. If we make this potential deeper the interaction between fermions becomes stronger, and for a strong enough attraction a bound molecular state will appear. This molecule will become more and more deeply bound as the potential becomes deeper.

The interaction in this system can be characterized by the $s$-wave scattering length $a$. The quantity $a$ comes out of studying two-body, low-energy scattering and is related to the $s$-wave collision cross section through $\sigma = 4\pi a^2$. The top of Fig. 2.2 shows a pictorial representation of $a$. Just before the bound state appears $a$ is large and negative, corresponding to a strong attractive interaction. As the bound state moves through threshold $a$ diverges and then becomes large and positive, corresponding to a strong repulsive interaction. When $a$ is much larger than $r_0$, the interaction is independent of the exact form of the potential, and $a > 0$ is universally related to the binding energy of the two-body bound state through $E_b = \frac{\hbar^2}{ma^2}$, where $m$ is the mass of a fermion [94].

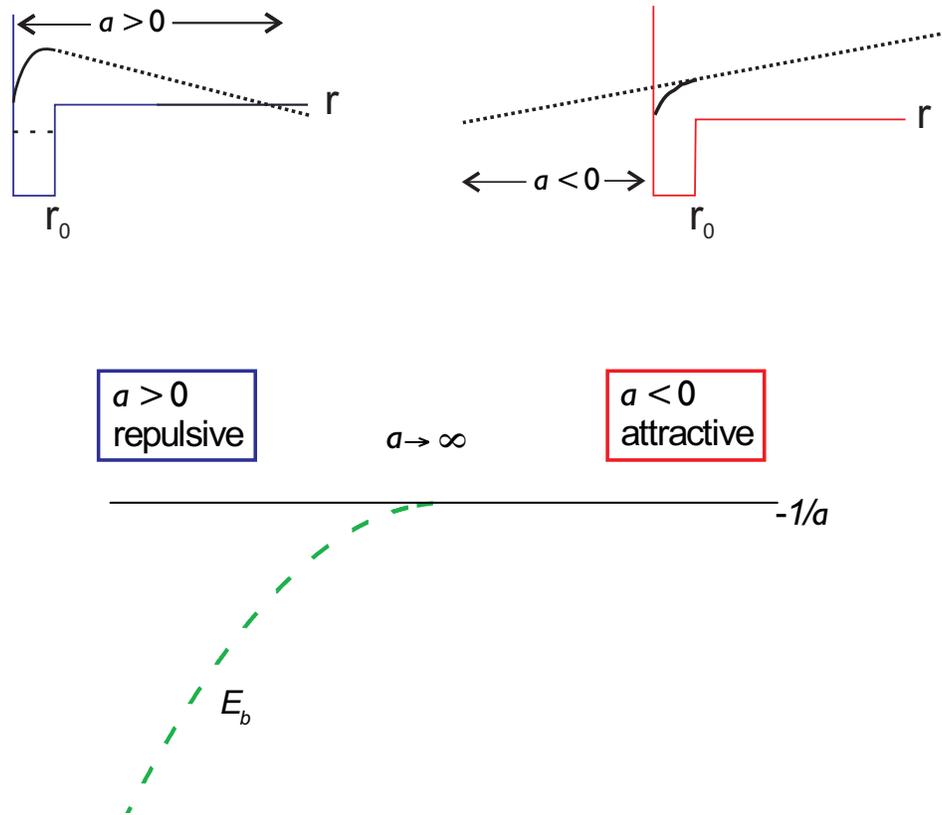

Figure 2.2: Scattering wavefunctions in the presence of an attractive potential (right) and a more deeply attractive potential (left), in a regime where a weakly bound state of the potential (dashed line) is near threshold. **r** here describes the relative position of two distinguishable fermions. The scattering length diverges as the bound state moves through threshold.

Now if we consider an ensemble of many fermions under the situation in Fig. 2.2, we have a system that can be tuned from BCS to BEC simply by tweaking the attractive potential. To the far right we have a small negative $a$ and thus the BCS limit. In the opposite limit we have an ensemble of diatomic molecules, hence the BEC limit. It is important to note that although the interaction between fermions in pairs is strongest in the BEC limit, from the point of view of collisions in the gas the BEC limit is actually weakly interacting because the molecule-molecule interaction is weak.[1] The most strongly interacting gas from the point of view of collisions occurs near the divergence of $a$. Here many-body calculations are difficult because there is no small expansion parameter. The precise point at which $a$ diverges is known as unitarity. Here the only length scale in the problem is $1/k_F$, giving this point many unique properties [96, 97, 98, 99].

## 2.3  Simple theory

BCS theory was originally applied in the limit where the interaction energy is extremely small compared to the Fermi energy. In this case the chemical potential, $\mu$, can be fixed at $E_F$, and many calculations become reasonably simple. Leggett pointed out that if the BCS gap equation is examined allowing $\mu$ to vary, the gap equation actually becomes precisely the Schrödinger equation for a diatomic molecule in the limit where $\mu$ dominates [47]. Below we briefly illustrate the key steps in an application of BCS theory to the entire crossover. This gives qualitatively correct results for the entire spectrum of pairing. The structure of the crossover theory below originates in the work of Nozieres and Schmitt-Rink (NSR) in Ref. [48] and of Randeria *et al.* in Ref. [49].

---

[1] In the limit $a \gg r_0$ the molecule-molecule scattering length $a_{mm}$ is predicted to be $0.6a_{aa}$, where $a_{aa}$ is the scattering length for atoms scattering above threshold [95].

Let us consider a homogeneous Fermi system in three dimensions in an equal mixture of two different states at $T = 0$. Application of usual BCS theory results in the gap equation

$$\Delta_k = -\sum_{\mathbf{k'}} U_{kk'} \frac{\Delta_{k'}}{2E_{k'}} \qquad (2.1)$$

where $E_k = \sqrt{\xi_k^2 + \Delta^2}$, $\xi_k = \epsilon_k - \mu$, and $\epsilon_k = \frac{\hbar^2 k^2}{2m}$. $U_{kk'} < 0$ is the attractive interaction for scattering of fermions with momenta $k'$ and $-k'$ to $k$ and $-k$. We also obtain the number equation

$$\langle N_{tot} \rangle = \sum_{\mathbf{k}} (1 - \frac{\xi_k}{E_k}) \qquad (2.2)$$

where $N_{tot}$ is the total number of fermions in both states.

To solve Eqn. 2.1 in the BCS limit the standard approach is to assume that the potential is constant at a value $U < 0$, which implies that the gap is constant as well, i.e., $\Delta_k = \Delta$. In this case the gap equation (Eqn. 2.1) becomes

$$-\frac{1}{U} = \sum_{\mathbf{k}} \frac{1}{2E_k}. \qquad (2.3)$$

One will find, however, that this equation diverges. For BCS superconductors this problem is resolved because the interaction can be limited to within the Debye energy, $\hbar\omega_D$, of $E_F$. This is a result of the nature of the phonon-mediated interaction between the electrons that gives rise to the attractive interaction. Further simplifications in the BCS limit are that $\mu = E_F$ and that, since $\hbar\omega_D \ll E_F$, the density of states is constant at the value $N(\xi = 0)$. The gap equation then becomes

$$-\frac{1}{N(0)U} = \int_{-\hbar\omega_D}^{\hbar\omega_D} \frac{d\xi}{2\sqrt{\Delta^2 + \xi^2}}. \qquad (2.4)$$

Solving Eqn. 2.4 produces the BCS result $\Delta \approx 2\hbar\omega_D e^{-1/N(0)|U|}$.

To extend this calculation to the crossover in atomic systems we can no longer apply the $\hbar\omega_D$ cutoff. The solution to the divergence problem in this case is nontrivial and requires a renormalization procedure, the full description of which we will not present here (but see, for example, Ref. [8] and references therein).

The result of such a procedure is a "renormalized" gap equation

$$-\frac{m}{4\pi\hbar^2 a} = \frac{1}{V}\sum_{\mathbf{k}}\left(\frac{1}{2E_k} - \frac{1}{2\epsilon_k}\right) \qquad (2.5)$$

where the interaction is now described by the s-wave scattering length $a$ instead of $U$ and $V$ is the volume of the system. Furthermore, in the crossover we cannot assume $\mu = E_F$; instead we must solve the gap equation (Eqn. 2.5) and number equation (Eqn. 2.2) simultaneously for $\mu$ and the gap parameter $\Delta$. We will solve for these parameters as a function of the dimensionless parameter $k_F a$, where $k_F = \sqrt{2mE_F}/\hbar$. As pointed out in a useful paper by M. Marini *et al.* this can actually be done analytically using elliptic integrals [100].

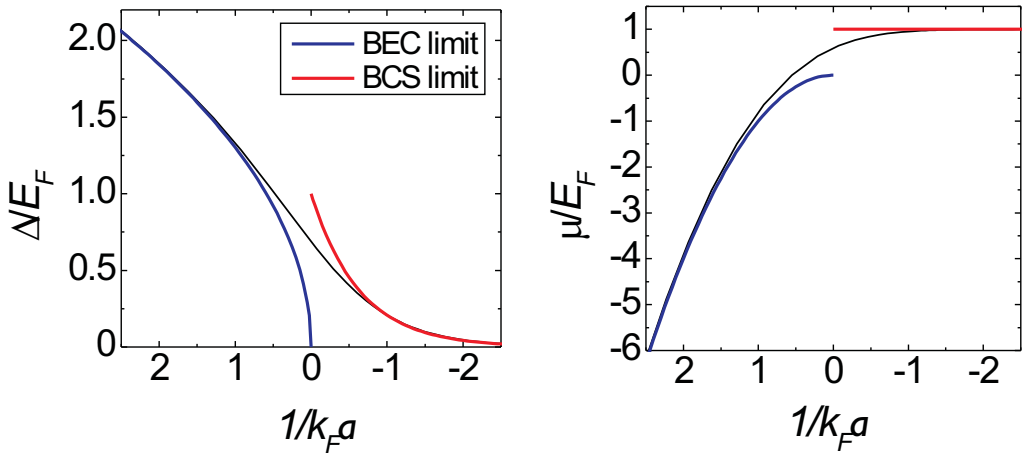

Figure 2.3: Gap parameter, $\Delta$, and chemical potential, $\mu$, of a homogeneous Fermi gas at $T = 0$ as determined through NSR theory. The red and blue lines show the BCS and BEC limits of the theory. Note that the limiting theories only deviate significantly from the full theory in approximately the range $-1 < \frac{1}{k_F a} < 1$.

The black lines in Fig. 2.3 show the result of this calculation of $\Delta$ and $\mu$. We also plot the values of both $\Delta$ and $\mu$ as they would be calculated in the BCS and BEC limits to find that the crossover occurs in a relatively small region of the parameter $1/k_F a$, namely from approximately $-1 < 1/k_F a < 1$. In typical crossover experiments with $^{40}$K or $^6$Li, this regime corresponds to varying $a$ from $-2000\ a_0$ through $\infty$ and to $2000\ a_0$, where $a_0$ is the Bohr radius (0.0529 nm).

It is useful to explicitly understand the value and meaning of both $\mu$ and $\Delta$ in the two limits. $\mu$ is $E_F$ in the BCS limit and $-E_b/2 = -(\frac{1}{k_F a})^2 E_F$ in the BEC limit. $\Delta$ is $e^{\frac{-\pi}{k_F|a|}} E_F$ in the BCS limit and $E_F \sqrt{\frac{16}{3\pi} \frac{1}{k_F a}}$ in the BEC limit [100]. Although referred to as the gap parameter, $\Delta$ only has meaning as the excitation gap, i.e., the smallest possible energy that can create a hole (remove a fermion) in the superfluid, in the BCS limit. In general the excitation energy is $E_{gap} = \min E_k = \min \left[ \sqrt{(\frac{\hbar^2 k^2}{2m} - \mu)^2 + \Delta^2} \right]$ [8]. This is $\Delta$ when $\mu$ is positive (as in the BCS limit), but becomes $\sqrt{\mu^2 + \Delta^2}$ when $\mu$ is negative.

## 2.4   Beyond $T = 0$

The phase transition temperature, $T_c$, is an important parameter for any super-fluid system. In the BCS-BEC crossover the transition temperature increases as the interaction is increased, i.e., it is lowest in the perturbative BCS regime and highest in the BEC limit (Fig. 1.1). In a homogeneous system, in the BCS limit $T_c/T_F = \frac{8}{\pi} e^{\gamma-2} e^{\frac{-\pi}{2k_F|a|}}$ where $\gamma = 0.58$ [51], and in the BEC limit $T_c/T_F = 0.22$ [101]. Note that BCS transition temperatures can be extremely small due to the exponential dependence upon $1/k_F a$. For example, at a typical interaction strength in alkali fermionic gases ($a = -100$ $a_0$) and a typical $k_F$ ($1/k_F = 2000$ $a_0$) the BCS transition temperature would be $\sim 10^{-14}$ $T_F$, which is a completely inaccessible temperature in contemporary atomic systems. Still, at $1/k_F a = -1$ where BCS theory nearly applies, the transition temperature is on the order of $0.1$ $T_F$, which is accessible in current atomic systems.

In the BCS limit pairing and the phase transition to a superfluid state occur at the same temperature. However, in the BEC limit this is not the case; because the constituent fermions are very tightly bound, pairs can form far above $T_c$. It is natural to expect that there would be a crossover between these two behaviors in the BCS-BEC crossover, i.e., at the cusp of the crossover the pairing temperature, $T^*$, would be distinct from $T_c$, yet not far from it. In a simple picture $T^*$ should be related to a pair dissociation temperature, which in the case of molecules is $\sim E_b/k_b$ [8]. It is important to differentiate between the superconducting order parameter, which exists below $T_c$, and the pairing gap, which exists below $T^*$ [18]. The pairing gap is associated with so-called preformed pairs, which are pairs that are not yet phase coherent. Aspects of the pseudogap observed in high-$T_c$ superconductors may be a manifestation of pre-formed pairs [18].

## 2.5 Modern challenges

The discussions and calculations above, while providing an introduction to basic crossover theory, are far from the state-of-the-art for theory in the field. There are noticeable problems with the NSR theory presented in Sec. 2.3. For example, the result for the chemical potential at unitarity ($1/k_F a = 0$) is significantly different from the result of more accurate calculations using full Monte Carlo simulations [102, 103]. The chemical potential is often written as $\mu = (1 + \beta)E_F$, and the NSR theory produced $\beta = -0.41$ at unitarity, while the Monte Carlo simulation of Ref. [103] finds $\beta = -0.58 \pm 0.01$ at unitarity. As another example, extension of the NSR theory predicts that $a_{mm} = 2a_{aa}$, while a full 4-body calculation in the BEC limit yields $a_{mm} = 0.6a_{aa}$ [95]. (Experiments at ENS have shown that $a_{mm} = 0.6^{+0.3}_{-0.2}a_{aa}$ [71], in agreement with the full calculation but not the NSR theory.) Both of these problems point to the fact that the NSR ground state, which only includes two-particle correlations, is not sufficient to accurately describe the system. Thus, it is clear that adding higher-order correlations to BCS-BEC crossover theory is necessary, yet not a simple task [104].

An even greater challenge is to extend accurate theories to nonzero temperature where predictions can be made about the critical temperatures and the role of pre-formed pairs. Furthermore, all of the calculations considered thus far are carried out for a homogeneous Fermi system. However, the experiments with ultracold gases take place in traps (most often harmonic traps). The use of an inhomogeneous density gas can lead to qualitative changes in the system that must be accounted for in theory: Strong interactions can modify the density of the trapped gas [105], and signals can become blurred as the density, and hence the gap, varies across the sample [80].

# Chapter 3

# Feshbach resonances

In the previous chapter we determined that varying the fermion-fermion interaction is the key to accessing BCS-BEC crossover physics. We also observed that to arbitrarily vary the interactions we could imagine using a variable attractive potential with a bound state near threshold. Amazingly atomic systems can achieve exactly such a physical situation. The attractive potential is provided by the van der Waals interaction between two atoms, and the knob to tune this potential is a homogeneous magnetic field. The sensitive magnetic-field dependence of the potential can be provided through a scattering resonance known as a Feshbach resonance [106, 107, 108, 109]. The goal of this chapter will be to understand exactly how a Feshbach resonance allows us to arbitrarily tune the interaction using a magnetic field. In this chapter I will be discussing only two-body physics, namely the problem of two fermions scattering in vacuum. We will find that the Feshbach resonance used for the experiments in this thesis approximates well the classic two-body system required for study of the BCS-BEC crossover problem of Ch. 2.

## 3.1   Description

Calculating the interaction between two ground state alkali atoms is a nontrivial problem that has been studied extensively in atomic physics [110]. Study of this problem shows that for S-state atoms the interatomic potential is repulsive for very small $r$ and has a weak attractive tail that goes as $-C_6/r^6$ as $r \to \infty$.

This weak attractive tail is a result of the interaction between mutually induced dipole moments of the atoms, which is known as the van der Waals interaction. The interatomic potentials are deep enough to contain a large number of bound vibrational states. A Feshbach resonance occurs when one of these bound states (often called the bare molecule state) coincides with the collision energy of two free atoms in a different scattering channel. Such a situation is depicted in Fig. 3.1(a). The interatomic potential of the two free atoms is often referred to as the open channel, while the potential containing the bare molecule state is referred to as the closed channel. When the closed and open channels describe atoms in different magnetic sublevels, they can be shifted with respect to each other through the Zeeman effect using an external magnetic field (Fig. 3.1(b)).

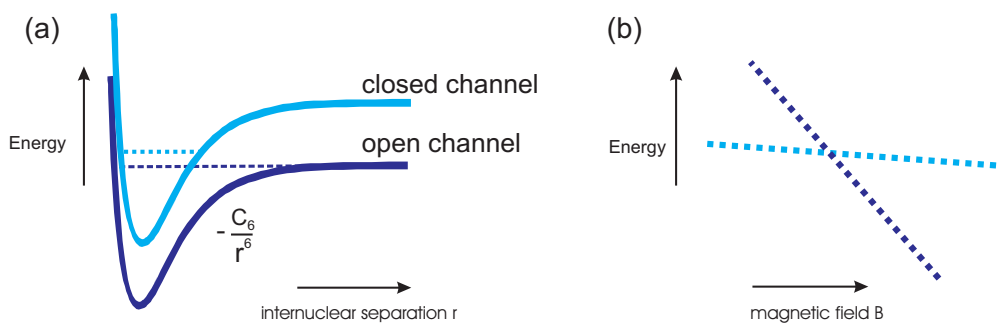

Figure 3.1: (a) Feshbach resonances are the result of coupling between a molecular state in one interatomic potential with the threshold of another. (b) The bare molecule state of the closed channel tunes differently with magnetic field than the open-channel threshold. This can lead to a crossing of the two levels.

Typically the effect of the coupling between the closed and open channels is small, but at a Feshbach resonance when the open-channel dissociation threshold is nearly degenerate with the bare molecule state, the effect of the coupling can be significantly enhanced. This coupling changes the effective interatomic potential, which we will refer to as the multichannel potential [111]. A bound state will be added to this multichannel potential at a magnetic field value near (but not exactly at) the magnetic field position of the crossing of the bare molecule and open-channel threshold. When I use the term "molecule" in this thesis I am always referring to this additional bound state of the multichannel potential, a so-called dressed molecule. The wavefunction of these molecules is generally a linear superposition of open-channel and closed-channel wavefunctions. As we will see in the next section, the open-channel component dominates for the molecules studied in this thesis.

As the magnetic field is tuned this multichannel bound state moves through threshold, and the scattering length between atoms in the open channel diverges. The scattering length near a Feshbach resonance varies with the magnetic field, $B$, according to the following equation [110].

$$a(B) = a_{bg} \left( 1 - \frac{w}{B - B_0} \right) \tag{3.1}$$

Here $a_{bg}$ is the triplet background (nonresonant) scattering length for atoms scattering in the open channel, $B_0$ is the magnetic field position at which the molecular bound state of the coupled system goes through threshold, and $w$ is the magnetic-field width of the Feshbach resonance, defined as the distance in magnetic field between $B_0$ and the magnetic field at which $a = 0$. Figure 3.2 shows how $a$ diverges according to Eqn. 3.1 for the $^{40}$K resonance described in the next section.

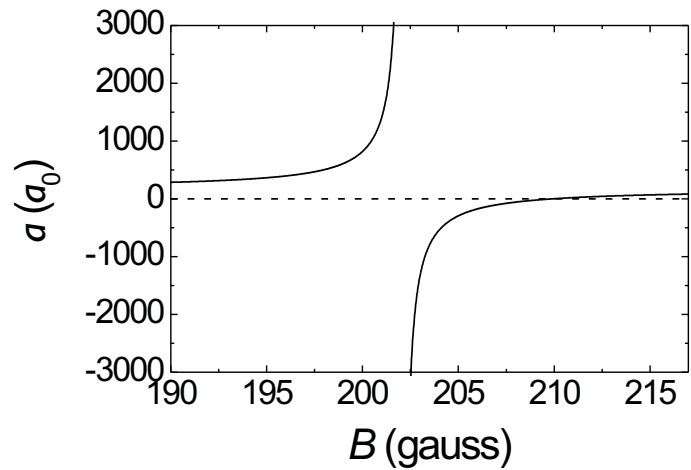

Figure 3.2: Behavior of the scattering length at a Feshbach resonance in $^{40}$K between the $f$=9/2 and $m_f$=−7/2, −9/2 states.

## 3.2 A specific example

To further illustrate Feshbach resonance physics it is useful to consider a specific atomic example. Here I will discuss the case of the $^{40}$K resonance that is used for most of the BCS-BEC crossover studies in this thesis. The open channel in this case describes the scattering of $|f, m_f\rangle = |9/2, -9/2\rangle$ and $|9/2, -7/2\rangle$ atoms, which are the two lowest energy states in the $^{40}$K system (see Fig. 4.1). $f$ describes the total atomic angular momentum and $m_f$ is the magnetic quantum number. For this particular problem the open channel couples to only one closed channel, $|f, m_f\rangle = |9/2, -9/2\rangle$ and $|7/2, -7/2\rangle$ atoms. We will compare the results of this example to the requirements for studying the classic BCS-BEC crossover problem of Ch. 2. In the classic problem instead of two coupled channels a single variable depth potential with $r_0 \ll a$ is considered (where $r_0$ is the range of the potential).

To calculate exactly the properties of our $^{40}$K resonance we would need to carry out a full coupled channels calculation using realistic potassium potentials. The description of such a calculation is beyond the scope of this thesis but is described nicely in, for example, Ref. [110].[1] Instead, for the demonstrative purposes of this chapter, we will examine the results of a simpler technique that is derived from K. Góral *et al.* in Ref. [112] and applied to the case of $^{40}$K by M. Szymańska *et al.* in Ref. [111]. We will solve the coupled Schrödinger equations that describe our two-channel system with a few simplifying assumptions. This technique reproduces the important physics of our $^{40}$K resonance using a small set of experimentally measurable parameters. The main approximation is the so-called "pole approximation" described in Ref. [112] that holds when the open channel is strongly coupled to only one bare molecule state.

Our main goal will be to use the simplified two-channel calculation to determine the binding energy of the molecular state in the multichannel potential as a function of the magnetic field. The Hamiltonian for our two-channel system is

$$H = \begin{pmatrix} -\frac{\hbar^2}{m}\nabla^2 + V_{oc} & W \\ W & -\frac{\hbar^2}{m}\nabla^2 + V_{cc} \end{pmatrix} \qquad (3.2)$$

---

[1] We have compared some of the data presented in this thesis to full coupled channels calculations carried out by Chris Ticknor in John Bohn's group at JILA. See Ch. 5 and Ch. 6.

where $V_{oc}$ is the uncoupled open-channel potential, $V_{cc}$ is the uncoupled closed-channel potential, the potential $W$ describes the coupling between the open and closed channels, and $m$ is the mass of $^{40}$K. It can be shown that the solution to this problem depends on only a few accessible parameters of the $^{40}$K system [112]. For the $^{40}$K resonance we are considering, these parameters include the background scattering length $a_{bg} = 174\ a_0$ [54], the van der Waals coefficient $C_6 = 3897$ a.u. [113], the resonance width $w = 7.8$ G [57], and the resonance position $B_0 = 202.1$ G [73]. Also useful is the binding energy of the first bound state in $V_{oc}$, $E_{-1} = 8.75$ MHz (which can be attained from $a_{bg}$ and $C_6$) [112, 111]. Finally, we need the difference in magnetic moment (change in energy with magnetic field) of the open channel threshold with respect to the closed channel bare molecule. In our simplified calculation we will assume this to be the linear value that best approximates the result of a full coupled channels calculation, $\mu_{co} = 1.679\ \mu_B = h\,2.35$ MHz/G [111, 114].[2] These parameters, which in the end come from experimentally measured values, of course all have uncertainties, which we will ignore for now.

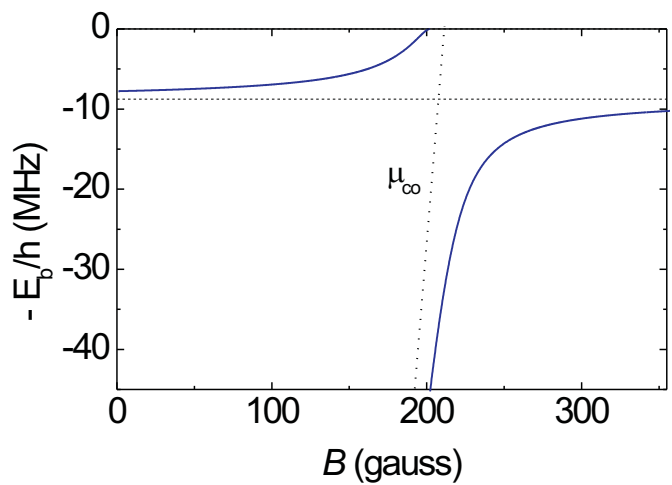

Figure 3.3: Multichannel bound state of a Feshbach resonance in $^{40}$K determined through the simplified calculation described in the text [112, 111].

---

[2]For $^{40}$K $\mu_{co}$ from the result of the full calculation comes quite close to the difference in magnetic moments of the closed and open channel thresholds, which is $\mu_{co} = h\,2.49$ MHz/G at 200 G according to the Breit-Rabi formula [115].

The solution to the coupled Schrödinger equations based on the Hamiltonian above using the pole approximation is outlined in Ref. [112]. The calculation is not computationally intensive and, after inserting the parameters above, provides the multichannel binding energy $E_b$, which I plot as the solid line in Fig. 3.3. Here $E_b$ is plotted with respect to the energy of the open channel threshold, i.e., the open channel threshold is zero for all values of $B$. The dotted line shows the movement of the bare molecule energy with magnetic field, and the flat dashed line is the value of $E_{-1}$. The multichannel bound state is the dressed state of the avoided crossing of these two levels. The bare molecule state crosses threshold about 9 G higher than the position where the multichannel bound state comes through threshold, $B_0$. The difference between these two crossings is proportional to the resonance width parameter $w$ and related to the interchannel coupling parameter in the Hamiltonian, $W$ [112]. Note that for $^{40}$K the multichannel bound state adiabatically connects to the the highest-lying vibrational state of the open channel at low field, rather than the bare molecule of the closed channel.

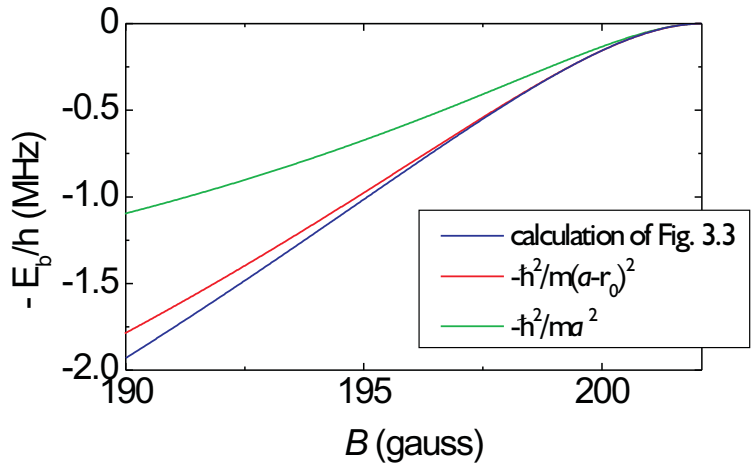

Figure 3.4: Binding energy near the Feshbach resonance peak using three calculations of varying degrees of approximation. These results agree with those in Ref. [111].

The physics we study in this thesis actually occurs over a very small region near threshold in Fig. 3.3. The blue line in Fig. 3.4 shows a closeup of the multichannel bound state near threshold. Within a few gauss range of the Feshbach resonance position the result obeys the expectation for $a >> r_0$, $E_b = \frac{\hbar^2}{ma^2}$ (where $a$ is determined through Eqn. 3.1), which is shown by the green line in Fig. 3.4. Farther away there is a clear deviation from quadratic behavior in $1/a$. This behavior can be estimated by subtracting the range of the van der Waals potential $r_0$ from $a$ in the calculation of the binding energy, $E_b = \frac{\hbar^2}{m(a-r_0)^2}$ (red line in Fig. 3.4) [112, 111, 116]. The range of the van der Waals potential is given by [117]

$$r_0 = \frac{1}{\sqrt{8}} \frac{\Gamma(3/4)}{\Gamma(5/4)} \left( \frac{mC_6}{\hbar^2} \right)^{1/4}. \tag{3.3}$$

The value of $r_0$ is $\sim$60 $a_0$ for $^{40}$K (and $\sim$30 $a_0$ for $^6$Li).

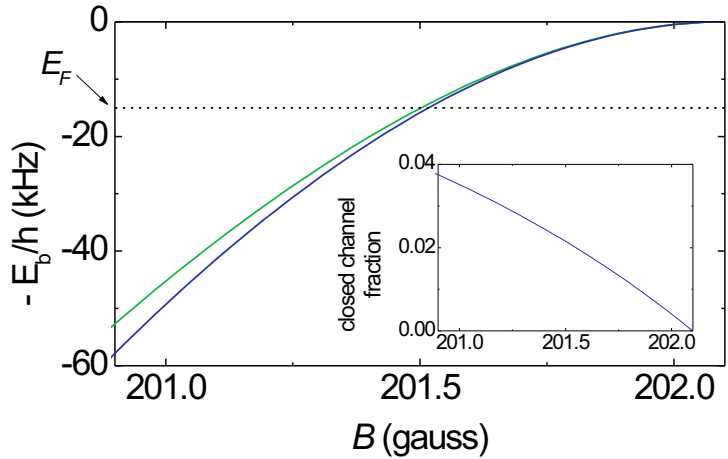

Figure 3.5: Binding energy in the magnetic field regime used for the BCS-BEC crossover studies of this thesis.

In Fig. 3.5 we examine the binding energy even closer to threshold. Here the result of the simplified two-channel calculation and $E_b = \frac{\hbar^2}{m(a-r_0)^2}$ (blue) are indistinguishable, while the $E_b = \frac{\hbar^2}{ma^2}$ prediction (green) only deviates slightly. We also plot the closed channel contribution to the molecule wavefunction, which is given by $\frac{dE_b}{dB}/\mu_{co}$. The Feshbach molecules that we will be interested in for the BCS-BEC crossover studies will have binding energies on order of or smaller than $E_F$. A typical value of $E_F$ in our experiments ($h$ 15 kHz) is shown by the dotted line in Fig. 3.5. At this point the closed channel fraction according to our two-channel calculation is only 2%; this result is in agreement with full coupled channels calculations [118, 114]. This open-channel dominance means that, for our crossover studies, the two coupled channels problem of the Feshbach resonance can be approximated by small effective changes to the open channel potential. A resonance for which this is true is often referred to as a "broad resonance," while the limit in which the closed channel dominates for $E_b \approx E_F$ is referred to as a "narrow resonance" [119, 120, 121]. In the case in which $E_b \approx \frac{\hbar^2}{m(a-r_0)^2}$ we can derive a simple equation that describes this criterion for a broad resonance in a Fermi system,

$$\frac{w^2 \mu_{co}^2}{4E_F \frac{\hbar^2}{ma_{bg}^2}} \gg 1. \tag{3.4}$$

For the $^{40}$K resonance we have been considering the left side of this expression has a numerical value of 43, and it has a value of 10,000 for the $^6$Li resonance at $B_0 = 834$ G [116, 122], indicating both Feshbach resonances used for BCS-BEC crossover studies thus far can be considered broad.

In conclusion, we have seen how in atomic systems a Feshbach resonance can be used to add an additional bound state to an interatomic potential, leading to a divergence in the zero-energy scattering cross section for atoms colliding through the open channel. The Feshbach resonances that have so far been used for BCS-BEC crossover studies are broad resonances. These resonances can be approximated by a single-channel problem and display **universal** properties, i.e., they have no dependence on the details of the atomic structure, but rather only on the parameters $a$ and $k_F$. For the $^{40}$K resonance there are slight deviations from the $a \gg r_0$ limit that must be taken into account for precise measurements (Fig. 3.5). Still the physics of a $^{40}$K gas at this resonance should basically reproduce the classic BCS-BEC crossover scenario envisioned by Leggett [47] and described in Sec. 2.2.

# Chapter 4

# Cooling a Fermi gas and measuring its temperature

To access the superfluid state of a Fermi gas in the crossover the gas must be cooled below the critical temperature, $T_c$. To achieve such a temperature with a trapped gas of fermions was one of our largest challenges. In this chapter I describe how we cool $^{40}$K and assess the success of this cooling through temperature measurements. More technical details related to the contents of this chapter can be found in Ch. 10.

## 4.1   Cooling $^{40}$K

The apparatus used to cool $^{40}$K for these BCS-BEC crossover experiments employs the strategy used for some of the first experiments with $^{87}$Rb BEC [22, 123, 124]. We perform the "usual" combination of trapping and cooling in a magneto-optical trap (MOT) followed by evaporative cooling [19, 20, 21]. The laser cooling uses light from semiconductor diode lasers on the $^{40}$K $D_2$ line (4S$_{1/2}$ to 4P$_{3/2}$ transition at 766.7 nm), and a two chamber apparatus allows for an ultra high vacuum region for evaporative cooling [83]. The major difference compared to the $^{87}$Rb experiments stems from the fact that elastic collisions between identical fermions are suppressed at ultracold temperatures. This is because quantum statistics require antisymmetry of the total wavefunction for two colliding fermions, which forbids

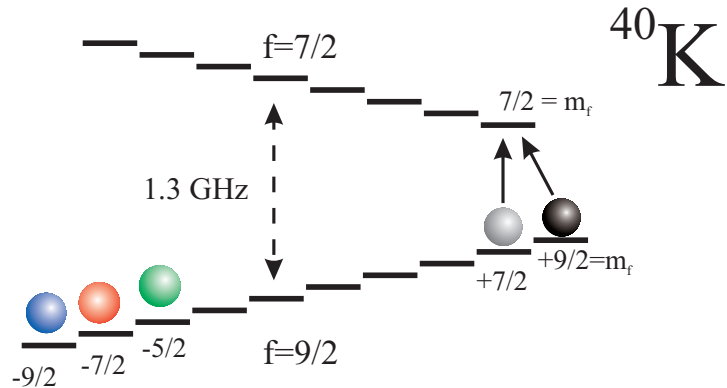

Figure 4.1: $^{40}$K ground state level diagram, with exaggerated Zeeman splittings. The two levels represent the hyperfine structure, which originates from the coupling of the nuclear spin ($I = 4$) with the electron spin ($S = 1/2$). Note the hyperfine structure of $^{40}$K is inverted.

$s$-wave collisions for identical fermions.  While odd partial wave collisions, such as $p$-wave, are allowed, these collisions are suppressed below $T \approx 100~\mu$K because of the angular momentum barrier [125].  Thus, since evaporative cooling requires collisions to rethermalize the gas, a mixture of two distinguishable particles is required to cool fermions.

$^{40}$K provides an elegant solution to this problem.  Figure 4.1 shows the ground state energy levels of $^{40}$K.  The large angular momentum of the lowest ground state hyperfine level ($f$=9/2) provides 10 distinct spin states.  The two highest energy states, $m_f$=+9/2 and $m_f$=+7/2, can both be held with reasonable spatial overlap in a magnetic trap, which is the type of trap most proven for evaporative cooling when starting from a MOT.  In this way an apparatus designed for only one atomic species could provide two distinguishable states for cooling.  To remove the highest energy atoms for evaporative cooling, microwaves at ∼1.3 GHz were used to transfer atoms to untrapped spin states in the upper hyperfine state [32, 83].  With this technique, quantum degeneracy was reached in 1999 and by 2001 two-component $^{40}$K Fermi gases at temperatures of 0.25 $T_F$ could be created.  A detailed description of the work to achieve these temperatures is recorded in Brian DeMarco's Ph.D. thesis (Ref. [83]).

One of the first steps in accessing the BCS-BEC crossover was to create a degenerate Fermi gas in an equal mixture of the $f{=}9/2$ Zeeman states between which a Feshbach resonance was predicted, $m_f{=}{-}9/2$ and $-7/2$ (Fig. 4.1). To accomplish this we needed to trap the high-field-seeking spin states of the $f{=}9/2$ manifold; these states cannot be trapped in a magnetic trap. Thus, we developed a far-off-resonance optical dipole trap (FORT) to confine the required spin states for the crossover experiments. Such a trap can be formed at the focus of a gaussian laser beam whose optical frequency is far detuned from the $^{40}$K transitions (Fig. 4.2).

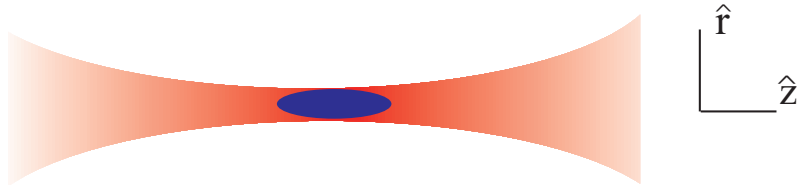

Figure 4.2: Optical trap at the focus of a gaussian laser beam.

We found that the best way to realize a cold gas in an optical dipole trap was to load a relatively hot gas of fermions after some evaporation in the magnetic trap. Typically we load the optical trap when the sample in the magnetic trap reaches $T/T_F \approx 3$. After transfer to the $m_f{=}{-}9/2, -7/2$ spin states the gas is evaporated in the optical trap simply by lowering the depth of the trap and allowing the hottest atoms to spill out. This evaporation and the subsequent temperature measurements are typically performed at a magnetic field of $\sim$235 G. At this field the scattering length is the background (nonresonant) value of 174 $a_0$, and from this field Feshbach resonances can be easily accessed.[1]

Evaporative cooling in the optical trap actually allowed us to achieve colder temperatures than previous records. We could cool to $T/T_F \approx 0.1$ (and possibly colder) with up to $10^6$ atoms in each spin state. At these temperatures another impediment to cooling fermions appears - Pauli blocking of collisions [126]. In a degenerate Fermi sea the Pauli exclusion principle forbids collisions for which the final state would place fermions in already occupied levels. This results in a suppression of the collision rate compared to the classical expectation and may make evaporative cooling more difficult in an all-fermion system. Many recent experiments have used a bosonic atom as their "second particle" as a possible approach to avoiding this problem [33, 34, 35, 37, 38].

---

[1] The background scattering length is reasonably large compared to many other alkali atoms, resulting in nice evaporative cooling. Yet the interaction it provides still allows us to approximate the gas as an ideal Fermi gas for the purpose of temperature measurement.

An important question is why colder temperatures could be achieved in the optical trap than had been previously achieved with the same apparatus in the magnetic trap. There are two main possibilities for the answer to this question. First, evaporation in a magnetic trap relies on the ability to drive spin-changing transitions with rf or microwaves. At cold temperatures the microwave frequencies required for the $m_f$=+9/2 and the $m_f$=+7/2 states differ, requiring the use of a more complicated two-frequency evaporation scheme [83]. The optical trap on the other hand evaporates both states equally, maintaining the optimum equal mixture automatically. Second, the history of the cooling attempts with the apparatus have shown that, not surprisingly, a low heating rate is essential to achieving a degenerate Fermi gas. We have found that the coldest achievable temperatures have been proportional to the heating rate in the trap. The heating rate in our optical trap is typically 5 nK/sec in weak traps (after solving the technical issues discussed in Ch. 10). For a similar atom number in the magnetic trap the heating rate is 20 nK/sec [83].

## 4.2 Measuring the temperature of a Fermi gas

In atomic gas experiments the standard technique for measuring the temperature of a gas is to suddenly turn off the trapping potential confining the atoms and then wait for an "expansion time" $t$. A two-dimensional image of the atomic distribution then reveals the velocity, or momentum, distribution of the gas [20]. This technique is often referred to as time-of-flight imaging. For a classical gas the velocity spread is directly related to the temperature of the gas through the Maxwell-Boltzmann distribution. In the quantum regime measuring temperature from such distributions becomes trickier. In this section I will describe techniques for measuring the temperature of an ideal Fermi gas using images of gas distributions.

### 4.2.1 An optically trapped Fermi gas

Before presenting temperature measurements, it is useful to describe the expected distribution of an optically trapped Fermi gas. First, we analyze the optical dipole potential created from a single gaussian laser beam using a harmonic approximation; this is the potential used for most of the BCS-BEC crossover experiments presented in this thesis. Then we present the most relevant formulas for describing the spatial and momentum distributions of a Fermi gas in such a harmonic trap. A more thorough description of a harmonically trapped Fermi gas can be found in Ch. 5.2 of Ref. [83].

#### Optical dipole trap

An optical dipole trap results from the interaction between a light field and the oscillating atomic electric dipole moment that is induced by the light field. This effect, known as the ac Stark shift, forms the conservative part of the interaction of atoms with light. The shift is proportional to the light intensity, and the sign depends upon the sign of the detuning of the light compared to the atomic resonance. When the light is detuned red of the atomic transition and the light intensity varies in space, an attractive potential well can be formed. While the photon scattering rate decreases quadratically with the frequency detuning between the light and the atomic transition, the ac Stark shift varies linearly with this detuning. Thus, a trap with a large detuning results in a significant dipole potential at a low photon scattering rate. This allows use of the dipole potential for atom traps with long storage times, as was first realized in Ref. [127].

The simplest optical dipole trap is formed by the focus of a single red-detuned gaussian laser beam (Fig. 4.2). The trap potential from the beam is proportional to the laser intensity, which is given by

$$I(r,z) = \frac{I_{pk}}{1 + (z/z_R)^2} e^{\frac{-2r^2/\mathrm{w}^2}{1+(z/z_r)^2}} \qquad (4.1)$$

where w is the beam waist ($1/e^2$ radius), $z_R = \pi \mathrm{w}^2/\lambda$ is the Rayleigh range, $I_{pk} = \frac{2p}{\pi \mathrm{w}^2}$ is the peak intensity, and $p$ is the total optical power.

Calculating the strength of the potential from atomic properties is straightfor-ward [128]. Let $\lambda$ be the laser wavelength and $\lambda_0$ be the wavelength of the atomic transition; correspondingly $\omega = 2\pi c/\lambda$ and $\omega_0 = 2\pi c/\lambda_0$. The dipole potential is then given by

$$U(r,z) = -\frac{3\pi c^2 \Gamma}{2\omega_0^3}\left(\frac{1}{\omega_0 - \omega} + \frac{1}{\omega_0 + \omega}\right) I(r,z) \qquad (4.2)$$

where $\Gamma$ is the linewidth of the atomic transition.

For our potassium trap $\lambda = 1064$ nm, $\Gamma = 2\pi \times 6.09$ MHz, and typically w=15 to 20 $\mu$m. Equation 4.2 is in principle only valid for a two-level system, while potassium is of course multi-level, with the most relevant structure being the fine structure giving the $D_1$ and $D_2$ lines at 770.1 nm and 766.7 nm, respectively. However, in this case where the fine structure splitting is small compared to $\omega_0 - \omega$ we can approximate the system as two-level with $\lambda_0$ given by the center of the two lines, 768.4 nm [128]. Note though that we do not make the rotating wave approximation (RWA) in the Eqn. 4.2; for our detuning the $\frac{1}{\omega_0 + \omega}$ term actually contributes a 14% effect.

Figure 4.3 shows cross-sections of the potential for a typical trap that would hold a degenerate Fermi gas in our system. In Eqn. 4.2 we have ignored the effect of gravity on the trap. Yet when the trap becomes shallow, gravity can play a large role. The y-direction in Fig. 4.3 shows the gravitational potential added to Eqn. 4.2. Gravity effectively lowers the trap depth in the y-direction, making evaporation in shallow traps one-dimensional.

At the bottom of the trap we can approximate the gaussian potential as a parabolic potential (dotted line in Fig. 4.3). In this case we have a harmonic trap with oscillator frequencies given by

$$\omega_r = 2\pi\nu_r = \sqrt{\frac{4U_0}{m\mathrm{w}^2}} \qquad \omega_z = 2\pi\nu_z = \sqrt{\frac{4U_0}{mz_R^2}}. \qquad (4.3)$$

where $U_0 = |U(0,0)|$. In shallow traps the harmonic approximation breaks down since atoms sample the potential near the top of the well where the gaussian rises less steeply than the parabola. We account for this effect in experiments by modelling the real trap as a harmonic trap with a lower effective frequency. We determine the effective frequency experimentally through observation of the

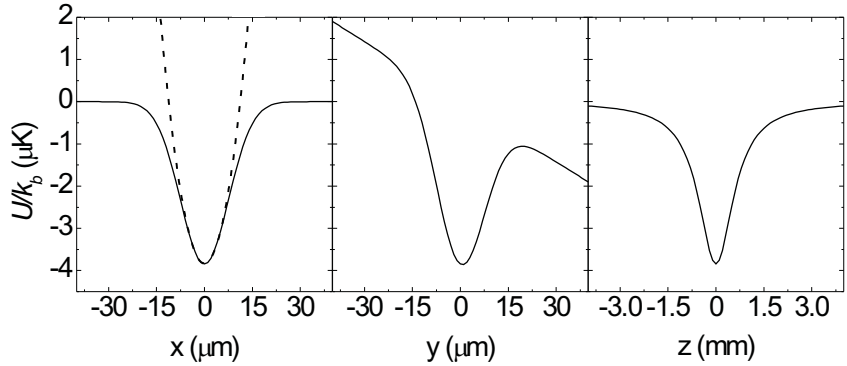

Figure 4.3: Potential for potassium atoms created by a single 1064 nm gaussian laser beam with w=15 $\mu$m and a power of 10 mW. The dotted line in the left-most graph shows the parabola that approximates the potential for small $x$. The y-direction includes the effect of gravity. Note that the horizontal scale in the third graph (z-direction) is a factor of 100 larger than the scale in the first two graphs.

Fermi gas excitation frequencies in the weakly interacting regime.

**Trapped, ideal Fermi gas distribution**

We now consider a Fermi gas in a harmonic potential. We start with the Hamiltonian for a particle in a harmonic well

$$H = \frac{p^2}{2m} + \frac{1}{2}m\omega_r^2\rho^2 \qquad (4.4)$$

where $p^2 = p_x^2 + p_y^2 + p_z^2$ and $\rho^2 = x^2 + y^2 + \lambda^2 z^2$, ($\lambda = \omega_z/\omega_r$). The density of states is $g(\epsilon) = \frac{\epsilon^2}{2(\hbar\bar{\omega})^3}$ where $\bar{\omega} = (\omega_r^2\omega_z)^{1/3}$ [101]. The statistics of the Fermi gas are described by the Fermi-Dirac distribution function

$$f(\epsilon) = \frac{1}{e^{\frac{\epsilon}{k_b T}}/\zeta + 1} \qquad (4.5)$$

where $\zeta = e^{\mu/k_b T}$ is the fugacity.

We can calculate the Fermi energy, which is defined as the energy of the highest occupied level of the potential at $T = 0$. We simply equate the integral over all states up to $E_F$ to the number of particles in one fermion spin state, $N$:

$$N = \int_0^{E_F} g(\epsilon) d\epsilon. \tag{4.6}$$

The result is

$$T_F = \frac{E_F}{k_b} = \frac{\hbar \bar{\omega}}{k_b} (6N)^{1/3}. \tag{4.7}$$

The temperature compared to $T_F$ describes the degeneracy of the Fermi gas and in the classical limit is related to peak phase space density through $PSD_{pk} = (T/T_F)^{-3}/6$.

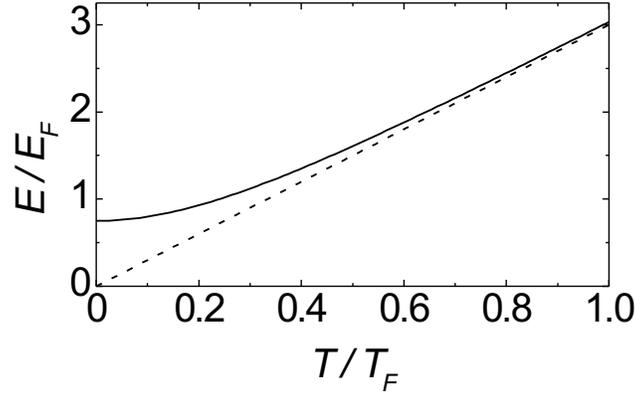

Figure 4.4: Energy of an ideal Fermi gas.

Also using the distribution function and the density of states we can obtain thermodynamic quantities such as the energy per particle, which will become important in Ch. 9.

$$E = \frac{U}{N} = \frac{\int_0^\infty \epsilon g(\epsilon) f(\epsilon) d\epsilon}{\int_0^\infty g(\epsilon) f(\epsilon) d\epsilon} = 3k_b T \frac{Li_4(-\zeta)}{Li_3(-\zeta)} \tag{4.8}$$

The function $Li_n(x) = \sum_{k=1}^\infty x^k/k^n$ appears often in the analysis of a harmonically trapped Fermi gas. $Li_n$ is the Poly-Logarithmic function of order $n$, sometimes written $g_n$. Figure 4.4 plots the result of Eqn. 4.8. In the classical regime the energy is proportional to the temperature, while in the Fermi gas limit the energy asymptotes to $\frac{3}{4}E_F$ ($\frac{3}{8}E_F$ kinetic energy and $\frac{3}{8}E_F$ potential energy).

Table 4.1: Distribution functions for a harmonically trapped Fermi gas.

| Validity | Spatial distribution where $\sigma_r^2 = \frac{k_b T}{m \omega_r^2}$ and $r_F^2 = \frac{2 E_F}{m \omega_r^2}$ |
|---|---|
| all $T/T_F$ | $n(\rho) = \frac{\lambda N}{(2\pi)^{3/2}\sigma_r^3} Li_{3/2}(-\zeta e^{-\rho^2/2\sigma_r^2})/Li_3(-\zeta)$ |
| $T/T_F \gg 1$ | $n_c(\rho) = \frac{\lambda N}{(2\pi)^{3/2}\sigma_r^3} e^{-\rho^2/2\sigma_r^2}$ |
| $T = 0$ | $n_0(\rho) = \frac{\lambda N}{r_F^3} \frac{8}{\pi^2} \left(1 - \frac{\rho^2}{r_F^2}\right)^{3/2}$ for $\rho < r_F$, 0 otherwise |
| Validity | Momentum distribution where $\sigma_p^2 = m k_b T$ and $p_F^2 = 2 m E_F$ |
| all $T/T_F$ | $\Pi(p) = \frac{N}{(2\pi)^{3/2}\sigma_p^3} Li_{3/2}(-\zeta e^{-p^2/2\sigma_p^2})/Li_3(-\zeta)$ |
| $T/T_F \gg 1$ | $\Pi_c(p) = \frac{N}{(2\pi)^{3/2}\sigma_p^3} e^{-p^2/2\sigma_p^2}$ |
| $T = 0$ | $\Pi_0(p) = \frac{N}{p_F^3} \frac{8}{\pi^2} \left(1 - \frac{p^2}{p_F^2}\right)^{3/2}$ for $p < p_F$, 0 otherwise |

Distribution functions in position and momentum can be determined through standard statistical mechanics techniques and the Thomas-Fermi approximation, which holds when many oscillator states are occupied (see Refs. [129, 83]). Table 4.1 shows the resulting Fermi-Dirac distribution functions in position and momentum space. The full expression is shown first, but it is also useful to understand the classical and $T = 0$ limits. In the classical limit the distribution is gaussian, and at $T = 0$ the distributions are only nonzero for values less than the Fermi radius $r_F$ or Fermi momentum $p_F = \hbar k_F$. It is useful to explicitly state the meaning of the Fermi wavevector, $k_F$, as it is a quantity we normalize to in the BCS-BEC crossover. $k_F$ is defined through $E_F = \frac{\hbar^2 k_F^2}{2m}$ and is related to the inverse distance between fermions at the center of the trap; at $T = 0$ the number density of one fermion spin state at the center of the trap is written in terms of $k_F$ as $n_{pk} = \frac{k_F^3}{6\pi^2}$.

## 4.2.2 Measuring temperature from the momentum distribution

In the experiment we access the distribution of the Fermi gas through absorption images of an expanded gas. Absorption images are aquired by illuminating the atoms with a resonant laser beam and imaging the shadow cast by the atoms onto a CCD camera. These images effectively integrate through one dimension to give a two-dimensional image (for example Fig. 4.6(a)). The appropriate function for this distribution is $n(\rho)$ or $\Pi(p)$ (see Table 4.1) integrated over one dimension [83]. Integrating through $x$ and writing the result in terms of the experimentally observed optical depth ($OD$), we find for the Fermi-Dirac distribution

$$OD_{FD}(y, z) = OD_{pk} \, Li_2(-\zeta e^{-\frac{y^2}{2\sigma_y^2}} e^{-\frac{z^2}{2\sigma_z^2}})/Li_2(-\zeta). \qquad (4.9)$$

In the classical limit this equation becomes a two-dimensional gaussian function:

$$OD_{gauss}(y, z) = OD_{pk} \, e^{-\frac{y^2}{2\sigma_y^2}} e^{-\frac{z^2}{2\sigma_z^2}}. \qquad (4.10)$$

These forms are applicable for both the spatial and momentum profiles, and for arbitrary expansion times through the relations $\sigma_y^2 = \frac{k_b T}{m\omega_r^2}[1 + (\omega_r t)^2]$ and $\sigma_z^2 = \frac{k_b T}{m\omega_z^2}[1 + (\omega_z t)^2]$, where $t$ is the expansion time.

Examples of theoretical cross sections of $OD_{FD}$ for various degeneracies, and constant $T_F$, are shown in Fig. 4.5. As the gas becomes colder the distribution becomes narrower and the shape of the distribution becomes flatter. Note that as the temperature is lowered far below $T_F$ the changes in the distribution become small compared to $T_F$. Still, down to $T/T_F \approx 0.1$ the temperature can be determined from least-squared fits to such distributions. Figure 4.6(a) is a sample absorption image of an expanded Fermi gas. The black points in Fig. 4.6(b) are the result of an azimuthal average of the image. The red line shows the result of a surface fit of the two-dimensional image to the Fermi-Dirac distribution (Eqn. 4.9), which reveals that the gas is at a temperature of 0.1 $T_F$. For comparison the blue line is the result of a fit appropriate for a classical distribution (Eqn. 4.10). Clearly the experimental distribution is flatter than a classical distribution.

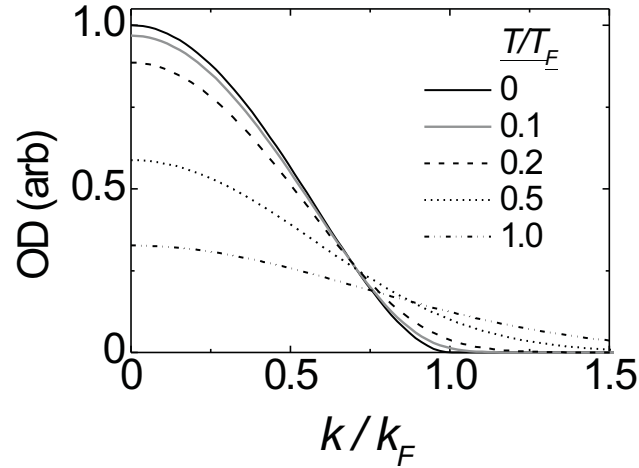

Figure 4.5: Theoretical cross sections of a harmonically trapped Fermi gas distribution integrated along one dimension. $T_F$ is held fixed as $T/T_F$ varies.

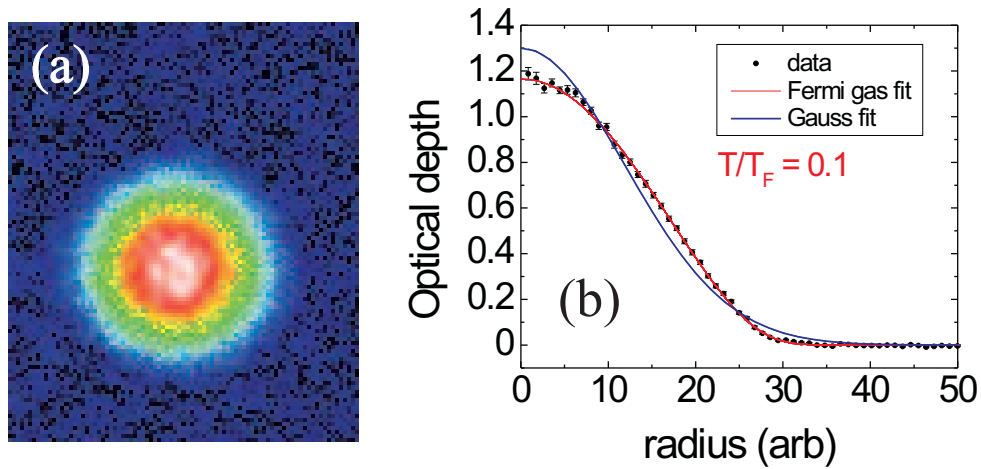

Figure 4.6: Nonclassical momentum distribution of Fermi gas. (a) Sample false color absorption image of the momentum distribution of a degenerate Fermi gas. Here the integration is along the $z$-direction. (b) Azimuthally averaged profile of the absorption image. The error bars represent the standard deviation of the mean of averaged points.

To evaluate this thermometry, we can examine the results of least-squared surface fits for gases at a variety of expected temperatures. In the fits $OD_{pk}$, $\sigma_y$, $\sigma_z$, and $\zeta$ are independent fit parameters. $\sigma_y$ and $\sigma_z$ tell us the temperature; $\zeta$ can be viewed as a shape parameter that is directly related to $T/T_F$ through $Li_3(-\zeta) = -(T/T_F)^{-3}/6$. As a check on the fits we compare the result for $\zeta$ to $T/T_F$ calculated through

$$\frac{T}{T_F} = \frac{\sigma_y^2 m \omega_r^2}{\hbar \bar{\omega}(6N)^{1/3}(1 + (\omega_r t)^2)}. \tag{4.11}$$

We use the measured trap frequencies for $\omega_r$ and $\omega_z$ and the number of atoms in each spin state $N$ calculated from the total absorption in the image. For most purposes we can extract $N$ from a gaussian fit (Eqn. 4.10), in which case the total number of atoms in the image is $\frac{2\pi OD_{pk}\sigma_r\sigma_z}{\sigma}$, where $\sigma$ is the photon absorption cross-section, $\frac{3\lambda_0^2}{2\pi}$. This result is exact for a classical gas and is only 7% off at $T/T_F = 0.1$.

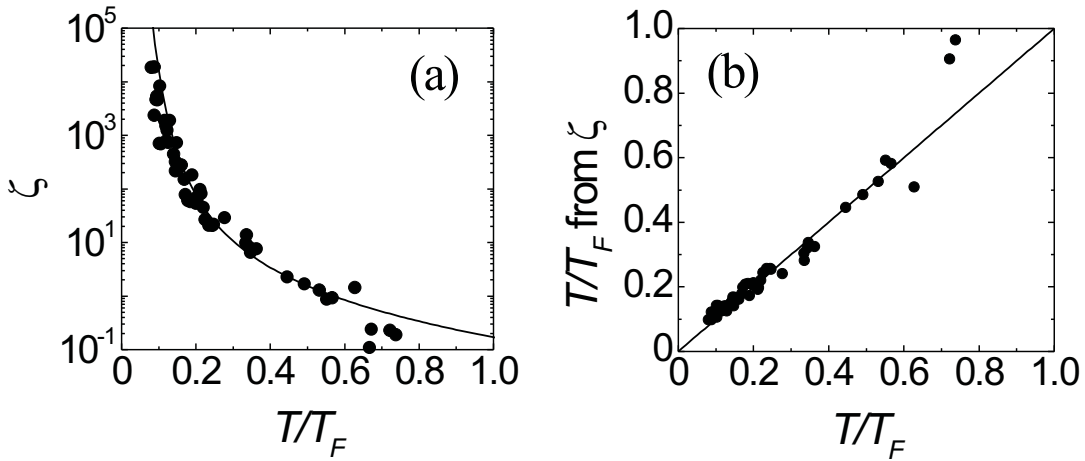

Figure 4.7: Analysis of fits to Eqn. 4.9 for expansion images of an optically trapped gas with an equal mixture of $m_f = -9/2, -7/2$ atoms [59]. For these data the integration was in the $x$-direction, and $T$ was extracted from $\sigma_y$.

Figure 4.7(a) shows a comparison between $T/T_F$ from Eqn. 4.11 and $\zeta$. The black line shows the expected relationship for an ideal Fermi gas. In Fig. 4.7(b) $\zeta$ is converted to $T/T_F$ for a more direct comparison. In general the two values

agree, indicating the fits are extracting the correct information. Note that the noise in the points becomes large at temperatures $> 0.5 \; T_F$. This is expected because the changes in the shape of the distribution become small in this limit. A similar effect occurs in the low temperature limit where the distribution changes little as the $T = 0$ Fermi gas limit is approached. However, the success of this thermometer in the $0.1 < T/T_F < 0.5$ range has made this method the workhorse of temperature measurements in the Jin lab since Brian DeMarco's work.

### 4.2.3 Measuring temperature using an impurity spin state

A second technique for measuring temperature that we have explored is impurity spin-state thermometry. Eric Cornell proposed this idea in the JILA hallway as a method to check the Fermi-Dirac surface-fit technique outlined in the previous section. A check is especially necessary for the coldest temperature gases at $0.1 \; T_F$ and below, due to the decrease in the sensitivity of the Fermi-Dirac fits at these temperatures. We have not done extensive work using impurity thermometer. However, as we will see here this thermometer works quite well and has great unexplored potential, in particular as a technique that could measure temperatures less than $0.1 \; T_F$.

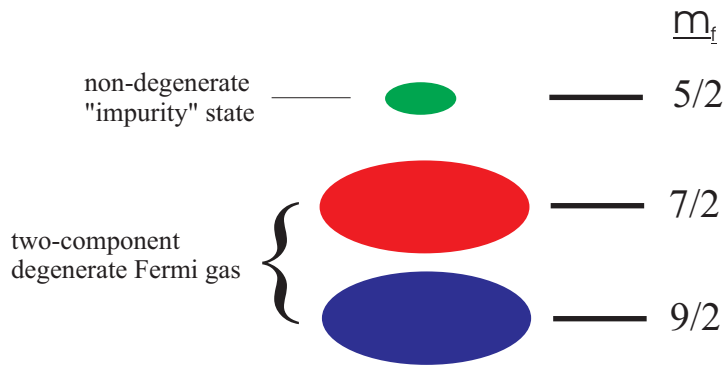

Figure 4.8: Measurement of $T$ through an embedded impurity spin state. All three components are overlapped in the optical trap.

The idea of the impurity spin-state technique is to embed a small number of atoms in a third state within the usual two-component gas (Fig. 4.8). In the limit where the number of atoms in the impurity spin, $N_{im}$, is small compared to the particle number in the original states, the Fermi energy of the impurity state will be low enough that the impurity gas will be nondegenerate. Provided all of the spin states in the system are in thermal equilibrium, the temperature of the system will be $\frac{m\sigma_{im}^2}{k_b t^2}$, where $\sigma_{im}$ is simply determined from a fit of the impurity gas momentum distribution to a gaussian distribution (Eqn. 4.10).

A difficulty with this method is that $E_F$ scales weakly with particle number. Suppose we originally have a gas with a particle number of $10^5$ at $T/T_F = 0.1$. For $T/T_F$ of the impurity to be 1, $N_{im}$ would need to be 100, and detecting the distribution of 100 atoms with good signal to noise is not trivial. However, the fully classical limit does not need to be reached to gain information about the temperature from the impurity state. It is only required that the $T/T_F$ be large enough that the energy of the impurity gas changes significantly with temperature. Figure 4.4 illustrates for the range of $T/T_F$ for which this is the case.

To see if an impurity spin-state thermometer was feasible we designed an experiment to test this thermometer against the surface-fit technique described previously. We started with a (not necessarily equal) mixture of atoms in the $m_f$=+7/2, +9/2 spin states. Part way through the evaporative cooling process a small fraction of the $m_f$=+7/2 atoms were transferred to the $m_f$=+5/2 state, which served as our impurity (Fig. 4.8). For this experiment the gas was prepared at a low magnetic field of a few gauss where the three-state mixture is fully stable. Here the scattering length between any pair of the three spin states is 174 $a_0$. The spin states were selectively imaged by applying a large magnetic field gradient of ∼80 G/cm during the expansion to spatially separate the spin states (Stern-Gerlach imaging) [83].

For analysis of the impurity spin-state data in the most general case where the impurity is not fully classical, it is useful to introduce the variable $T_{gauss}$. For a momentum space distribution $T_{gauss}$ is defined as $\frac{m\sigma_{gauss}^2}{k_b t^2}$, where $\sigma_{gauss}$ is the result from a least-squared fit to the gaussian distribution of Eqn. 4.10. As we noted in Fig. 4.6 the Fermi distributions is not well fit by a gaussian, but the result is a well-defined quantity. Figure 4.9 displays $T_{gauss}/T_F$ as a function of degeneracy, as determined through least-squared fits to theoretical distributions. $T_{gauss}$ provides much the same information as the Fermi gas energy $E$ (Fig. 4.4). However, it is more useful for our current purposes because it can be extracted from real images with better signal to noise than $E$, and it is convenient because in the classical limit it becomes precisely the real temperature of the system.

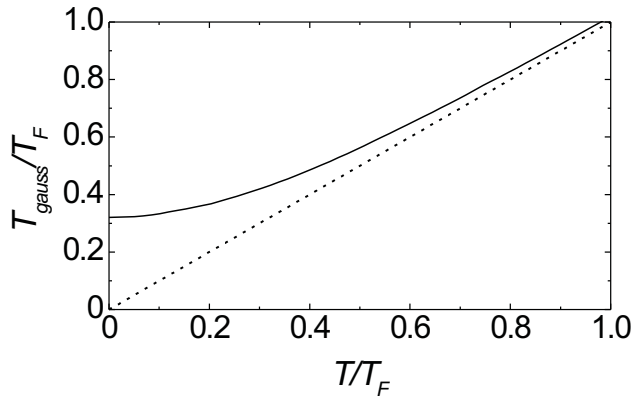

Figure 4.9: Dimensionless plot of the variable $T_{gauss}$ (defined in the text) versus temperature.

In Fig. 4.10 is $T_{gauss}$ as a function of $N$ for a cold gas of atoms distributed among the three states $m_f$=9/2, 7/2, and 5/2. $N$ is the measured number of atoms in the spin state from which $T_{gauss}$ is extracted. The seven sets of points come from multiple iterations of the experiment for which the temperature was expected to be constant. Since the trap strength was held constant $N$ uniquely defines $E_F$, and given that the temperature $T$ is constant for all points on the plot, we can translate the dimensionless theoretical result of Fig. 4.9 to the situation of Fig. 4.10. The black line in Fig. 4.10 shows the best-fit curve to the data, in which the only free parameter is the real temperature $T$. Note $T$ is the value of $T_{gauss}$ as $N$ goes to zero. If the gas were fully classical, $T_{gauss}$ would be constant as a function of $N$; it is a manifestation of Pauli blocking that different components of an equilibrium gas can have the same temperature but different energies [126].

Figure 4.11 shows the results of four experiments like the one shown in Fig. 4.10. The temperature result from the impurity measurement is compared to the result from the surface fits described in Sec. 9.1 (applied to the $m_f$=9/2 cloud). We see that both methods agree to within the uncertainty for clouds in the $T/T_F = 0.1 - 0.2$ range.

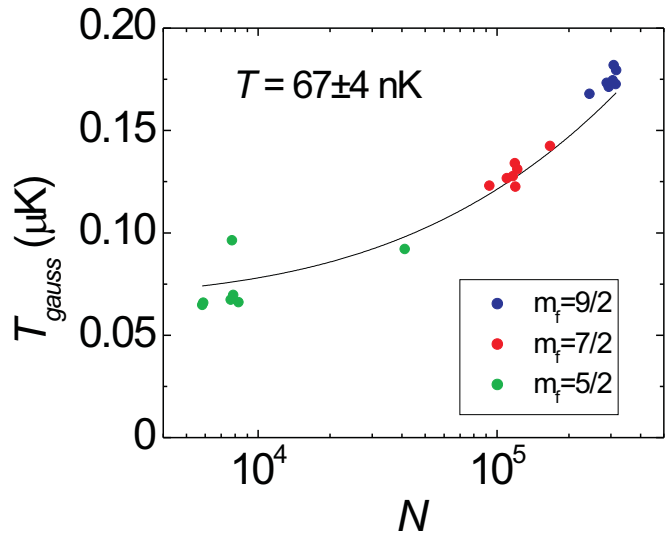

Figure 4.10: Impurity spin state thermometry. $T_{gauss}$ is plotted for all of the states in the gas, for multiple experiment iterations in which $T$ was held fixed. The $m_f$=9/2 points have an average degeneracy of $T/T_F = 0.13$.

### 4.2.4 Temperature in the BCS-BEC crossover

Our focus thus far has been on measuring the temperature of an ideal, normal Fermi gas. For the BCS-BEC crossover experiments we would like to know the temperature of our gas at any interaction strength. To this point there have been no experiments that have directly measured the temperature of a gas throughout the entire crossover. The main technique that has been used thus far is to instead measure the entropy of the system, which, through theory, can be translated to a temperature at any point in the crossover. The entropy, $S$, can be determined through temperature measurements in the weakly interacting regime. For example, in the Fermi gas limit the entropy is given by $S = (\frac{4}{3}E - \mu)/k_b T$ [130]. Then if experiments are performed using adiabatic magnetic-field ramps to the crossover regime, the entropy will be held constant at the weakly-interacting value. As we have seen in $^{40}$K it is convenient to measure the temperature of the weakly interacting gas above the Feshbach resonance with a Fermi gas, while in most of the $^6$Li experiments the temperature is measured through the condensate fraction close to the BEC limit. In either case, the limitation of this technique is that it relies upon theory to convert between entropy and temperature in the crossover. Some initial work on this theoretical problem can be found in Ref. [131].

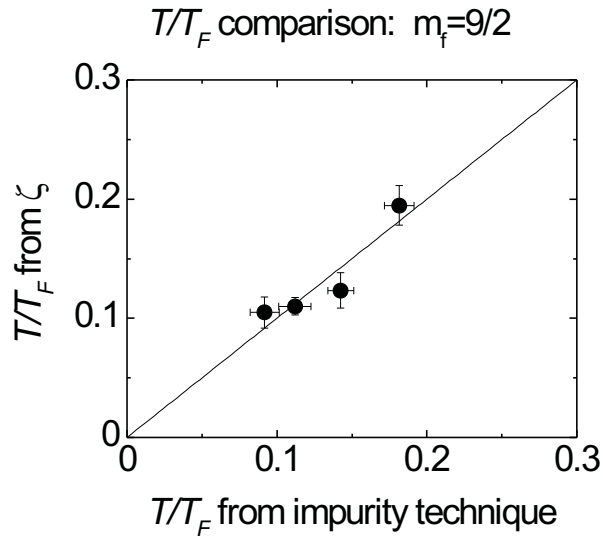

Figure 4.11: Comparison of thermometers. The *y*-axis shows the result of the Fermi-Dirac fits to the $m_f$ =9/2 distribution, and the *x*-axis shows the result of the impurity spin-state technique.

An alternative temperature measurement in the crossover was applied in the group of J. Thomas where they measured the temperature at unitarity by using fits to the momentum distribution in the hydrodynamic limit [79]; here the distribution is simply a rescaled Fermi distribution and thus amenable to the same fitting procedure as the weakly interacting regime [132]. The impurity spin-state thermometer described in Sec. 4.2.3 is another possible, but as of yet unexplored, direct thermometer in the crossover regime. Since the $m_f$=−5/2 state is weakly interacting with $m_f$=−9/2, −7/2 states even at the $m_f$=−9/2, −7/2 Feshbach resonance, the $m_f$=−5/2 distribution could be used to measure $T$ on resonance. However, at high magnetic field values the $m_f$ = −5/2, −9/2 mixture is only metastable; experimentally we have found this mixture has an exponential decay time of ∼1 sec at ∼200 G for a typical density of ∼$10^{13}$ cm$^{-3}$. This short lifetime makes a possible measurement more difficult, but not impossible.

# Chapter 5

# Elastic scattering near Feshbach resonances between fermionic atoms

Some of the first signatures of the presence of fermionic Feshbach resonances were the observation of magnetic-field dependent changes in the elastic-scattering properties of a normal Fermi gas [54, 58, 55, 57, 59, 60, 61, 62]. Here I present three different techniques that we used to probe changes in scattering properties at a fermionic Feshbach resonance. Two techniques probed the collision cross section, which reveals the magnitude of $a$. In the third measurement we observed the sign of $a$ change and saw evidence for unitarity-limited interactions.

## 5.1   Measuring the elastic collision cross section

The first Feshbach resonance we searched for experimentally is the resonance described in Ch. 3 that occurs between the $m_f=-9/2$ and $m_f=-7/2$ spin states. The original theoretical prediction for the location of this resonance was $B_0 = 196^{+9}_{-21}$ G, based on available potassium potentials [53]. We first experimentally measured the position of this resonance using the technique of cross-dimensional rethermalization, which measures the collision cross section [133]. This was a technique that had provided much information about a Feshbach resonance in bosonic $^{85}$Rb gas [30, 134].

For this measurement we started with a gas of fermions in the $m_f = -7/2$, $-9/2$ spin states at $T \approx 2\,T_F$. The gas was taken out of thermal equilibrium by modulating the optical trap intensity at $2\,\nu_y$, which caused selective heating in the $y$-direction. (We could selectively modulate one radial direction because for this measurement our optical trap was not cylindrically symmetric, $\nu_x = 1.7$ $\nu_y$.) The exponential time constant for energy transfer between the two radial directions, $\tau$, was measured as a function of magnetic field. $\tau$ is related to the $s$-wave collision cross section through $1/\tau = 2\langle n\rangle\sigma v/\alpha$. $v = 4\sqrt{k_b T/\pi m}$ is the mean relative speed between colliding fermions and $\langle n\rangle = \frac{1}{N_{tot}}\int n_7(\mathbf{r})\,n_9(\mathbf{r})\,d^3\mathbf{r}$ is the density weighted density. $\alpha$ is the calculated number of binary $s$-wave collisions required for rethermalization [125].

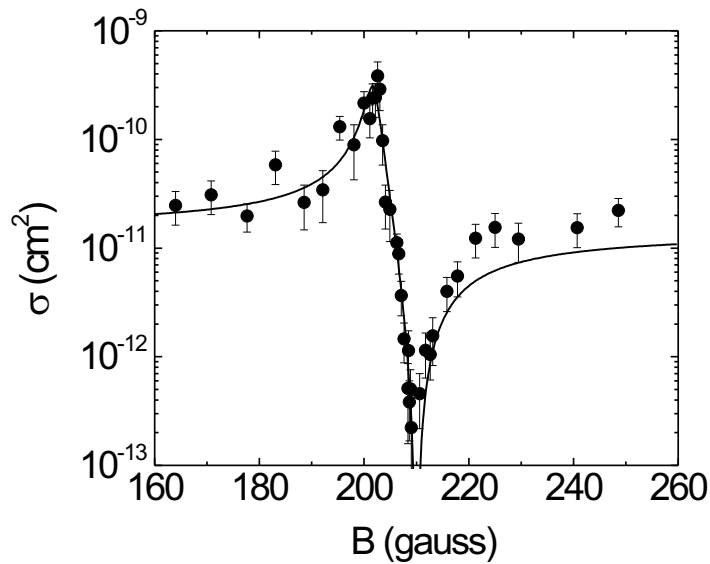

Figure 5.1: Collision cross section measured near an $s$-wave Feshbach resonance between $^{40}$K atoms in the $m_f = -7/2, -9/2$ spin states at $T = 4.4\ \mu$K [57]. In between the peak and dip in $\sigma$ the interaction is attractive; everywhere else it is repulsive.

Figure 5.1 plots the result of this measurement as a function of the magnetic field, $B$. The magnetic field was calibrated through radio-frequency (rf) transitions between $m_f$ levels in the $^{40}$K system. An advantage of the cross-dimensional rethermalization technique is that it allows measurements of $\sigma$ over a large range. Through cross section measurements that extend over four orders of magnitude, both the position of the divergence of the scattering length, $B_0$, and the position of the zero crossing could be measured (Fig. 5.1). This told us the magnetic-field width of the resonance $w$, which as we saw in Ch. 3 describes the coupling strength.

The line in Fig. 5.1 is the result of a full coupled channels calculation of $\sigma$ carried out by C. Ticknor and J. Bohn, in which the parameters of the potassium potential were adjusted to achieve a best fit to our data from two different $^{40}$K resonances [57, 135]. This calculation took into account the distribution of collision energies in the gas by thermally averaging over a gaussian distribution defined by a temperature of 4.4 $\mu$K. The fit result placed the Feshbach resonance at $B_0 = 201.6 \pm 0.6$ G and the zero crossing at $209.9 \pm 0.6$ G.

## 5.2 Anisotropic expansion

A disadvantage of the cross-dimensional rethermalization method is that it only provides a valid measurement of $\sigma$ in the so-called collisionless regime. A trapped gas is considered collisionless if the trap oscillator period $1/\nu$ is much shorter than the mean time between collisions in the gas, $1/\Gamma$. In the opposite limit where $\Gamma \gg \nu$ the gas is collisionally hydrodynamic, obeying the classic hydrodynamic equations. If a gas is hydrodynamic the cross-dimensional rethermalization time will be determined by the oscillator period $1/\nu$ instead of the mean time between collisions.

At the peak of the Feshbach resonance the fermion-fermion interactions can easily become strong enough to make the gas collisionally hydrodynamic. In this regime a technique more suited to measuring changes in the elastic cross section is anisotropic expansion. In the hydrodynamic limit collisions during the expansion transfer kinetic energy from the elongated axial ($z$) gas dimension into the radial ($r$) direction. This changes the aspect ratio of the expanded gas ($\sigma_z/\sigma_r$) compared to the collisionless expectation (see Ch. 4). This effect was first observed in a $^6$Li Fermi gas in Ref. [58].

Figure 5.2 presents a measurement of anisotropic expansion in a $^{40}$K gas at $T/T_F = 0.34$. Here we enhanced interactions between the $m_f = -9/2, -5/2$ spin states using a Feshbach resonance between these states at ∼224 G. At the position of the Feshbach resonance where $\sigma$ is large, the aspect ratio $\sigma_z/\sigma_y$ decreases. As the gas becomes collisionless away from the resonance the aspect ratio smoothly evolves to the collisionless value. The key to observing anisotropic expansion is to hold the magnetic field at the value near the Feshbach resonance during the beginning of expansion. In these experiments the magnetic field remained high for 5 ms of expansion and a resonant absorption image was taken after a total expansion time of 20 ms. In this measurement the $m_f = -9/2, -5/2$ gas was created at the field $B$ by starting with a $m_f = -9/2, -7/2$ gas and applying a $\pi$ pulse between the -5/2 and -7/2 states 0.3 ms before expansion. This technique avoids complications due to atom loss but creates a nonequilibrium gas.

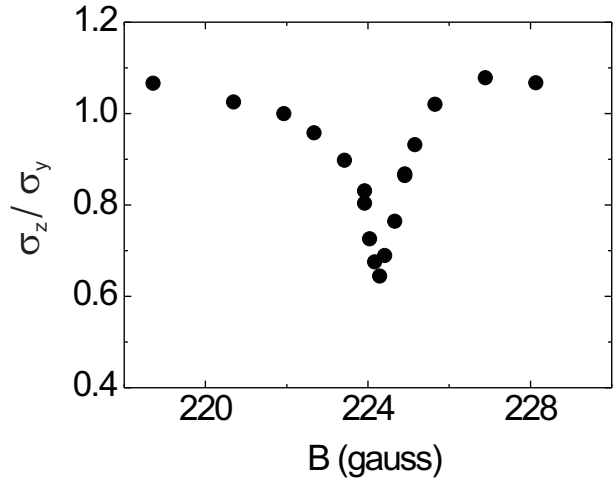

Figure 5.2: Anisotropic expansion of a strongly interacting Fermi gas [59].

In general the expected aspect ratio in the regime between collisionless and hydrodynamic behavior is difficult to calculate. We can however check to see if some degree of hydrodynamic expansion of the normal gas is expected. We can calculate the elastic collision rate $\Gamma = 2\langle n \rangle \sigma v$ in the gas, using an elastic collision cross section given by $\sigma = 4\pi a_{59}^2$ and $|a_{59}| = 2000\ a_0$ (as was measured near the resonance peak [59]) to find $\Gamma = 46$ kHz. Comparing this rate to the trapping frequencies we find $\Gamma/\nu_r = 37$ and $\Gamma/\nu_z = 2400$. Hence, it is not surprising that we observe anisotropic expansion. For a gas that was fully hydrodynamic, with $\Gamma \gg \nu_r, \nu_z$, we would expect our measured aspect ratio to reach 0.4 [136, 137].

In ultracold gas experiments anisotropic expansion is often associated with superfluidity of a BEC. This is because a typical ultracold Bose gas is collisionless, while below the superfluid transition the gas obeys superfluid hydrodynamic equations. Near a Feshbach resonance however where the gas can be collisionally hydrodynamic, anisotropic expansion can be observed both above and below the superfluid transition temperature. Still anisotropic expansion has been put forth as a possible signature of superfluidity in the BCS-BEC crossover regime [136, 137]. In pursuing this signature it is important to carefully distinguish between collisional and superfluid hydrodynamics and take into account that changes during expansion will affect the many-body state [138]. Such analysis is possible and has been considered in Refs. [58, 71].

## 5.3 Measuring the mean-field interaction

Measurements of the collision cross section are useful for detecting the strength of the interaction but are not sensitive to whether the interactions are attractive or repulsive. The mean-field energy, on the other hand, is a quantum mechanical, many-body effect that is proportional to $na$. For Bose-Einstein condensates with repulsive interactions the mean-field energy (and therefore $a$) can be determined from the size of the trapped condensate [29, 139], while attractive interactions cause condensates with large atom number to become mechanically unstable [140, 141]. For an atomic Fermi gas the mean-field interaction energy has a smaller impact on the thermodynamics. Here I discuss a novel spectroscopic technique that measures the mean-field energy of a two-component Fermi gas directly [59, 60].

In this measurement we again used the Feshbach resonance between the $m_f = -5/2$ and $-9/2$ spin states. At magnetic fields near the resonance peak, we measured the mean-field energy in the Fermi gas using rf spectroscopy (Fig. 5.3(a)). First, optically trapped atoms were evaporatively cooled in a 72/28 mixture of the $m_f = -9/2$ and $m_f = -7/2$ spin states. After the evaporation the optical trap was recompressed to achieve a larger density, and the magnetic field was ramped to the desired value near the resonance. We then quickly turned on the resonant interaction by transferring atoms from the $m_f=-7/2$ state to the $m_f=-5/2$ state with a 73 $\mu$s rf $\pi$-pulse. The fraction of $m_f=-7/2$ atoms remaining after the pulse was measured as a function of the rf frequency.

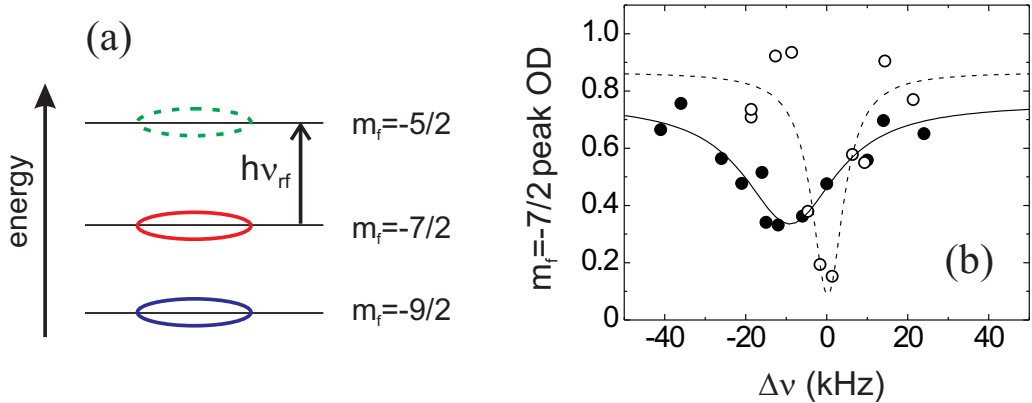

Figure 5.3: Radio-frequency spectra. (a) Transition of interest. (b) Rf lineshapes with (solid line) and without (dotted line) interactions [59].

The relative number of $m_f=-7/2$ atoms was obtained from a resonant absorption image of the gas taken after 1 ms of expansion from the optical trap. Atoms in the $m_f=-7/2$ state were probed selectively by leaving the magnetic field high and taking advantage of nonlinear Zeeman shifts. Sample rf absorption spectra are shown in Fig. 5.3(b). At magnetic fields well away from the Feshbach resonance we could transfer all of the $m_f=-7/2$ atoms to the $m_f=-5/2$ state and the rf lineshape had a Fourier width defined by the rf pulse duration. At the Feshbach resonance there were two changes to the rf spectra. First, the frequency for maximum transfer was shifted relative to the expected value from a magnetic field calibration. Second, the maximum transfer was reduced and the measured lineshape is wider.

Both of these effects arise from the mean-field energy due to strong interactions between $m_f=-9/2, -5/2$ atoms at the Feshbach resonance. The mean-field energy produces a density-dependent frequency shift given by

$$\Delta\nu = \frac{2\hbar}{m} n_9 (a_{59} - a_{79}),\tag{5.1}$$

where $n_9$ is the number density of atoms in the $m_f = -9/2$ state, and $a_{59}$ ($a_{79}$) is the scattering length for collisions between atoms in the $m_f = -9/2$ and $m_f = -5/2$ ($m_f = -7/2$) states [142]. Here the nonresonant interaction term proportional to the population difference between the $m_f = -7/2$ and $m_f = -5/2$ states is ignored; this term equals 0 for a perfect $\pi$-pulse. For our spatially inhomogeneous trapped gas, the density dependence broadens the lineshape and lowers the maximum transfer. This effect occurs on both sides of the Feshbach resonance peak. In contrast, the frequency shift for maximum transfer reflects the scattering length and changes sign across the resonance.

We measured the mean-field shift $\Delta\nu$ as a function of $B$ near the Feshbach resonance peak. The rf frequency for maximum transfer was obtained from Lorentzian fits to spectra like those shown in Fig. 5.3(b). The expected resonance frequency was then subtracted to yield $\Delta\nu$. The scattering length $a_{59}$ was obtained using Eqn. 5.1 with $n_9 = 0.5\, n_{pk}$ and $a_{79} = 174\, a_0$ [54]. The peak density of the trapped $m_f = -9/2$ gas $n_{pk}$ was obtained from gaussian fits to absorption images. The numerical factor 0.5 multiplying $n_{pk}$ was determined by modelling the transfer with a pulse-width limited Lorentzian integrated over a gaussian density profile.

The measured scattering length as a function of $B$ is shown in Fig. 5.4. This plot, which combines data taken for two different gas densities, shows that we were able to realize both large positive and large negative values of $a_{59}$ near the Feshbach resonance. The solid line in Fig. 5.4 shows a fit to the expected form for a Feshbach resonance (Eqn. 3.1). Data within $\pm 0.5$ G of the peak were excluded from the fit. With $a_{bg} = 174\, a_0$ we found that the Feshbach resonance peak occurs at $224.21 \pm 0.05$ G and the resonance has a width $w$ of $9.7 \pm 0.6$ G.

When $B$ is tuned very close to the Feshbach resonance peak, we expect the measured $a_{59}$ to have a maximum value on the order of $1/k_F$ due to the unitarity limit. This saturation can be seen in the data shown in Fig. 5.4. Two points that were taken within $\pm 0.5$ gauss of the Feshbach resonance peak, one on either side of the resonance, clearly lie below the fit curve. The unitarity-limited point on the attractive interaction side of the resonance (higher $B$) has an effective scattering length of $\sim 2/k_F$. (Here $\hbar k_F$ is the Fermi momentum for the $m_f = -9/2$ gas.)

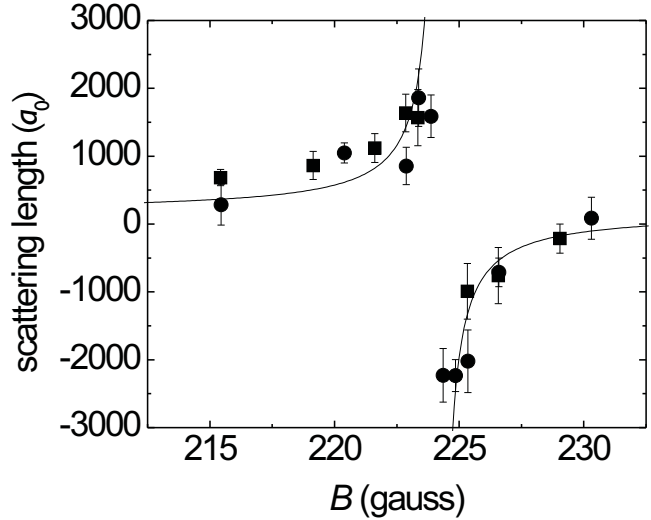

Figure 5.4: Scattering length as measured through the mean-field interaction [59]. These data were taken for a normal Fermi gas at $T/T_F = 0.4$ and at two different densities: $n_{pk} = 1.8 \times 10^{14}$ cm$^{-3}$ (circles) and $n_{pk} = 0.58 \times 10^{14}$ cm$^{-3}$ (squares).

Table 5.1: Observed Feshbach resonances in $^{40}$K.

| open channel $|f, m_f\rangle$ | l | $B_0$ (G) | $w$ (G) | reference |
|---|---|---|---|---|
| $|9/2, -9/2\rangle + |9/2, -7/2\rangle$ | s | $202.10 \pm 0.07$ | $7.8 \pm 0.6$ | [57, 73] |
| $|9/2, -9/2\rangle + |9/2, -5/2\rangle$ | s | $224.21 \pm 0.05$ | $9.7 \pm 0.6$ | [59] |
| $|9/2, -7/2\rangle + |9/2, -5/2\rangle$ | s | $\sim 174$ | $\sim 7$ | unpublished |
| $|9/2, -7/2\rangle$ | p | $\sim 198.8$ | | [57, 143] |

## 5.4 $^{40}$K Feshbach resonance summary

Table 5.1 lists the Feshbach resonances we have studied experimentally in $^{40}$K. All of these resonances were originally located by measuring scattering properties using the techniques described above. I include the states between which the resonance occurs, the partial wave of the resonant collision $l$, our most precise measurement of the resonance position $B_0$, and the resonance width $w$.

# Chapter 6

# Creating molecules from a Fermi gas of atoms

After locating Feshbach resonances in our $^{40}$K system, we wanted to observe evidence of a molecular bound state near threshold on the low-field side of the Feshbach resonances. Creating molecules in this bound state, referred to as "Feshbach molecules," would be the first step towards achieving the BEC limit of the crossover problem. We were motivated to believe that it would be possible to create Feshbach molecules by experiments carried out in the Wieman group at JILA [144, 145]. By pulsing a magnetic field quickly towards a Feshbach resonance they were able to observe coherent oscillations between atoms and Feshbach molecules in a $^{85}$Rb BEC. We hoped to employ a slightly different approach to creating molecules in which we would ramp the magnetic field fully across the Feshbach resonance. In this chapter I will present how, using this technique, we were able to efficiently and reversibly create Feshbach molecules from a Fermi gas of atoms. Another of our contributions to the study of Feshbach molecules was a spectroscopic detection technique that firmly established that Feshbach molecules had been created. I also present our current understanding of the physics of conversion of atoms to molecules using adiabatic magnetic-field ramps; this understanding was gained through a study of the conversion dependences led by the Wieman group [146]. Lastly, I describe how the dissociation of molecules in a sample at low density provided our most precise measurement of the Feshbach resonance position.

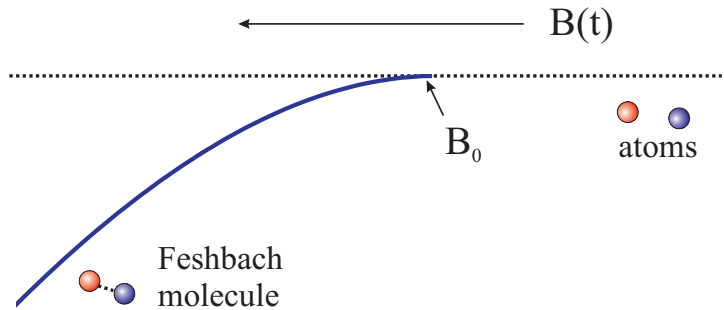

Figure 6.1: Creating molecules using magnetic-field ramps across a Feshbach resonance.

## 6.1 Magnetic-field association

Figure 6.1 shows the behavior of the bound molecular state at a Feshbach resonance presented in Ch. 3. Given this picture, one would expect that atoms could be converted to molecules simply by ramping the magnetic field in time across the Feshbach resonance position $B_0$ [147, 148, 149]. The only requirement to creating molecules in this way is that the magnetic-field ramp must be slow enough to be adiabatic with respect to the two-body physics of the Feshbach resonance (two-body adiabatic). To a very good approximation the Feshbach molecules have twice the polarizability of the atoms [150] and therefore would be confined in the optical dipole trap along with the atoms.[1]

We performed such an experiment using a magnetic-field ramp across the $m_f = -5/2, -9/2$ resonance introduced in the previous chapter. We started with a nearly equal mixture of the two spin states $m_f = -5/2, -9/2$ at a magnetic field of 227.81 G. The field was ramped at a rate of $(40 \ \mu s/G)^{-1}$ across the resonance to various final values. The number of atoms remaining following the ramp was determined from an absorption image of the gas at $\sim 4$ G after expansion from the optical trap. Since the light used for these images was resonant with the atomic transition, but not with any molecular transitions, we selectively detected only the atoms. Figure 6.2 shows the observed total atom number in the $m_f = -5/2, -9/2$ states as a function of the final magnetic-field value of the ramp. We found that the atoms disappeared abruptly at the Feshbach resonance peak (dashed line). We also found in similar experiments that we could recover the lost atoms with an immediate magnetic-field ramp back above the Feshbach resonance. This ruled out many atom loss processes and strongly suggested that all of the lost atoms were converted to Feshbach molecules. We were surprised at the efficiency of the conversion of our Fermi gas of atoms to a Bose gas of molecules; we could easily create hundreds of thousands of Feshbach molecules.

---

[1]In fact the atoms and molecules have the same trapping frequency, but the molecule trap depth is twice as large as the atom trap depth.

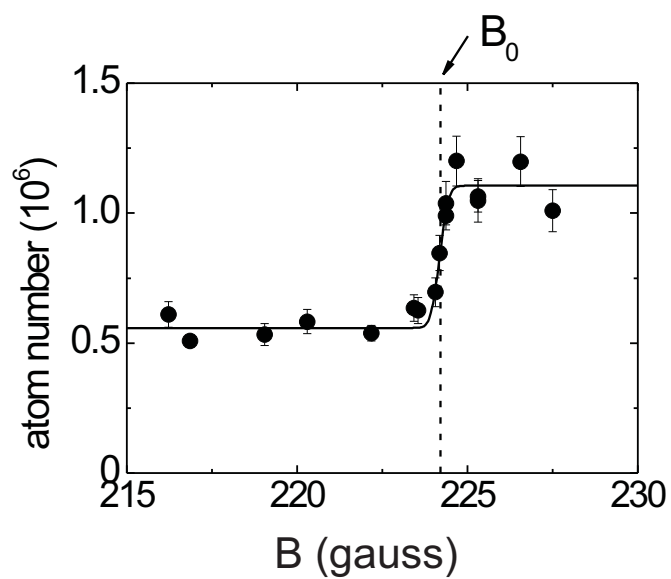

Figure 6.2: Creation of molecules as seen through atom loss [63]. A fit to an error function provides a guide to the eye. Atom loss occurs at precisely the expected position of the Feshbach resonance given a previous measurement of the scattering length divergence [59].

## 6.2 Detecting weakly bound molecules

### 6.2.1 Rf spectroscopy

While suggestive of molecule creation, the measured atom loss was not conclusive proof for the existence of Feshbach molecules. We thus employed a spectroscopic technique to probe the molecules. First, we created the molecules with a magnetic-field ramp across the Feshbach resonance and stopped at a magnetic field $B_{hold}$. At $B_{hold}$ a 13 $\mu$s radio frequency (rf) pulse was applied to the gas; the rf frequency was chosen so that the photon energy was near the energy splitting between the $m_f = -5/2$ and $m_f = -7/2$ atom states (see Fig. 5.3(a)). The resulting population in the $m_f = -7/2$ state, which was initially unoccupied, was then probed selectively either by separating the spin states spatially using a strong magnetic-field gradient during free expansion (Stern-Gerlach imaging) or by leaving the magnetic field high (215 G) and taking advantage of nonlinear Zeeman shifts.

Figure 6.3(a) shows a sample rf spectrum at $B_{hold} = 222.49$ G; the resulting number of atoms in the $m_f = -7/2$ state is plotted as a function of the frequency of the rf pulse. We observed two distinct features in the spectrum. The sharp, symmetric peak was very near the expected $m_f = -5/2$ to $m_f = -7/2$ transition frequency for free atoms. With the Stern-Gerlach imaging we saw that the total number of $m_f = -5/2$ and $m_f = -7/2$ atoms was constant, consistent with transfer between these two atom states. The width of this peak was defined by the Fourier width of the applied rf pulse. Nearby was a broader, asymmetric peak shifted lower in frequency. Here we found that after the rf pulse the total number of observed atoms ($m_f = -5/2 + -7/2$) actually increased. Also, the resulting

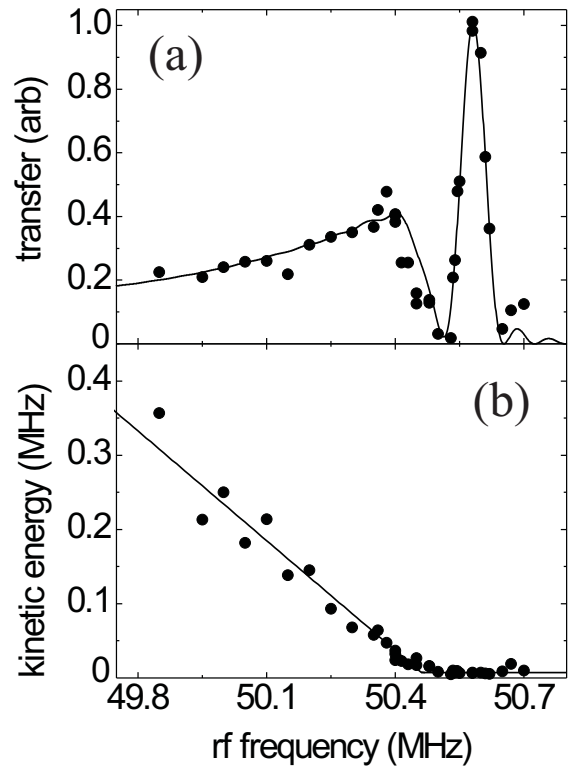

Figure 6.3: Rf spectrum for an atom/Feshbach molecule mixture [63]. (a) Transfer to the $m_f=-7/2$ states as a function of rf frequency. The left feature is the molecule dissociation spectrum and the right feature represents the transfer of atoms between $m_f=-5/2$ and $m_f=-7/2$. (b) Corresponding kinetic energy of the $m_f=-7/2$ state.

$m_f{=}{-}7/2$ gas in this region had a significantly increased kinetic energy, which grew linearly for larger frequency shifts from the atom peak (Fig. 6.3(b)).

The asymmetric peak corresponds to the dissociation of molecules into free $m_f{=}7/2$ and $m_f{=}{-}9/2$ atoms. Since the applied rf pulse stimulates a transition to a lower energy Zeeman state, we expected $h\nu_{rf} = h\nu_0 - E_b - \Delta E$, where $E_b$ is the binding energy of the molecule, $\nu_0$ is the atom-atom transition frequency for noninteracting atoms, and we have ignored mean-field interaction energy shifts. The remaining energy, $\Delta E$, must be imparted to the dissociating atom pair as kinetic energy. Two separate linear fits were applied to the kinetic energy data in Fig. 6.3(b) to determine the threshold position. The slope beyond threshold for the data is $0.49 \pm 0.03$; this indicates that the atom pair ($m_f{=}{-}7/2 + m_f{=}{-}9/2$) does indeed receive the additional energy, $\Delta E$, beyond the binding energy when the molecule is dissociated.

The observed lineshape of the asymmetric peak in Fig. 6.3(a) should depend upon the Franck-Condon factor, which gives the overlap of the molecular wavefunction with the atomic wavefunction. C. Ticknor and J. Bohn calculated this multichannel Franck-Condon overlap as a function of energy. The resulting transition rate, convolved with the frequency width of the applied rf and scaled vertically, is shown as the solid line in Fig. 6.3(a). The agreement between theory and experiment for the dissociation spectrum is quite good. This well-resolved spectrum provides much information about the molecular wavefunction. A useful discussion of the theoretical aspects of these dissociation spectra and their relation to the wavefunction of the initial and final states can be found in Ref. [151].

In Fig. 6.4 is the magnetic-field dependence of the frequency shift $\Delta\nu = \nu_{rf} - \nu_0$, which to first order should correspond to the molecular binding energy. While $\Delta\nu$ could in principle be obtained directly from the transfer spectrum (Fig. 6.3(a)), we used the appearance of the threshold in the energy of the $m_f{=}{-}7/2$ gas, as it is more clear (Fig. 6.3(b)). We compared the position of this energy threshold to the expected atom-atom transition frequency $\nu_0$ based upon a calibration of the magnetic-field strength. The data are consistent with a theoretical calculation of the binding energy (solid line) based upon a full coupled channels calculation with no free parameters carried out by C. Ticknor and J. Bohn. This measurement of $E_b$ accentuates the fact that these Feshbach molecules are not typical molecules.

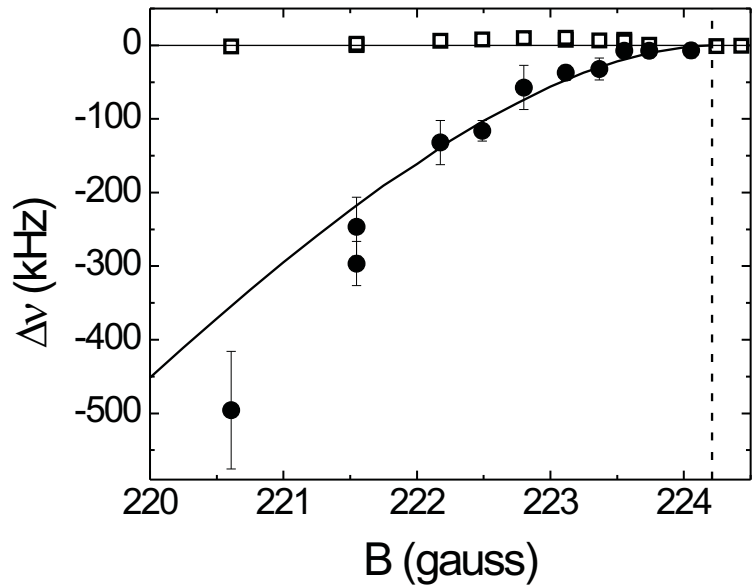

Figure 6.4: The frequency shift ($\Delta\nu$) from the expected $m_f = -5/2 \rightarrow -7/2$ transition plotted versus magnetic field for the $m_f = -7/2$ atoms (squares) and the molecules (circles). The line corresponds to a calculation of the binding energy of the molecules as a function of detuning from the Feshbach resonance [63].

With binding energies on the order of $h$ 100 kHz ($4 \times 10^{-10}$ eV) they are extremely weakly bound compared to molecules chemists are accustomed to studying.

The excellent agreement with theory in Fig. 6.4 left no doubt that efficient creation of Feshbach molecules is possible. In addition, our rf spectroscopy technique was extended for a variety of other measurements in paired systems. It was proposed that rf spectroscopy could be used to measure the excitation gap in a superfluid Fermi gas [152, 153]; such a measurement is published in Ref. [80]. Our rf spectroscopy technique was also extended to detect confinement induced molecules in a one-dimensional Fermi gas [154]. Molecule dissociation via rf spectroscopy has proven useful for giving atoms in a molecule a large relative momentum, as in Ref. [155]. Dissociating the molecules far above threshold produces a fun absorption image. The dissociated atoms fly out in a spherical shell, and the resulting absorption image is a ring structure (Fig. 6.5).

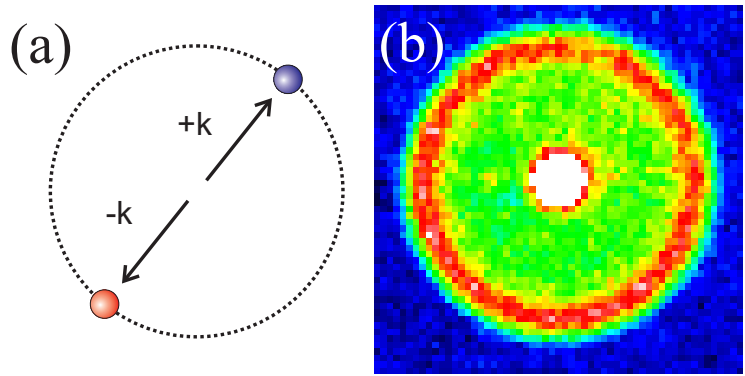

Figure 6.5: Dissociation of molecules with radio frequencies [155]. (a) The atoms that result from the dissociation have equal and opposite momenta. (b) False color absorption image of a dissociated molecular gas.

## 6.2.2 Dissociation through magnetic-field modulation

A very similar technique to rf spectroscopy that we later employed to study fermionic atom pairs such as molecules was magnetic-field modulation [81]. After creating Feshbach molecules at a magnetic field $B_{hold}$, we modulated the homogeneous magnetic field at a frequency near $E_b/h$, as shown in Fig. 6.6(a). To quantify the response of the gas, we determined the increase in the gas temperature after the perturbation was applied. We found that a sensitive way to detect this increase in temperature was to measure the number of atoms that escaped from a shallow optical dipole trap due to evaporation [81]. By varying the frequency of the modulation we could map out a molecule dissociation spectrum and measure the dissociation threshold $\Delta\nu$. The result of this measurement for a variety of magnetic fields $B_{hold}$ is shown in Fig. 6.6(b).

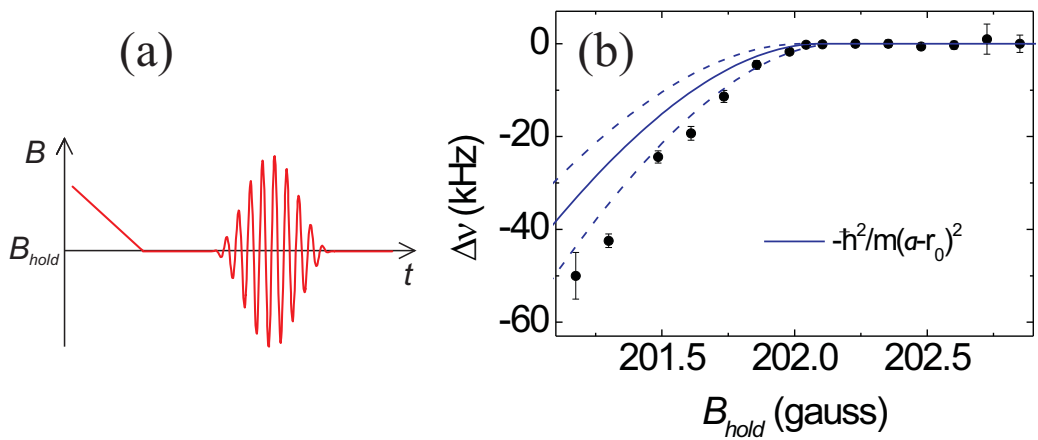

Figure 6.6: (a) Magnetic field modulation to dissociate molecules. (b) Resulting binding energy.

Note that this measurement is considerably more precise than the rf spectroscopy measurement. One reason for this is that the $B$-modulation technique does not change the magnetic moment of the atoms, meaning the measurement is not sensitive to magnetic-field fluctuations. The lines in Fig. 6.6(b) show the result of the calculation of Ch. 3, with the dashed lines indicating the uncertainty on the curve due to the uncertainty in the measured Feshbach resonance parameters $B_0$ and $w$.[2] The data agree with the calculation to within the uncertainty, but note in this measurement many-body effects may result in systematic shifts. This magnetic-field modulation method of dissociating molecules was extended to associate free $^{85}$Rb atoms into Feshbach molecules in Ref. [156].

---

[2]Most of this uncertainty originates in $B_0$, and the value and uncertainty in $B_0$ comes from the measurement in Sec. 6.4. Fig. 6.6(b) suggest that the resonance position is actually at the upper limit of the measurement of Sec. 6.4.

## 6.3 Understanding molecule conversion efficiency

While using magnetic-field ramps to create molecules was very successful, there were many outstanding questions about the physics of the process. For example, what parameters define the conversion efficiency from atoms to molecules? The first important parameter turns out to be the rate of the magnetic-field ramp across the resonance. If the ramp is too fast no molecules will be created because the ramp will be diabatic with respect to the atom-molecule coupling. As the ramp is made slower, however, atoms will pair to form molecules. This effect is shown in Fig. 6.7 where molecule creation through atom loss is shown. Theoretical predictions find that this effect can be well modelled by the Landau-Zener formula for the transition probability at a two-level crossing

$$f = f_m(1 - e^{-\delta_{LZ}}) \tag{6.1}$$

where $f$ is the fraction of atoms converted to molecules, $f_m$ is the maximum fraction of atoms that can be converted to molecules, and $\delta_{LZ}$ is the Landau-Zener parameter [149, 112]. Let $\delta_{LZ}$ be $\beta(dB/dt)^{-1}$ where $(dB/dt)^{-1}$ is the inverse magnetic-field ramp rate across the resonance. We can fit Fig. 6.7 with $\beta$ as the fitting parameter to find in this case $\beta \approx 20$ $\mu$s/G [63, 146]. Reference [112] predicts that $\beta = \alpha\,n\,w\,a_{bg}$ where $n$ is the atomic gas density, $w$ is the width of the Feshbach resonance, $a_{bg}$ is the background scattering length, and $\alpha$ is a proportionality constant. A study by Hodby *et al.* verified the linear dependence of $\beta$ upon the density $n$, but the proportionality constant $\alpha$ for this expression is still under investigation [146].

Notice in Fig. 6.7 that even at rates a few times slower than $\beta$ not all the atoms are converted to molecules. One would expect that if the atom-molecule system were in chemical equilibrium and the temperature of the molecular sample were much less than $E_b/k_b$ then 100% of the atoms should be converted to molecules. An important point to recognize for all of the experiments in this chapter is that we are **not** operating in chemical equilibrium. At the final magnetic field values in these experiments the time scale for chemical equilibrium is significantly longer than time scales in Fig. 6.7, and we routinely work on time scales intermediate between the time scale of $\beta$ and the chemical equilibrium time scale. Thus, the observed saturation in molecule conversion is important to understand. Reference [146] studies this phenomenon both for a bosonic gas of $^{85}$Rb and our fermionic gas.

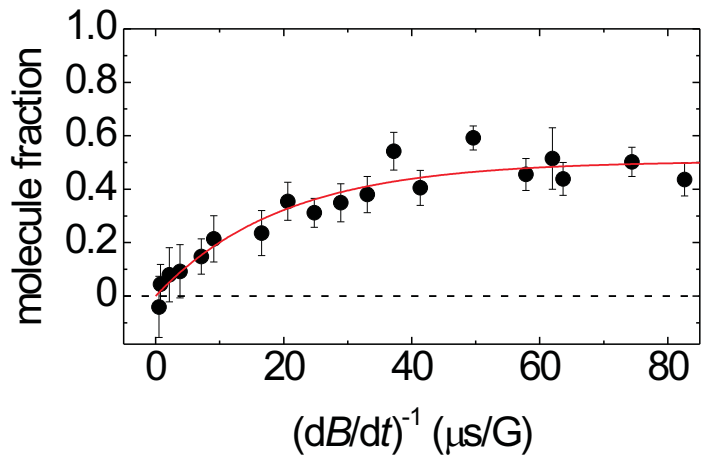

Figure 6.7: Time scale for two-body adiabaticity [63].

Eric Cornell suggested that the saturation in molecule conversion was likely related to phase space density (or $T/T_F$ in our case) based on intuitive arguments. An adiabatic process smoothly alters the wavefunction describing atom pairs but does not change the occupation of states in phase space. Thus to form a molecule a pair of atoms must initially be sufficiently close in phase space that their combined wavefunction can evolve smoothly into the Feshbach molecule state as the resonance is crossed. In other words one would expect a molecule to form if a pair of atoms has a relative position $r_{rel}$ and relative velocity $v_{rel}$ such that

$$|\delta r_{rel} m \delta v_{rel}| < \gamma h \qquad (6.2)$$

where $\gamma$ is an experimentally determined constant.

Figure 6.8 shows the result of a measurement of the saturation in molecule conversion ($f_m$) for a $^{40}$K gas as a function of $T/T_F$, which is monotonically related to the phase space density of the gas. We see that indeed the conversion fraction does increase as $T/T_F$ decreases (phase space density increases) with a maximum conversion at our lowest temperatures of about 90%. $^{85}$Rb data show a similar dependence. To quantify the conversion expected for the many-body problem given the two-body criterion above (Eqn. 6.2) an algorithm described in Ref. [146] was developed. The line in Fig. 6.8 is the result of this algorithm for the best fit value of $\gamma$. We found that for the fermion data $\gamma = 0.38 \pm 0.04$ and for the boson data $\gamma = 0.44 \pm 0.03$, indicating that a similar process is at work in both the Fermi and Bose cases.

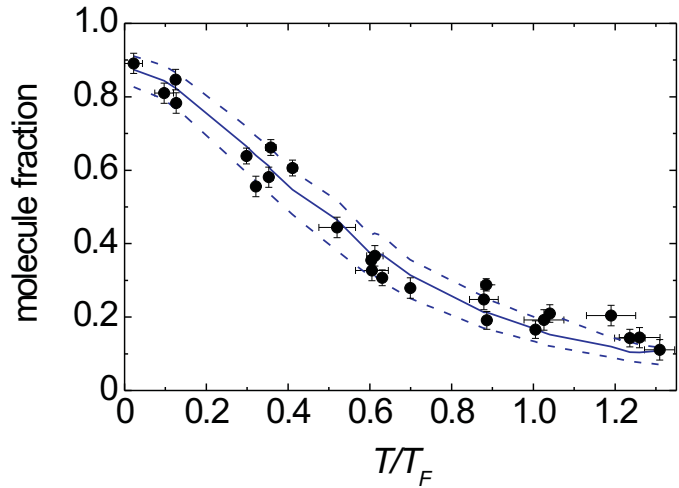

Figure 6.8: Dependence of molecule conversion on the initial $T/T_F$ of a two-component Fermi gas [146].

## 6.4    A precise measurement of $B_0$

A precise determination of the magnetic-field location of the Feshbach resonance $B_0$ is an essential ingredient for exploring the BCS-BEC crossover regime. Knowledge of the position and the width of the resonance allows a precise calculation of the interaction strength at a particular magnetic field (see Ch. 3). As we saw in Ch. 5, $B_0$ can be measured via scattering properties of the resonance, but our most precise measurement of $B_0$ actually comes from the study of Feshbach molecules. In particular we looked for dissociation of Feshbach molecules in a low density gas as a function of magnetic field. To determine if the molecules had been dissociated or not we probed the gas at low magnetic field; here atoms not bound in molecules can be selectively detected.

Figure 6.9 shows the result of such a measurement. Molecules created by a slow magnetic-field ramp across the resonance were dissociated by raising the magnetic field to a value $B_{probe}$ near the Feshbach resonance (inset to Fig. 6.9). Note that to avoid many-body effects, we dissociated the molecules after allowing the gas to expand from the trap to much lower density. This plan also allowed us to be certain we would not create molecules in the ramp of the magnetic field to near zero field for imaging. The measured number of atoms increased sharply at $B_0 = 202.10 \pm 0.07$ G. This measurement of the resonance position agreed well with previous less precise results [57, 67].[3]

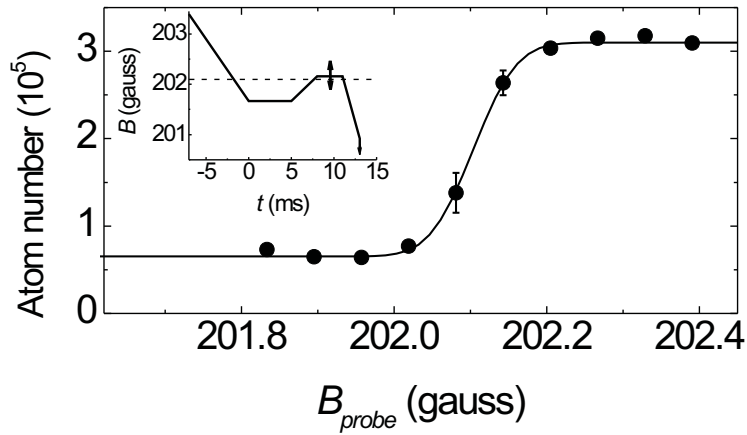

Figure 6.9: Determination of the position of the Feshbach resonance via molecule dissociation in a low-density Fermi gas [73]. A fit of the data to an error function reveals $B_0 = 202.10 \pm 0.07$ G, where this uncertainty is given by the 10%-90% width.

---

[3]Note that this measurement of the resonance position is quite similar to the molecule association result of Fig. 6.2. However, the disadvantages of the association method are: (1) The molecule creation must take place in a high-density sample; hence many-body effects may play a role. (2) The details of the magnetic-field ramp to low field for imaging, in particular its initial rate, are crucial.

# Chapter 7

# Inelastic collisions near a fermionic Feshbach resonance

Elastic collisions between atoms (discussed in Ch. 5) are often referred to as good collisions. These collisions allow rethermalization in the gas but do not change the internal state of the atoms or molecules. In atomic gas experiments a constant concern is inelastic collisions, often referred to as bad collisions. In these collisions the products often are particles in lower energy internal states. The difference in energy between the incoming particles and the products of the collision must be carried away in the form of kinetic energy. When this energy difference is large compared to energy scales such as the trap depth particles can be lost from the trap and significant heating of the sample can occur.

Near a Feshbach resonance inelastic collisions can be enhanced along with elastic collisions. To accomplish the work in this thesis we spent a large fraction of our time understanding the inelastic processes near $^{40}$K Feshbach resonances and designing experiments that minimize the effect of inelastic collisions. In this chapter I will discuss the inelastic collisions near a $^{40}$K Feshbach resonance and present measurements of relevant inelastic collision rates. We observed clear evidence of inelastic processes near the fermionic Feshbach resonance, but found that despite these inelastic processes the lifetime of the sample was long enough to study BCS-BEC crossover physics.

## 7.1 Expected inelastic decay processes

Let us first consider the stability of free fermionic atoms on the BCS side of a fermionic Feshbach resonance. In particular consider the Feshbach resonance between the $|f, m_f\rangle = |9/2, -9/2\rangle$ and $|9/2, -7/2\rangle$ states that is used for many of the experiments in $^{40}$K. Since these are the two lowest energy states of $^{40}$K the only inelastic collision involving two of these fermions that is energetically favorable is

$$|9/2, -9/2\rangle + |9/2, -7/2\rangle \rightarrow |9/2, -9/2\rangle + |9/2, -9/2\rangle. \qquad (7.1)$$

This process however is forbidden due to the fermionic nature of the particles [53]. Thus, any inelastic collision with these states must involve at least three fermions. A three-body inelastic collision in a two-component Fermi gas with components $X$ and $Y$ would take the form

$$X + X + Y \rightarrow X + (XY)_-. \qquad (7.2)$$

Here the subscript $-$ represents a lower-energy molecular state. Such lower-energy molecular states are always present in these atomic gas systems as there are many vibrational levels of the interatomic potential. To conserve energy and momentum in the collision the products $X$ and $(XY)_-$ carry away the binding energy of the $(XY)_-$ molecule in the form of relative kinetic energy. Theory predicts that this three-body collision process will be suppressed for $s$-wave interactions between fermions because it requires that two identical fermions approach each other [90, 91, 95, 157]. However, while the rate of this inelastic collision process is suppressed, it is not forbidden, making it an important process near a Feshbach resonance [57, 158].

As we cross the Feshbach resonance to the BEC side with a cold $^{40}$K Fermi gas we must consider the stability of a mixture of fermionic atoms and Feshbach molecules. An isolated Feshbach molecule for the $m_f = -7/2, -9/2$ $^{40}$K resonance will be stable for the same reason as the two fermion mixture is stable. Note that the case in which the two atoms in the molecule are not in the lowest energy internal states, such as molecules created using $^{85}$Rb, is quite different [144, 146].

These $^{85}$Rb Feshbach molecules will spontaneously dissociate as observed in Ref. [159]. For our $^{40}$K molecules however we again expect that any decay processes will require more than two fermions, for example [95, 160]

$$X + XY \rightarrow X + (XY)_- \tag{7.3}$$

$$XY + XY \rightarrow XY + (XY)_-. \tag{7.4}$$

The first process is reminiscent of Eqn. 7.2 above. These processes are often referred to as collisional quenching of vibrations [161, 162, 163]. Again we expect some suppression of these decay channels due to Fermi statistics since two identical fermions must approach each other, as shown schematically in Fig. 7.1.

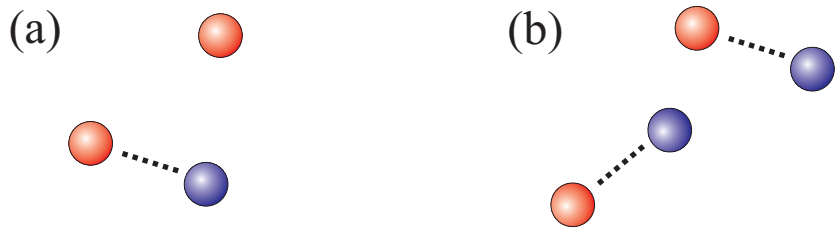

Figure 7.1: Particles involved in inelastic collisions in a Feshbach molecule/atom mixture. (a) Illustration of Eqn. 7.3 (b) Illustration of Eqn. 7.4.

## 7.2 Lifetime of Feshbach molecules

In this section I present experimental data on the stability of a mixture of atoms and Feshbach molecules. To obtain these data we created a molecule sample at the $m_f = -7/2, -9/2$ Feshbach resonance in which typically 50% of original atom gas was converted to molecules. We then measured the molecule number as a function of time while holding the molecule/atom mixture in a relatively shallow optical dipole trap [67]. Figure 7.2 shows the result of this measurement at a variety of magnetic fields on the BEC side of the Feshbach resonance. The plot shows $\dot{N}/N$ versus the atom-atom scattering length $a$. Here $N$ is the number of molecules and $\dot{N}$ is the initial linear decay rate.

We found that far from resonance the molecules decay quickly, but the decay rate changes by orders of magnitude as the Feshbach resonance is approached. Physically this effect is partially related to the overlap between the wavefunctions of the $XY$ molecule and the $(XY)_-$ molecule. As the Feshbach resonance is approached the $XY$ molecules become extremely weakly bound and quite large, and hence they have less overlap with the small $(XY)_-$ molecules.

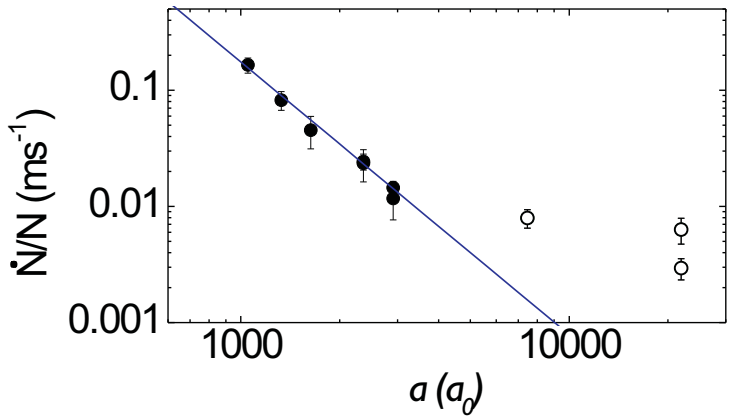

Figure 7.2: Feshbach molecule loss rate as a function of the atom-atom scattering length near a Feshbach resonance in $^{40}$K [67]. $N$ here is the number of molecules. The line is a fit of the closed circles ($\bullet$) to a power law. The open circles ($\circ$) are data for which the pair size is expected from two-body theory to be larger than the interparticle spacing.

A scaling law for the dependence of the molecule decay rate upon the atom-atom scattering length $a$ was found in [95] and later in [157] for the processes described by Eqns. 7.3 and 7.4 (Fig. 7.1). The scaling law was found by solving the full few-body problem in the limit where the molecules are smaller than the interparticle spacing, yet $a \gg r_0$. Physical effects important to the result are the Fermi statistics and the wavefunction overlap. The prediction for Eqn. 7.3 (molecule-atom collisions) is that the decay rate should scale with $a^{-3.33}$ and for Eqn. 7.4 (molecule-molecule collisions) with $a^{-2.55}$.

Since our measurement was carried out with thermal molecules the density of the molecule gas remains approximately constant over the $a = 1000\ a_0$ to $3000\ a_0$ range. (The total peak atom density in one spin state in the weakly interacting regime was $n_{pk}^0 = 7.5 \times 10^{12}$ cm$^{-3}$.) Thus, we could measure the power law by fitting the data in Fig. 7.2 to the functional form $Ca^{-p}$, where $C$ and $p$ are constants. We included only points for which the interatomic spacing in the center of the sample was larger than the expected size of an isolated molecule, $a/2$. We found $p = 2.3 \pm 0.4$, consistent with the predicted power law for molecule-molecule collisions. A similar power law was observed in a gas of $Li_2$ molecules at the 834 G Feshbach resonance [164].

In general we find that the lifetime of the molecules is surprisingly long near the Feshbach resonance. The molecule lifetime for magnetic fields at which $a > 3000$ $a_0$ is greater than 100 ms. This is much longer than lifetimes observed in bosonic systems for similar densities and internal states [165, 166]. 100 ms is actually a long time compared to many other time scales in our Fermi gas such as the time scale for two-body adiabaticity, the average time between elastic collisions, and the radial trap period. This comparison suggested that it would indeed be possible to study BCS-BEC crossover physics using atomic $^{40}K$ gases.

## 7.3 Three-body recombination

We also observe inelastic decay of fermionic atoms on the BCS side of the Feshbach resonance, where the decay is due the process described by Eqn. 7.2 [57]. These collisions cause both particle loss and heating. The heating can result from a combination of two processes: First, the density dependence of the three-body process results in preferential loss in high density regions of the gas [167]. Second, the products of the inelastic collision, which have kinetic energies on order the binding energy of the $XY_-$ molecules, can collide with other particles on the way out of the sample causing transfer of energy to the gas.

We found that one relevant measure of the effect of inelastic decay processes on our ability to study BCS-BEC crossover physics is the heating of our Fermi gas during magnetic-field ramps that are sufficiently slow to be adiabatic compared to many-body time scales (see Ch. 8). We performed an experiment in which we approached the Feshbach resonance at rate of $(6 \text{ ms/G})^{-1}$, waited 1 ms, and then ramped back at the same slow rate to the weakly interacting regime. The result of this experiment is shown in Fig. 7.3. If we started with a gas initially at $T/T_F = 0.10$, $T/T_F$ upon return increased by less than 10% for a ramp to $1/k_F^0 a = 0$ (yet by 80% for a ramp to $1/k_F^0 a = 0.5$). $k_F^0$ is the Fermi wavevector measured in the weakly interacting regime. For this measurement the initial peak density in one spin state was $n_{pk}^0 = 1.2 \times 10^{13}$ cm$^{-3}$.

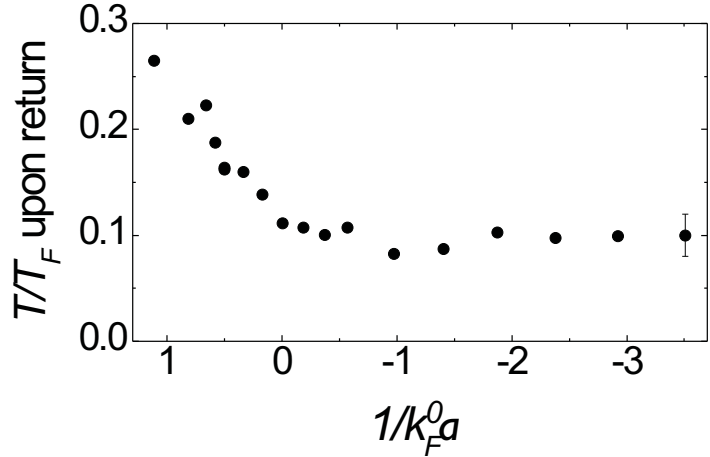

Figure 7.3: $T/T_F$ measured in the weakly interacting regime as a function of the final magnetic field in an adiabatic ramp towards the Feshbach resonance and back [78]. The magnetic field is represented through the dimensionless parameter $1/k_F^0 a$.

## 7.4 Comparison of $^{40}$K and $^6$Li

In both $^{40}$K and $^6$Li gases the scaling law of the decay rate as a function of molecule size was found to be the same [67, 164]. However, comparison of the absolute decay rates for similar densities and $a \approx 5000\ a_0$ shows that the inelastic decay of $^{40}$K occurs more than two orders of magnitude faster than at the broad $^6$Li Feshbach resonance [39, 164, 122]. This difference has so far not been fully explained theoretically. However, it must be related in some way to the difference in the full three-body potentials for $^{40}$K versus $^6$Li atoms. This difference in lifetime between the two atomic species affects how experiments in the BCS-BEC crossover are approached in $^{40}$K versus $^6$Li. For example, in our experiments we obtain the coldest gases in $^{40}$K by evaporating a Fermi gas in the weakly interacting regime where inelastic decay is negligible. Experimenters using $^6$Li, however, obtain their coldest gases through evaporation near the Feshbach resonance or slightly to the molecular side of the resonance [168, 69].

In our discussions thus far we have only considered inelastic collisions in which the final state is a more deeply bound molecule. In the $^6$Li systems experimenters make use of inelastic collisions that interconvert atoms on the repulsive side of the Feshbach resonance and Feshbach molecules [168, 69]. The rate of these inelastic collisions in the $^6$Li system is faster than inelastic decay to more deeply bound molecules, and the binding energy of the Feshbach molecules $E_b \approx \frac{\hbar^2}{ma^2}$ is small enough to not cause significant problems in the cooling process. Hence, they typically start with a hot gas of atoms on the BEC side of the resonance and after evaporative cooling observe a pure sample of Feshbach molecules [66, 69]. Thus, in contrast to the $^{40}$K experiments in this thesis, $^6$Li experiments can operate with chemical equilibrium between atoms and Feshbach molecules.

# Chapter 8

# Making condensates from a Fermi gas of atoms

So far we have considered the normal state of the Feshbach resonance/Fermi gas system and found that it has all the elements necessary to study BCS-BEC crossover physics. However, the true test of whether we could access BCS-BEC crossover physics with our atomic gas would be to observe a phase transition. The phase transition could be distinguished through observation of the onset of either superfluid behavior or condensation. Due to the linked nature of these phenomena one would necessarily imply the other. Just as with alkali BEC in $^{87}$Rb and $^{23}$Na (Fig. 1.3), the observable of choice for the first experiments to observe this phase transition in the BCS-BEC crossover was condensation.

In this chapter I will discuss how we were able to show condensation of fermionic atom pairs in the BCS-BEC crossover regime. This demonstration relied heavily upon our previous knowledge of the normal state of a Fermi gas at a Feshbach resonance. First, I present our observation of condensation of Feshbach molecules to create one of the first molecular BECs [68]. This work led the way to observation of condensation of fermionic atom pairs in the crossover regime. Then I present a phase diagram of the BCS-BEC crossover regime attained through measurements of condensate fraction.

## 8.1 Emergence of a molecular condensate from a Fermi gas of atoms

We have seen that a Feshbach resonance can be used to create a large number of ultracold molecules starting with a Fermi gas of atoms. After observing that these molecules can be long lived, the creation of a BEC from these bosonic Feshbach molecules was an obvious goal. Previously we had created molecules by applying a magnetic-field ramp just slow enough to be two-body adiabatic; in the experiments here the idea will be to apply a magnetic-field ramp that is not only two-body adiabatic, but also slow with respect to the many-body physics timescale in our gas (many-body adiabatic). With such a magnetic-field ramp across the Feshbach resonance the entropy of the original quantum Fermi gas should be conserved [65, 169]. For an initial atom gas with a sufficiently low $T/T_F$ the result should be a low entropy sample of bosonic molecules, which for a low enough entropy is a BEC.

To pursue this idea experimentally we again used the Feshbach resonance between the $m_f = -9/2$ and $m_f = -7/2$ spin states starting with a Fermi gas at temperatures below quantum degeneracy. We applied a time-dependent ramp of the magnetic field starting above the Feshbach resonance and ending below the resonance. The magnetic field was typically ramped in 7 ms from $B = 202.78$ G to either $B = 201.54$ G or $B = 201.67$ G, where a sample of 78% to 88% Feshbach molecules was observed. A critical element of this experiment is that the lifetime of the Feshbach molecules can be much longer than the typical collision time in the gas and longer than the radial trapping period (see Ch. 7). The relatively long molecule lifetime near the Feshbach resonance allows the atom/molecule mixture to achieve thermal equilibrium during the magnetic-field ramp. Note however that since the optical trap used for these experiments is strongly anisotropic ($\nu_r/\nu_z \approx 80$) we may attain only local equilibrium in the axial direction.

To study the resulting atom-molecule gas mixture after the magnetic-field ramp, we measured the momentum distribution of both the molecules and the residual atoms using time-of-flight absorption imaging. After typically 10 to 20 ms of expansion we applied a radio frequency (rf) pulse that dissociated the molecules into free atoms in the $m_f = -5/2$ and $m_f = -9/2$ spin states [63]. Immediately af-ter this rf dissociation pulse we took a spin-selective absorption image. The rf pulse had a duration of 140 $\mu$s and was detuned 50 kHz beyond the molecule dissociation threshold where it did not affect the residual unpaired atoms in the $m_f=-7/2$ state. We selectively detected the expanded molecule gas by imaging atoms transferred into the previously unoccupied $m_f=-5/2$ state by the rf dissociation pulse. Alternatively we could image only the expanded atom gas by detecting atoms in the $m_f=-7/2$ spin state.

Close to the Feshbach resonance, the atoms and molecules are strongly interacting with effectively repulsive interactions. The scattering length for atom-molecule and molecule-molecule collisions was calculated by Petrov *et al.* to be 1.2 $a$ and 0.6 $a$ respectively, for values of $a$ much larger than $r_0$ and smaller than the interparticle spacing [95]. During the initial stage of expansion the positive interaction energy is converted into additional kinetic energy of the expanding gas. Therefore, the measured momentum distribution is very different from the original momentum distribution of the trapped gas. In order to reduce the effect of these interactions on the molecule time-of-flight images we used the magnetic-field Feshbach resonance to control the interparticle interaction strength during expansion. We could significantly reduce momentum kick due to the interaction energy by rapidly changing the magnetic field before we switched off the optical trap for expansion. The field was typically lowered by 4 G in 10 $\mu$s. At this magnetic field farther away from the resonance the atom-atom scattering length $a$ was reduced to $\sim$500 $a_0$. We found that this magnetic-field jump resulted in a loss of typically 50% of the molecules, which is a result of the reduced molecule lifetime away from the Feshbach resonance.

To attempt to observe condensation of molecules we monitored the molecule momentum distribution while varying the temperature of the initial weakly interacting Fermi gas, $(T/T_F)^0$. Below $(T/T_F)^0$ of 0.17 we observed the sudden onset of a pronounced bimodal momentum distribution. Figure 8.1 shows such a bimodal distribution for an experiment starting with an initial temperature of 0.1 $T_F$; for comparison the figure shows the resulting molecule momentum distribution for an experiment starting at 0.19 $T_F$. The bimodal momentum distribution is a striking indication that the gas of weakly bound molecules has undergone a phase transition to a BEC [22, 23]. To obtain thermodynamic information about the molecule gas the momentum distribution is fit with a two-component distribution. The fit function is the sum of an inverted parabola describing the Thomas-Fermi momentum distribution of a bosonic condensate and a gaussian momentum distribution describing the noncondensed component of the molecule gas. In Fig. 8.2 the measured condensate fraction is plotted as a function of the fitted temperature of the molecular thermal component in units of the critical temperature for an ideal Bose gas $T_c = 0.94(N\nu_r^2\nu_z)^{1/3}h/k_B$. In this calculated $T_c$, $N$ is the total number of molecules measured without changing the magnetic-field for the expansion. The critical temperature for the strongly interacting molecules measured from Fig. 8.2 is $0.8 \pm 0.1$ $T_c$. Such a decrease of the critical temperature relative to the ideal gas prediction is expected for a strongly interacting gas [170].

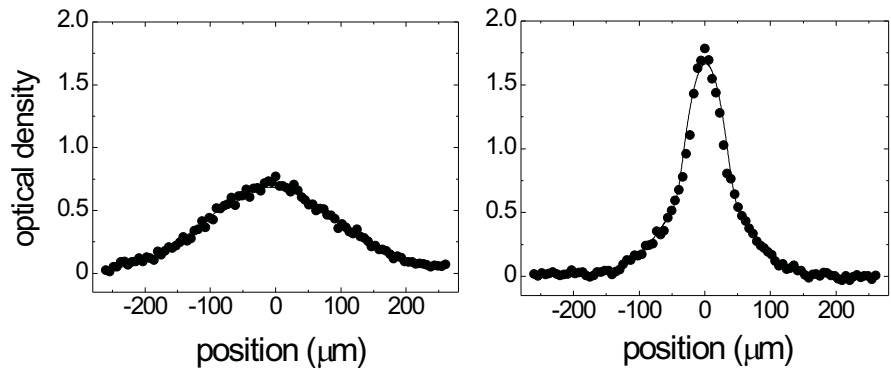

Figure 8.1: Momentum distribution of a molecule sample created by applying a magnetic-field ramp to an atomic Fermi gas with an initial temperature of 0.19 $T_F$ (0.1 $T_F$) for the left (right) picture [68]. In the right sample the molecules form a Bose-Einstein condensate. The lines illustrate the result of bimodal surface fits.

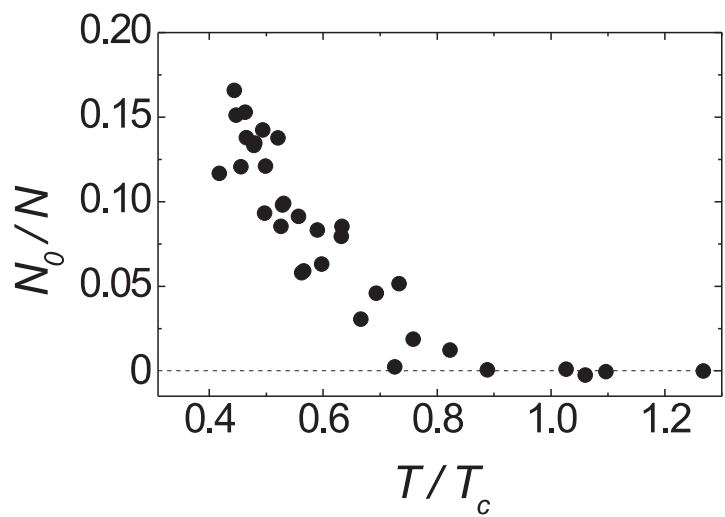

Figure 8.2: Molecular condensate fraction $N_0/N$ versus the scaled temperature $T/T_c$ [68]. The temperature of the molecules is varied by changing the initial temperature of the fermionic atoms prior to the formation of the molecules, yet measured through the momentum distribution of the molecular thermal gas.

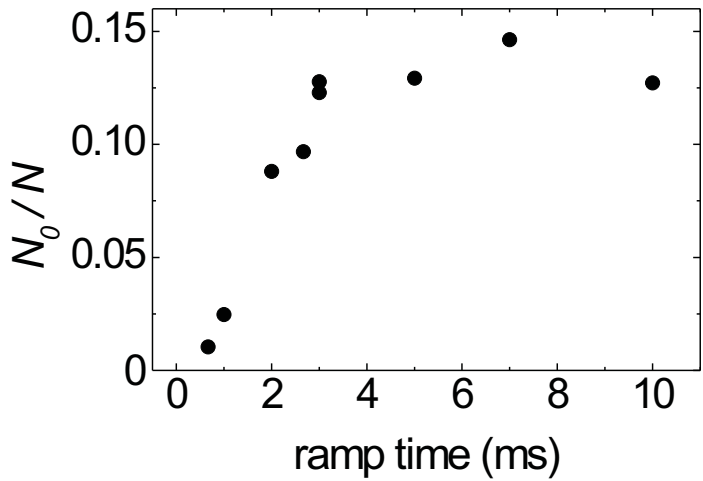

Figure 8.3: Time scale for many-body adiabaticity [68]. The fraction of condensed molecules is plotted versus the time in which the magnetic field is ramped across the Feshbach resonance from 202.78 G to 201.54 G.

As expected we found that the creation of a BEC of molecules requires that the Feshbach resonance be traversed sufficiently slowly to be many-body adiabatic. This many-body time scale should be determined by the time it takes atoms to collide and move in the trap. In Fig. 8.3 the measured condensate fraction is plotted versus the ramp time across the Feshbach resonance starting with a Fermi gas at a temperature $\sim 0.1$ $T_F$. Our fastest ramps resulted in a much smaller condensate fraction while the largest condensate fraction appeared for a magnetic-field ramps slower than ~3 ms/G.

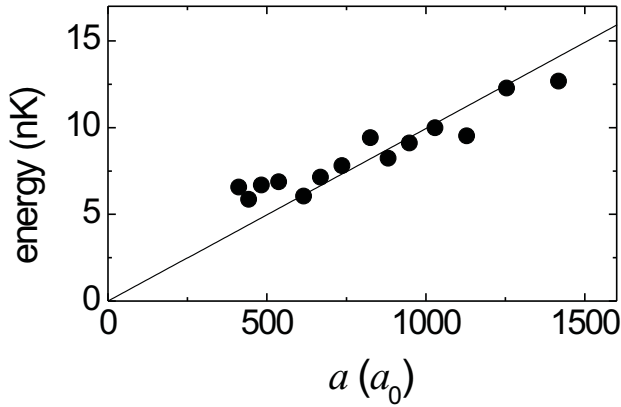

Figure 8.4: Expansion energy of the molecular condensate versus the interaction strength during expansion. In this measurement the BEC is created in a regime where the atom-atom scattering length is ~3300 $a_0$. For time-of-flight expansion the magnetic field is rapidly changed to different final values, characterized by the atom-atom scattering length $a$.

Rapidly changing the interaction strength for time-of-flight expansion of the condensate allowed us to measure the interaction energy in the molecular sample. Figure 8.4 shows a plot of the expansion energy of the molecule BEC for various interaction strengths during time-of-flight expansion. Here the condensate was created at a fixed interaction strength, and thus the initial peak density $n_{pk}$ was constant. The linear dependence of the energy upon $a$ suggests that the molecule-molecule scattering length is proportional to the atom-atom scattering length as predicted in Ref. [95]. In addition the expansion energy extrapolates to near zero energy for $a$=0. This is consistent with a Bose-Einstein condensate of molecules. Assuming the molecule-molecule interaction strength calculated in Ref. [95], the peak density of the strongly interacting condensate was $n_{pk} = 7 \times 10^{12}$ cm$^{-3}$.

In conclusion, I have discussed the creation of a BEC of weakly bound molecules starting with a gas of ultracold fermionic atoms. With a relatively slow ramp of an applied magnetic field that converts most of the fermionic atoms into bosonic molecules and an initial atomic gas below $T/T_F = 0.17$, we observed a molecular condensate in time-of-flight absorption images taken immediately following the magnetic-field ramp. Our experiment approaches the BEC limit, in which super-fluidity occurs due to BEC of essentially local pairs whose binding energy is larger than the Fermi energy. Strikingly, our $^{40}$K molecular condensate is not formed by any active cooling of the molecules, but rather merely by traversing the BCS-BEC crossover regime. At the same time as these experiments in $^{40}$K, experiments using $^{6}$Li created a similar BEC of Feshbach molecules. Their approach however was direct evaporative cooling of the Feshbach molecules [69, 70].

## 8.2   Observing condensates in the crossover

To create a molecular condensate we started with a quantum Fermi gas, slowly traversed the BCS-BEC crossover regime, and ended up with a BEC of molecules. An obvious question was whether condensation also had occurred in the crossover regime that we had passed through. To answer this question we needed to overcome a number of challenges. First, we required a probe of the momentum distribution of pairs in the crossover. In the BEC limit the momentum distribution of the molecules could be measured using standard time-of-flight absorption imaging. However, this method is problematic in the crossover because the pairs depend on many-body effects and are not bound throughout expansion of the gas. Second, to prove observation of condensation in the BCS-BEC crossover regime we had to show that we were not simply seeing condensation of pairs in the two-body bound state (two-body pairs), but rather condensation of pairs requiring many-body effects to form (many-body pairs). A clear example of condensation of many-body pairs would be condensation on the $a < 0$, or BCS, side of the Feshbach resonance. Here the two-body physics of the resonance no longer supports the weakly bound molecular state; hence, only many-body effects can give rise to a condensation of fermion pairs.

To solve the problem of measuring the momentum distribution of pairs in the crossover we introduced a technique that took advantage of the Feshbach resonance to pairwise project the fermionic atoms onto Feshbach molecules. We were able to probe the system by rapidly ramping the magnetic field to the BEC side of the resonance, where time-of-flight imaging could be used to measure the momentum distribution of the weakly bound molecules. The projecting magnetic-field ramp was completed on a timescale that allowed molecule formation but was still too brief for particles to collide or move significantly in the trap. This is possible due to the clear separation of the two-body and many-body time scales. The timescale for many-body adiabaticity in Fig. 8.3 is two orders of magnitude longer than the timescale for two-body adiabaticity shown in Fig. 6.7.

The key to the second problem, verifying condensation of many-body pairs, came from careful understanding of the two-body physics. As discussed in Ch. 6 we were able to precisely measure the magnetic-field position above which a two-body bound state no longer exists, $B_0$ (Fig. 6.9). If we observed condensation of fermionic atom pairs at $B > B_0$ we could be assured that these were pairs that were the result of many-body effects.

To perform experiments making use of these ideas, we continued with the same experimental setup as the last section where we discussed the creation of molecular condensates. We initially prepared the ultracold two-component atom gas at a magnetic field of 235.6 G, far above the Feshbach resonance. Here the gas is not strongly interacting, and we measured $(T/T_F)^0$ through surface fits to time-of-flight images of the Fermi gas (Ch. 4). The field was then slowly lowered at the many-body adiabatic rate of 10 ms/G to a value of $B_{hold}$ near the resonance. Whereas before we had considered only values of $B_{hold}$ below $B_0$ on the BEC side we now explore the behavior of the sample when ramping slowly to values of $B_{hold}$ on either side of the Feshbach resonance.

To probe the system we projected the fermionic atoms pairwise onto molecules and measured the momentum distribution of the resulting molecular gas. This projection was accomplished by rapidly lowering the magnetic field by $\sim$10 G at a rate of $(40~\mu s/G)^{-1}$ while simultaneously releasing the gas from the trap (Fig. 8.5). This put the gas far on the BEC side of the resonance, where it was weakly interacting. After a total of typically 17 ms of expansion the molecules were selectively detected using rf photodissociation immediately followed by spin-selective absorption imaging. To look for condensation, these absorption images were again surface fit to a two-component function that was the sum of a Thomas-Fermi profile for a condensate and a gaussian function for noncondensed molecules.

Figures 8.6 and 8.7 present the main result of this section. In Fig. 8.6(a) is the measured condensate fraction $N_0/N$ as a function of the magnetic-field detuning

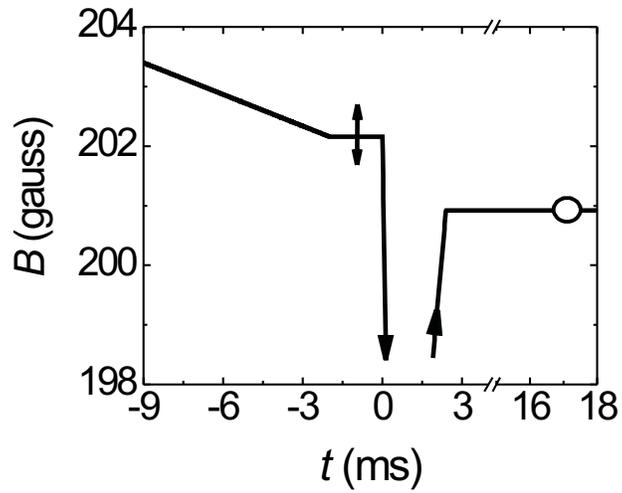

Figure 8.5: A typical magnetic-field ramp used to measure the fraction of condensed fermionic atom pairs [73]. The system is first prepared by a slow magnetic-field ramp towards the resonance to a variable position $B_{hold}$ (two-sided arrow). After a time $t_{hold}$ the optical trap is turned off at $t = 0$ and the magnetic field is quickly lowered by $\sim$10 G. After expansion, the molecules are imaged on the BEC side of the resonance ($\circ$).

from the resonance, $\Delta B = B_{hold} - B_0$. The data in Fig. 8.6(a) were taken for
a Fermi gas initially at $T/T_F = 0.08$ and for two different wait times at $B_{hold}$.
Condensation was observed on both the BCS ($\Delta B > 0$) and BEC ($\Delta B < 0$)
sides of the resonance. We further found that the condensation that occurs
on the BCS side of the Feshbach resonance was distinguished by its longer lifetime.
The lifetime was probed by increasing $t_{hold}$ to 30 ms (triangles in Fig. 8.6(a));
this time was much longer than the previously measured lifetime of the molecular
condensate in the BEC limit [68]. Not surprisingly then we found that for the
BEC side of the resonance no condensate was observed for $t_{hold} = 30$ ms except
very near the resonance. However, for all data on the BCS side of the resonance
the observed condensate fraction was still $> 70\%$ of that measured for $t_{hold} = 2$
ms.

An essential aspect of these measurements is the fast magnetic-field ramp that
projects the fermionic atoms pairwise onto molecules. It is a potential concern
that the condensation might occur during this ramp rather than at $B_{hold}$. To
verify that condensation did not occur during the ramp we studied the measured
condensate fraction for different magnetic-field ramp rates. Figure 8.6 compares
the condensate fraction measured using the 40 $\mu$s/G (circles) rate to that using
a ramp that was $\sim$7 times faster (open diamonds). We found that the measured
condensate fraction was identical for these two very different rates, indicating
that this measurement constitutes a projection with respect to the many-body
physics. The validity of the magnetic-field projection technique was also explored
in studies of a $^6$Li gas at MIT. Researchers there first reproduced the observation of
condensation using the pairwise projection technique with a $^6$Li gas [74]. They also
monitored the delayed response of the many-body system after modulating the
interaction strength [171]. They found that the response time of the many-body
system was slow compared to the rate of the rapid projection magnetic-field ramp.
There have been a number of theoretical papers on the subject of the pairwise
projection technique for measuring condensate fraction in the crossover [172, 173,
174]. Work thus far has established that observation of condensation of molecules
following a rapid projection ramp indicates the pre-existence of condensation of
fermionic atom pairs before the projection ramp.

To summarize, in this section we have introduced a method for probing the momentum distribution of fermionic atom pairs and seen how this technique could be employed to observe condensation near a Feshbach resonance. By projecting the system onto a molecule gas, we observed condensation of fermionic pairs as a function of the magnetic-field detuning from the resonance as shown in Fig. 8.7. While Fig. 8.7 is reminiscent of Fig. 1.3 where condensation was observed a function of $T/T_c$, note that the condensate here actually appears as a function of interaction strength, not temperature.

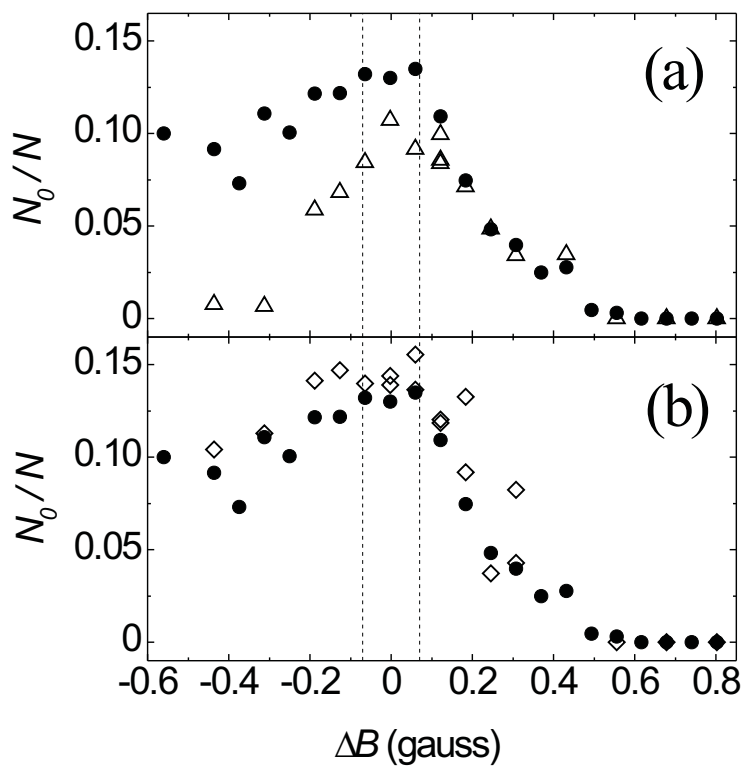

Figure 8.6: Measured condensate fraction as a function of detuning from the Feshbach resonance $\Delta B = B_{hold} - B_0$ [73]. (a) Data for $t_{hold} = 2$ ms ($\bullet$) and $t_{hold} = 30$ ms ($\triangle$) with an initial gas at $T/T_F = 0.08$. (b) Data for two different projection magnetic-field ramp rates: 40 $\mu$s/G ($\bullet$) and $\sim$6 $\mu$s/G ($\diamond$). The dashed lines $\Delta B = 0$ reflect the uncertainty in the Feshbach resonance position.

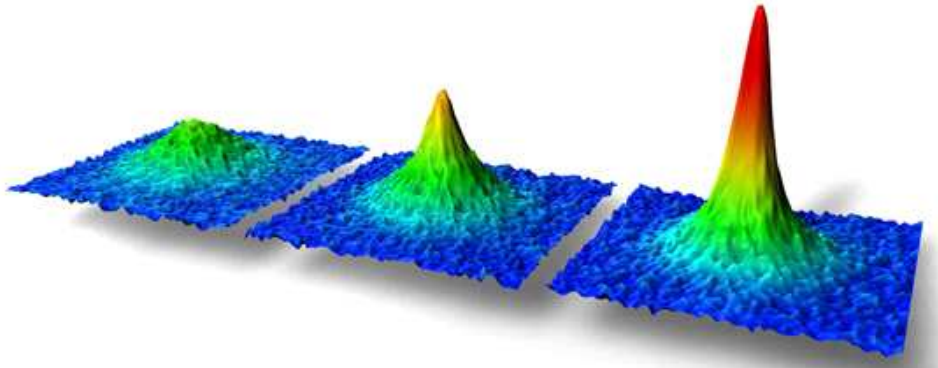

Figure 8.7: Time-of-flight images showing condensation of fermionic atom pairs. The images, taken after the projection of the fermionic system onto a molecule gas, are shown for $\Delta B = 0.12$, 0.25, and 0.55 G (right to left) on the BCS side of the resonance. The original atom gas starts at $(T/T_F)^0 = 0.07$. The 3D artistry is courtesy of Markus Greiner.

## 8.3 Measurement of a phase diagram

In addition to varying $\Delta B$ and measuring the condensate fraction, we can also vary the initial temperature of the Fermi gas. Figure 8.8 is a phase diagram created from data varying both $\Delta B$ and $(T/T_F)^0$. $\Delta B$ is converted to the dimensionless parameter $1/k_F^0 a$, where $a$ is calculated directly from $\Delta B$ through Eqn. 3.1 and $k_F^0$ is extracted from the weakly interacting Fermi gas. The colors represent the measured condensate fraction using the projection technique. The boundary between the light and dark blue regions shows where the phase transition occurs in the BCS-BEC crossover. On the BCS side of the resonance the condensate forms for higher initial $T/T_F$ as $\Delta B$ decreases, as expected based upon BCS-BEC crossover theories (Ch. 2).

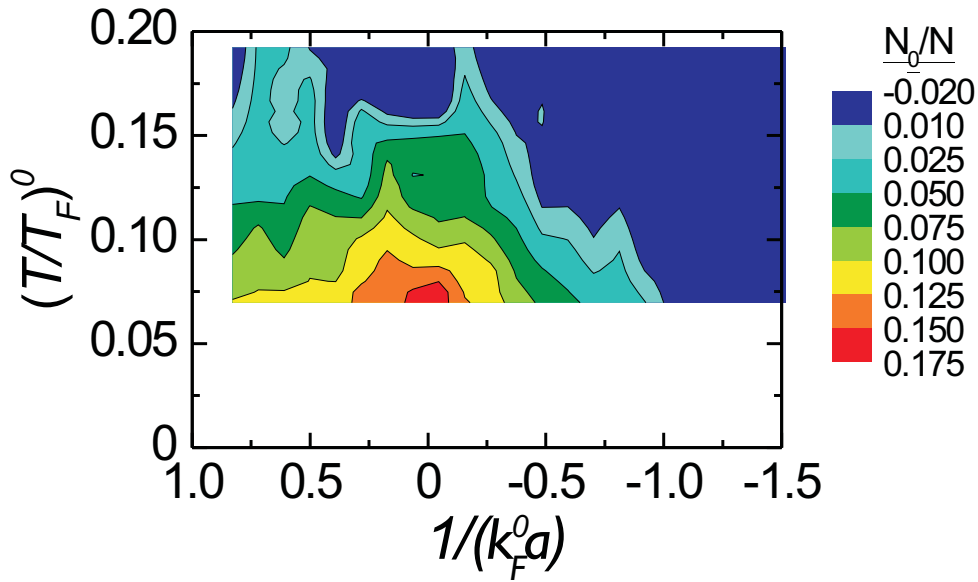

Figure 8.8: Transition to condensation as a function of both $\Delta B$ and $(T/T_F)^0$ [73]. The data for this phase diagram were collected using the same procedure as in Fig. 8.5 with $t_{hold} \approx 2$ ms. The false color surface and contour plot were obtained using a Renka-Cline interpolation of approximately 200 distinct data points.

These data lie precisely in the regime that is neither described by BCS nor by BEC physics, $-1 < 1/k_F a < 1$ (see Fig. 2.3). The condensed pairs in these experiments are pairs with some qualities of diatomic molecules and some qualities of Cooper pairs. Thus, these experiments realized a phase transition in the BCS-BEC crossover regime and initiated experimental study of this physics.

Finally, as in our previous measurements performed in the BEC limit, the mea-sured condensate fraction in Fig. 8.8 always remains well below one [68]. This is not observed in the case of $^6$Li experiments [74], suggesting that technical issues particular to $^{40}$K may play a role. As part of our probing procedure the magnetic field is set well below the Feshbach resonance where the molecule lifetime is only on the order of milliseconds [67, 95]. This results in a measured loss of 50% of the molecules and may also reduce the measured condensate fraction. Furthermore, temperature measurements using the surface-fitting procedure become less accurate below $T/T_F = 0.1$ (see Sec. 4). It also is expected that the isentropic ramp from a weakly interacting Fermi gas to the crossover results in a larger value of $T/T_F$ than $(T/T_F)^0$. The extent of this adiabatic heating is currently under theoretical investigation [131]. Density-dependent loss processes could also play a role in heating of the sample, especially as the BEC limit is approached (see Ch. 7). Whether any or all of these processes play a role in the small condensate fraction is a subject of current study in the Jin group.

# Chapter 9

# The momentum distribution of a Fermi gas in the crossover

The measurements of condensation in the crossover described in the last chapter probed the phase coherence between fermionic atom pairs. As discussed in Ch. 2, it is only in the BCS limit that the pairs are always coherent, and when the interaction becomes large so-called pre-formed pairs are predicted to exist above the phase transition temperature $T_c$ [8, 18]. To verify such theories it is important that techniques for detecting pairing be developed alongside studies probing the phase transition. There have been a number of probing techniques used to detect pairing including rf spectroscopy [80], magnetic-field modulation [81], spectroscopic probes [72], and measurements of the atomic momentum distribution [62, 78]. This chapter focuses on measurements of the atomic momentum distribution in the BCS-BEC crossover using $^{40}$K [78].

The classic characteristic change of the momentum distribution of a superfluid Fermi system is a broadening of the Fermi surface (see for example [92]). Figure 9.1 (inset) shows the expected momentum distribution of a homogeneous, zero-temperature Fermi system. In the BCS limit ($1/k_F a \rightarrow -\infty$) the amount of broadening is small and associated with $\Delta$. As the interaction increases this effect grows until at unitarity ($1/k_F a = 0$) the effect is on order of $E_F$, and in the BEC limit ($1/k_F a \rightarrow \infty$) the momentum distribution becomes the square of the Fourier transform of the molecule wavefunction (see for example [48]). This kinetic energy increase can be interpreted as a cost of pairing. Amazingly, the total energy of the system is lowered despite this large kinetic energy increase.

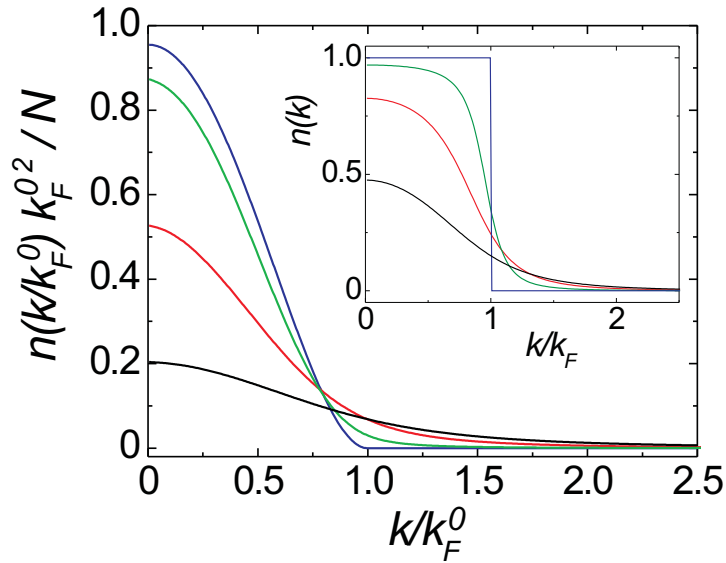

Figure 9.1: Theoretical column integrated momentum distributions of a trapped Fermi gas $n(k)$ calculated for $T = 0$ using NSR theory [175]. The normalization is given by $2\pi \int n(k)k\,dk = N$. The lines correspond to $a = 0$ (blue), $1/k_F^0 a = -0.66$ (green), $1/k_F^0 a = 0$ (red), and $1/k_F^0 a = 0.59$ (black). (inset) Corresponding distributions for a homogeneous system.

Here I will discuss the atom momentum distribution of a trapped Fermi gas. The case of an inhomogeneous trapped gas is more complicated than the homogeneous case. Nonetheless, in the strongly interacting regime the momentum distribution probes the pair wavefunction and consequently the nature of the pairs: Small, tightly bound pairs will broaden the momentum distribution more than large, weakly bound pairs. Similarly a fully paired gas will broaden the distribution more than a gas with a small fraction of paired atoms. It is expected that this measurement will probe the pairs independent of whether they have formed a condensate. Throughout this chapter I will compare our data to theory of the momentum distribution of a trapped gas. This theoretical work is the result of a collaboration between our group, Murray Holland of JILA, and Stefano Giorgini, a JILA visiting fellow from Trento, Italy.

# 9.1   Measuring the momentum distribution

We found that the momentum distribution of a Fermi gas in the crossover could actually be measured using the standard technique of time-of-flight expansion followed by absorption imaging [22]. The key to measuring the **atom** momentum distribution is that the gas must expand freely without any interatomic interactions; to achieve this we quickly changed the scattering length to zero for the expansion. This was particularly convenient using $^{40}$K because the zero crossing of the scattering length occurs only 7.8 G above $B_0$. Bourdel *et al.* pioneered this type of measurement using a gas of $^6$Li atoms at $T/T_F \approx 0.6$, where $T_F$ is the Fermi temperature [62]. In this work we carried out measurements down to $T/T_F \approx 0.1$, where pairing becomes a significant effect and condensates have been observed [73, 74].

To understand what we expect for our trapped atomic system, we can predict the atomic momentum distribution using a local density approximation and the results for the homogeneous case. In the trapped gas case, in addition to the local broadening of the momentum distribution due to pairing, attractive interactions compress the density profile and thereby enlarge the overall momentum distribution. Figure 9.1 shows a calculation of an integrated column density from the result of a mean-field calculation at $T = 0$ as described in Ref. [175].

First, I will discuss the atomic momentum distribution measured with a low temperature Fermi gas. We started with a weakly interacting $m_f = -7/2, -9/2$ gas at $T = 0.12\ T_F$ in a trap with a radial frequency of $\nu_r = 280$ Hz and $\nu_z/\nu_r = 0.071$.[1] We then adiabatically increased the interaction strength by ramping the magnetic field at a rate of $(6.5\ \text{ms/G})^{-1}$ to near the $m_f = -7/2, -9/2$ Feshbach resonance. After a delay of 1 ms, the optical trap was switched off and simultaneously a magnetic-field ramp to $a \approx 0$ ($B = 209.6$ G) at a rate of $(2\ \mu\text{s/G})^{-1}$ was initiated. The rate of this magnetic-field ramp was designed to be fast compared to typical many-body timescales as determined by $\frac{h}{E_F} = 90\ \mu$s. The gas was allowed to freely expand for 12.2 ms, and then an absorption image is taken. The imaging beam propagated along $\hat{z}$ and selectively probed the

---

[1]For this measurement we introduced the use of a crossed dipole trap configuration. In addition to the usual beam with a 15 $\mu$m waist, we focused a w=200 $\mu$m beam oriented parallel to the force of gravity ($\hat{y}$) on the atoms.

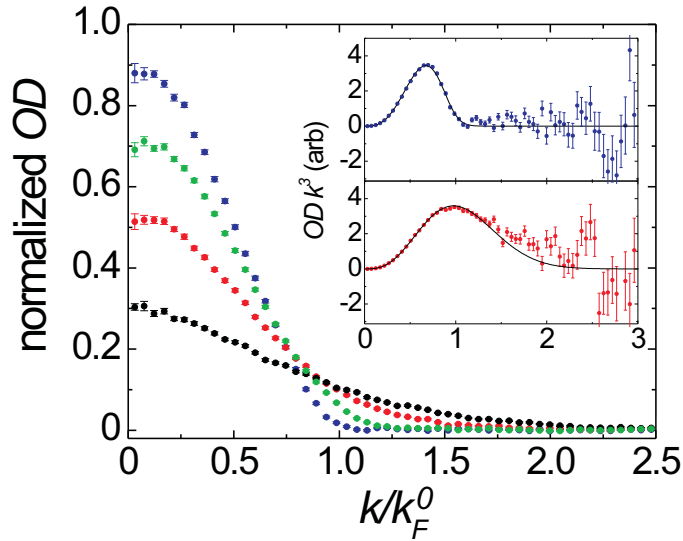

Figure 9.2: Experimental, azimuthally averaged, momentum distributions of a trapped Fermi gas at $(T/T_F)^0 = 0.12$ normalized such that the area under the curves is the same as in Fig. 9.1 [78]. The curves correspond to $1/k_F^0 a = -71$ (blue), -0.66 (green), 0 (red), and 0.59 (black). Error bars represent the standard deviation of the mean of averaged pixels. (inset) Curves for $1/k_F^0 a = -71$ (top) and 0 (bottom) weighted by $k^3$. The lines are the results of a fit to Eqn. 9.1.

$m_f = -9/2$ state.

Samples of these absorption images, azimuthally averaged, are shown in Fig. 9.2 for various values of $1/k_F^0 a$. There is a dramatic change in the distribution that is qualitatively very similar to the prediction in Fig. 9.1. Some precautions need to be taken in the quantitative comparison of Figs. 9.1 and 9.2. First, the magnetic-field ramp to the Feshbach resonance, while adiabatic with respect to most time scales, is not fully adiabatic with respect to the axial trap period. Second, in the experiment an adiabatic field ramp keeps the entropy of the gas, not $T/T_F$, constant. However, the resulting change in $T/T_F$ should have a minimal effect on the distribution for $1/k_F^0 a < 0$ [131]. Third, the theory assumes $T = 0$ and does not include the Hartree term, thus underestimating the broadening on the BCS side compared to a full theory [103].

## 9.2 Extracting the kinetic energy

It is natural now to consider extracting the kinetic energy from the momentum distribution. While the momentum distribution should be universal for small momenta, for large momenta it is influenced by details of the interatomic scattering potential. In the extreme case of a delta potential, which we used for the calculation in Fig. 9.1, the momentum distribution has a tail with a $1/k^4$ dependence, giving rise to a divergence of the kinetic energy. In the experiment we avoid a dependence of the measured kinetic energy on details of the interatomic potential because the magnetic-field ramp is never fast enough to access features on the order of the interaction length of the Van der Waals potential, $r_0 \approx 60 \ a_0$ for $^{40}$K [117]. Thus, the results presented here represent a universal quantity, independent of the details of the interatomic potential. Although universal in this sense, the measured kinetic energy is intrinsically dependent on the dynamics of the magnetic-field ramp, with faster ramps corresponding to higher measured energies.

To obtain the kinetic energy from the experimental data exactly we would need to take the second moment of the distribution, which is proportional to $\sum k^3 OD / \sum k OD$. As illustrated in Fig. 9.2 (inset) this is difficult due to the decreased signal-to-noise ratio for large $k$. Thus, our approach was to apply a 2D surface fit to the image and extract an energy from the fitted function. In the limit of weak interactions the appropriate function is that for an ideal, harmonically trapped Fermi gas (see Ch. 4)

$$OD(x,y) = OD_{pk} \, Li_2(-\zeta e^{-\frac{x^2}{2\sigma_x^2}} e^{-\frac{y^2}{2\sigma_y^2}})/Li_2(-\zeta). \tag{9.1}$$

The kinetic energy per particle is then given by

$$E_{kin} = \frac{3}{2} \frac{m\sigma_x \sigma_y}{t^2} \frac{Li_4(-\zeta)}{Li_3(-\zeta)} \tag{9.2}$$

where $t$ is the expansion time.[2] Although Eqn. 9.1 is not an accurate theoretical description of the Fermi gas in the crossover, empirically we found it to be a fitting function that describes the data reasonably well throughout the crossover, as illustrated in Fig. 9.2 (inset).

---

[2]This calculation gives the total $E_{kin}$ in all three dimensions. Since the momentum distribution was only measured in two directions, Eqn. 9.2 requires the assumption that the third dimension reveals an identical distribution.

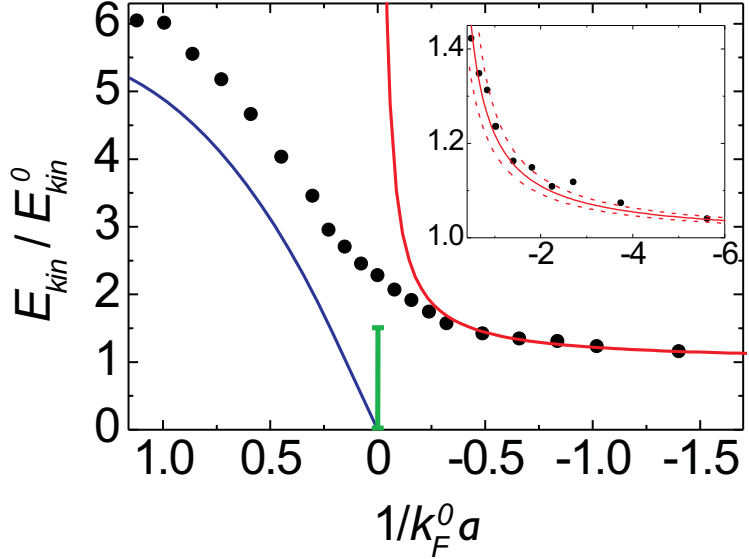

Figure 9.3: The measured energy $E_{kin}$ of a Fermi gas at at $(T/T_F)^0 = 0.12$ normalized to $E_{kin}^0 = 0.25 \ k_b \ \mu K$. The red line is the expected energy ratio from a calculation only valid in the weakly interacting regime ($1/k_F^0 a < -1$). The green bar represents the expected value of $E_{kin}/E_{kin}^0$ at unitarity just due to density profile rescaling. In the molecule limit ($1/k_F^0 a > 1$) we calculate the expected energy for an isolated molecule (blue line). (inset) A focus on the weakly interacting regime with the same axes definitions as the main graph.

Figure 9.3 shows the result of extracting $E_{kin}$ as a function of $1/k_F^0 a$; we see that $E_{kin}$ more than doubles between the noninteracting regime and unitarity. Using the fitting function of Eqn. 9.1 we could also extract information about the shape of the distribution through the parameter $\zeta$. Since $\zeta$ can range from -1 to $\infty$ it is convenient to plot the quantity $\ln(1 + \zeta)$ (Fig. 9.4). The shape evolves smoothly from that of an ideal Fermi gas at $T/T_F \approx 0.1$ in the weakly interacting regime, to a gaussian near unitarity, and to a shape more peaked than a gaussian in the BEC regime. These qualitative features are predicted by the mean-field calculation of the distributions in Fig. 9.1.

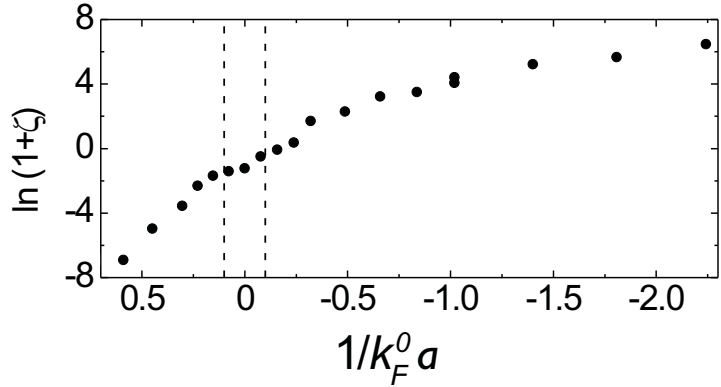

Figure 9.4: Shape of the momentum distribution as described through the parameter $\zeta$ (Eqn. 9.1). $\ln(1 + \zeta) = 0$ corresponds to a gaussian distribution, and for an ideal Fermi gas $\ln(1+\zeta)^{-1} \approx T/T_F$ in the limit of low $T/T_F$. The dashed lines show the uncertainty in the Feshbach resonance position [73].

## 9.3 Comparing the kinetic energy to theory

As mentioned earlier $E_{kin}$ of a trapped gas is affected both by the broadening due to pairing (Fig. 9.1 (inset)) and by changes in the trapped gas density profile. In the BCS limit, the broadening due to pairing scales with $e^{-\pi/2k_F^0|a|}$ and is thus exponentially small compared to density profile changes, which scale linearly with $k_F^0|a|$. In this limit we could calculate $E_{kin}/E_{kin}^0$ using a mean-field calculation in the normal state [105]; to lowest order in $k_F^0|a|$ we found $E_{kin}/E_{kin}^0 = \frac{2048}{945\pi^2}k_F^0|a|+1$. This result is plotted in Fig. 9.3 (inset) and shows good agreement for the weakly interacting regime ($1/k_F^0 a < -1$). However, caution must be taken with the apparent agreement for $-0.5 < 1/k_F^0 a < -1$. The agreement could be explained by the lack of pairing in this theory being compensated by the theory's overestimation of the density profile change when $|a|$ becomes larger than $1/k_F$.

In the crossover regime where the pairs are more tightly bound, pairing provides a significant contribution to the change in the momentum distribution. At unitarity a full Monte Carlo calculation predicts the radius of the Fermi gas density profile will become $(1 + \beta)^{1/4}R_0 = 0.81R_0$, where $R_0$ is the Thomas-Fermi radius of a noninteracting Fermi gas [103]. Just this rescaling would result in $E_{kin}/E_{kin}^0 = 1.54$ (green bar in Fig. 9.3). Thus, at unitarity, pairing effects on the momentum distribution must account for a large fraction of the measured value of $E_{kin}/E_{kin}^0 = 2.3 \pm 0.3$ (Fig. 9.3) and all of the observed change in distribution shape (Fig. 9.4).

In the BEC limit and at $T = 0$ the measured energy should be that of an isolated diatomic molecule after dissociation by the magnetic-field ramp. Provided the scattering length associated with the initial molecular state, $a(t = 0)$, is much larger than $r_0 \approx 60\ a_0$, the wavefunction for the molecule is given by $\psi = Ae^{-r/a(t=0)}/r$ where $r$ is the internuclear separation and $A$ is a normalization constant. M. Holland and S. Giorgini calculated the measured energy from the solution of the Schrödinger equation with a time-dependent boundary condition on the two-particle wavefunction $\frac{d \log(r\psi)}{dr}\big|_{r=0} = -\frac{1}{a(t)}$, where $a(t)$ is the scattering length fixed by the magnetic field at time $t$. In Fig. 9.3 is the result of this calculation for a pure gas of molecules with zero center of mass momentum and a $(2\ \mu s/G)^{-1}$ ramp rate. We found reasonable agreement considering that there is a large systematic uncertainty in the theory prediction resulting from the experimental uncertainty in the magnetic-field ramp rate and also that this two-body theory should match the data only in the BEC limit ($1/k_F^0 a \gg 1$).

A greater theoretical challenge is to calculate the expected kinetic energy for all values of $1/k_F^0 a$ in the crossover. This is a difficult problem because it requires an accurate many-body wavefunction at all points in the crossover and the ability to time-evolve this wavefunction. Recent work in Ref. [176] has carried out this calculation using the NSR ground state (see Ch. 2 and [48]). Qualitatively the result is sensible, but in the strongly interacting regime does not accurately reproduce the measured kinetic energy; this suggests that more sophisticated crossover theories are necessary.

## 9.4   Temperature dependence

We also studied the dependence of the momentum distribution on $(T/T_F)^0$. To vary the temperature of our gas, we recompressed the optical dipole trap after evaporation and heated the gas through modulation of the optical trap power. The experimental sequence for measuring the momentum distribution was the same as in Sec. 9.1 except the ramp rate to $a = 0$ for expansion was $\sim (8 \ \mu s/\mathrm{G})^{-1}$. Figure 9.5 shows the measured kinetic energy change $\Delta E_{kin} = E_{kin} - E_{kin}^0$. Ideally for this measurement we would have liked to have varied the temperature while maintaining a fixed density. However, for these data the density does vary some. For example, for the coldest dataset (black) the peak density, for atoms in one spin state, in the weakly interacting regime is $n_{pk}^0 = 1.4 \times 10^{13} \ \mathrm{cm}^{-3}$ and $E_F^0 = 0.56 \ \mu\mathrm{K}$. For the hottest dataset (magenta) $n_{pk}^0$ decreases to $6 \times 10^{12} \ \mathrm{cm}^{-3}$ and $E_F^0 = 0.79$ $\mu\mathrm{K}$.

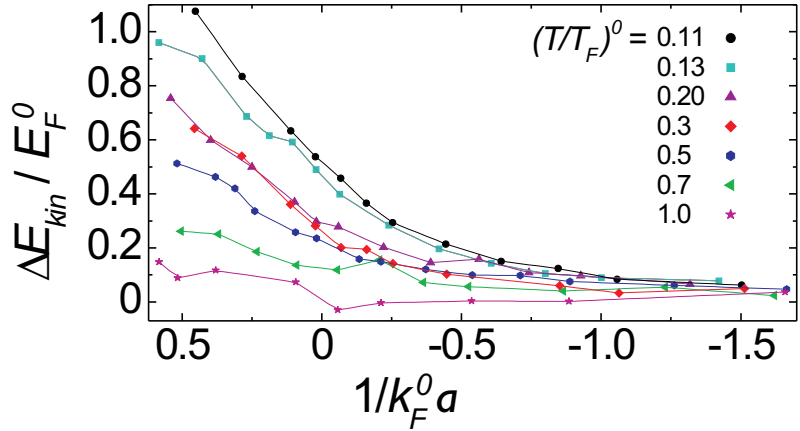

Figure 9.5: Temperature dependence of $\Delta E_{kin} = E_{kin} - E_{kin}^0$ normalized to the Fermi energy at $a = 0$. $(T/T_F)^0$ is the temperature of the noninteracting gas.

On the BEC side of the resonance, $\Delta E_{kin}$ decreases dramatically with $(T/T_F)^0$. Because $\Delta E_{kin}$ should be proportional to the molecule fraction, this result is closely related to the recent observation that the molecule conversion efficiency scales with $T/T_F$ [146]. In the strongly interacting regime we also observed a decrease in $\Delta E_{kin}$ with increasing $(T/T_F)^0$. However, the decrease is not as steep as would be expected if the phase transition, which occurs at $(T/T_F)^0 \approx 0.15$ as measured through $N_0/N$ (Fig. 8.8), were responsible for the change in the momentum distribution. Instead the temperature dependence of $\Delta E_{kin}$ is consistent with the expectation that the changes in the kinetic energy are caused by pairing and not condensation [73, 80, 18].

# Chapter 10

# The apparatus

The apparatus used for the experiments described in this thesis originated during the graduate career of Brian DeMarco; the elements constructed prior to 2001 are described in his thesis "Quantum behavior of an atomic Fermi gas" [83]. The technical work I completed involved, through the years, fixing a great deal of the apparatus components and inserting additional capabilities required to access BCS-BEC crossover physics. I will not document every change made to the apparatus. Instead, I will first provide an overview of the current apparatus (as of December 2005), namely a description of the components required for each step of the cooling and trapping sequence and a report on the final cold gas. (A portion of this information was already discussed in Ch. 4.) Then I will focus on six specific topics that were particularly important to the BCS-BEC crossover experiments or could be useful pieces of information for building future $^{40}$K experiments. These topics include our diode laser system, our optical dipole trap, the setup we use to create a large magnetic field, our circuit for fast magnetic-field control, our rf delivery system, and our absorption imaging configurations.

## 10.1   Procedure for making an ultracold $^{40}$K gas

The cycle to create an ultracold potassium gas with our apparatus begins with a vapor cell magneto-optical-trap (MOT). Since $^{40}$K is a very rare isotope of potassium (0.012% natural abundance), creating a large $^{40}$K MOT required development of enriched potassium sources [178]. These sources, which we refer to as "getters," are based on the design of SAES Getters Inc. but are made in the JILA shop. The JILA-made getters contain commercially available enriched KCl ($\sim$5% $^{40}$K) and a reducing agent, Ca. Enriched potassium is released through ohmic heating of the getter inside an evacuated glass cell; this forms a vapor from which $^{40}$K can be captured in a MOT. Because of the relatively low room-temperature vapor pressure of potassium, we heat our vapor cell to $\sim$50 °C to prevent potassium from sticking to the walls of the cell. Originally four potassium getters were placed in the vacuum system. We currently use the third of these getters. This particular getter has provided a constant $^{40}$K vapor pressure per amount of ohmic heating for over three years.

Table 10.1: Hyperfine constants for the $^{40}$K levels of interest in our experiments [177].

| $I$ | 4 |
|---|---|
| $g_I$ | -1.298099/4 |
| $A\ 4S_{1/2}$ | -285.731 MHz |
| $A\ 4P_{3/2}$ | -7.59 MHz |
| $B\ 4P_{3/2}$ | -3.5 MHz |

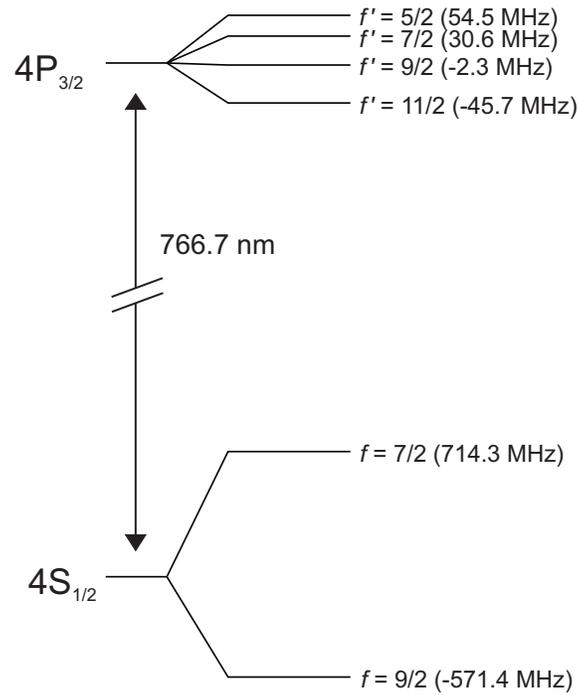

Figure 10.1: Hyperfine structure of relevant ground and excited state levels of $^{40}$K [177].

The $^{40}$K MOT uses light red detuned of the $f$=9/2 $\rightarrow$ $f'$=11/2 transition (trap light) and of the $f$=7/2 $\rightarrow$ $f'$=9/2 transition (repump light). Due to the small excited-state hyperfine splitting of $^{40}$K (Table 10.1 and Fig. 10.1), a relatively large repump intensity is required. Our vapor cell MOT is aligned in a retro-reflected configuration using laser beams with a 1.4 inch diameter. The total trap power is $\sim$90 mW and the repump power $\sim$50 mW. Setting the potassium vapor pressure such that the exponential MOT loading time is $\sim$1 sec, the result is a $^{40}$K MOT containing over $10^9$ atoms.

The vapor cell where we collect our potassium atoms we refer to as the "collection cell" (Fig. 10.2). The background pressure in this cell is too high to evaporatively cool the atoms. Thus, for the next stage of the experiment we transfer the atoms from the collection cell to another glass cell, the "science cell," which is separated from the collection cell by a long narrow tube (the transfer tube) [123]. The limited conductance of the transfer tube isolates the collection and science cells (Fig. 10.2), such that the vacuum pressure in the science cell is characterized by an exponential atom lifetime of $\sim$150 sec. The UHV components required to achieve this pressure are shown in Fig. 10.2; the main components of the original system were two ion pumps and a titanium sublimation pump (TSP). The ion pump along the transfer tube (P2) was turned off when it began to adversely affect the vacuum. Now the system is only pumped by one 40 l/s ion pump (P1). It has not been necessary to fire the titanium sublimation pump since 2001.

To transfer the atoms from the collection cell to the science cell we use a resonant laser beam to push the atoms down the transfer tube [83]. The atoms are guided by hexapole magnets that surround the tube. In the science cell the atoms are caught by a second MOT, the "science MOT". The push beam, the collection MOT, and the science MOT are currently all left on continuously as the atoms are transferred. With a well-optimized alignment of the push beam, this technique results in an exponential science MOT loading time of ∼10 sec. In the science cell the atoms are held in what is known as a "dark spot MOT"; here the repump beam intensity is significantly decreased at the location of the atom gas, resulting in fewer collisional losses [179].

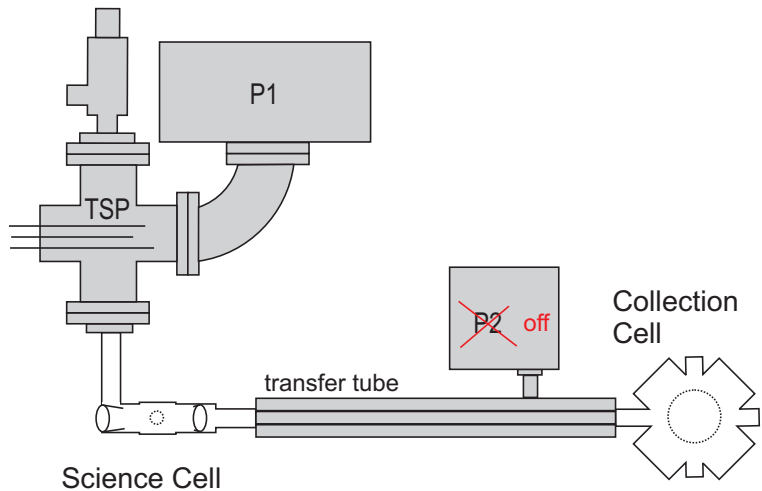

Figure 10.2: The vacuum system for our experiments as viewed from above; the system has remained essentially unchanged since 1998. Figure adapted with permission from Ref. [83].

The next step is to load atoms from the science MOT into a magnetic trap where they can be cooled using microwave evaporation. The magnetic trap we use is a cloverleaf design Ioffe-Pritchard style trap (Fig. 10.3). Before loading the atoms into the magnetic trap the MOT is moved to a position optimized for loading into the center of the magnetic trap, the trap light detuning is moved closer to resonance to provide as much Doppler cooling as possible (CMOT stage),[1] and the atoms are optically pumped towards the $f$=9/2, $m_f$=9/2 state (Fig. 4.1). We rely upon imperfect optical pumping to create the mixture of atoms in the $m_f$=9/2 and $m_f$=7/2 spin states required for evaporative cooling. We succeed in loading ∼55% of the atoms from the science MOT into a weak magnetic trap. Following the load we compress the magnetic trap for evaporation to a trap with a radial frequency of ∼250 Hz and an axial frequency of ∼20 Hz.

---

[1]Note this procedure is opposite of the requirement for sub-Doppler cooling. While sub-Doppler cooling has been observed in other potassium MOTs [180], we have never found it beneficial in our apparatus.

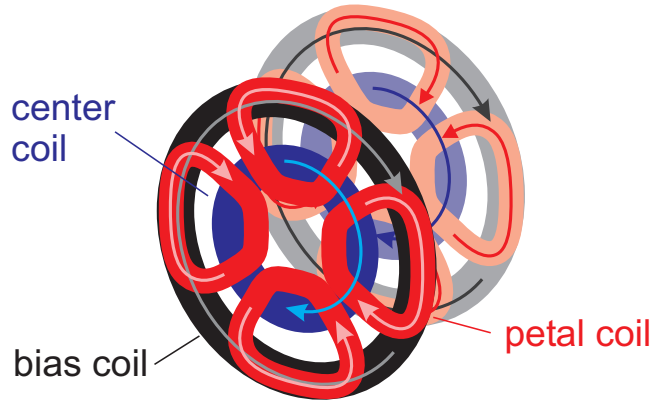

Figure 10.3: Coils of the cloverleaf Ioffe-Pritchard magnetic trap. Figure adapted with permission from Ref. [83]. The coils are wound with hollow copper tubing through which cooling water flows. These coils are reused to create a homogeneous magnetic field for accessing Feshbach resonances.

To remove the highest energy atoms for evaporative cooling in the magnetic trap, microwaves between 1200 and 1300 MHz are used to transfer atoms to untrapped spin states in the $f$=7/2 hyperfine state (Fig. 4.1) [32, 83]. The microwaves are delivered to the atoms using a stub-tuned microwave coil designed to deliver radiation in the 1200 - 1300 MHz range [83]. In the current procedure only one microwave frequency is required at a given time for evaporation; this is in contrast to the two-frequency evaporation scheme used to achieve two-component quantum degenerate gases in Ref. [83]. Typically we only cool the $^{40}$K sample in the magnetic trap to $T/T_F \approx 3$ ($T \approx 5$ $\mu$K with a few $10^7$ atoms) before loading the atoms into an optical dipole trap.

The procedure for obtaining a cold gas of atoms in our optical trap in the Feshbach resonance spin states includes the following. After loading the optical trap we immediately begin forced evaporation by lowering the optical trap power.

(This process occurs continuously throughout the next steps.) We then transfer the atoms from the $m_f=+9/2, +7/2$ spin states to the $m_f=-9/2, -7/2$ spin states using a rf frequency sweep across all the Zeeman transitions at $B \approx 20$ G. Next we move to a magnetic field above a Feshbach resonance, typically 235 G. This large magnetic field is supplied mainly by the "bias coils" of Fig. 10.3. Here we apply a $\pi/2$ rf pulse on the $m_f=-9/2 \leftrightarrow m_f=-7/2$ transition to achieve an equal mixture of the two spin states. We follow this spin transfer by further cooling at high magnetic field. We then hold the atoms for typically 100 ms at a final optical trap power that defines both the final evaporative cut and the parameters of the final trap.

Experiments studying BCS-BEC crossover physics happen in a relatively short amount of time at the end of the experiment cycle. The crossover experiments mostly involve changing the magnetic field, either slowly with the large magnetic field coils of Fig. 10.3 or very quickly with a low-inductance auxiliary coil pair (Sec. 10.5). The final step in an experiment cycle is to acquire an absorption image of the gas. It is important for the success of our experiments to have the ability to take absorption images of the gas with high resolution and in many different circumstances, i.e., at high and low magnetic fields, from multiple directions, and with the ability to distinguish the different spin states.

Many of the tasks during the experiment cycle that I have now described require precise timing. The heart of the system used for this timing is described in Brian DeMarco's thesis [83]. The software for the system is quite old, dating back to the JILA experiment of Ref. [124]. A QuickBASIC program is used to control TTL, DAC, and GPIB boards, operating with a clock now provided by a function generator operating at 3 kHz. The TTL and 12-bit DAC boards were made by Keithley Instruments. We have added additional precise timing that is controlled by the main timing system via TTL triggers. Precise timing instruments we have added include function generators (both the Agilent 33220A and the SRS DS345), two Quantum Composers pulse generators (Model 9614), and a National Instruments 6733 analog board.

Analysis of the absorption images shows that our apparatus can cool a Fermi gas to $T/T_F \approx 0.1$ or possibly colder (see Ch. 4). One figure of merit of the apparatus, besides the final $T/T_F$ it can reach, is the number of atoms achieved at a phase space density near quantum degeneracy. A large atom number (and hence a large density) is not necessarily always a desirable feature; however, producing a large enough atom number for ample signal has contributed to the success of many of our experiments. Under optimum conditions we can achieve a two-component gas at $T/T_F = 0.52$ with $2 \times 10^6$ atoms per spin state. Thus, if our atoms were bosons, we would have two Bose gases at $T_c$ with $2 \times 10^6$ atoms in each gas. This result is comparable to or better than many BEC experiments.

## 10.2 Diode laser system

The laser light for the trapping, cooling, and probing of $^{40}$K in our apparatus is provided by a semiconductor-diode laser system. The laser design is outlined in Fig. 10.4; this schematic diagram does not represent the actual geometric layout on our optics table and omits, for example, all mirrors and polarization rotation optics. The two main components are external cavity diode lasers (ECDLs), which provide narrow linewidth light ($\sim$1 MHz) near the repump and trap transitions. The first of these lasers, the "repump laser," is stabilized to the $f$=2 transition of $^{39}$K by locking to the peak of the corresponding saturated absorption spectroscopy peak (Fig. 10.5). The locking light is shifted using an acousto-optic modulator (AOM) 448 MHz blue of the main beam, which is near the repump transition. The "trap laser" is stabilized using a dichroic-atomic-vapor laser lock (DAVLL) [181]; this lock provides a wide dispersive signal that allows the laser to be locked anywhere near the potassium transitions [83]. Beating the trap laser against locking light from the repump laser using a 1 GHz photodetector provides a frequency reference for setting the DAVLL lock point. Using the DAVLL we can typically tune the trap laser by 220 MHz in an experiment cycle.

Constructing external cavity diode lasers for the potassium transition (767 nm) is more difficult than, for example, for the rubidium transition (780 nm). This is because the readily available diodes in this wavelength range are centered at 780 to 785 nm. Since laser diodes typically decrease their wavelength with cooling by 1 nm per 5 °C, historically an approach to making lasers for our apparatus was to cool laser diodes. A diode centered at 780 nm at room temperature will reach 767 nm when cooled to around -40 °C; selecting nominally 780 nm diodes with anomalously low wavelengths can help make the cooling requirement less severe [83]. Nonetheless, building and maintaining a stable cold laser requires considerable time and effort; thus, we have moved towards alternatives. These alternatives include, for example, buying ECDLs made by New Focus Inc. (Vortex lasers) designed for 767 nm. The Vortex laser option has the disadvantage that the frequency noise on these lasers is often larger than JILA-built lasers. Another alternative is to use diodes that are anti-reflection (AR) coated. AR coating tends to pull the gain profile of the diode blue as well as allow a much wider tuning range within an external-cavity configuration. A particularly good source of AR coated diodes that has emerged recently is Eagleyard Photonics.

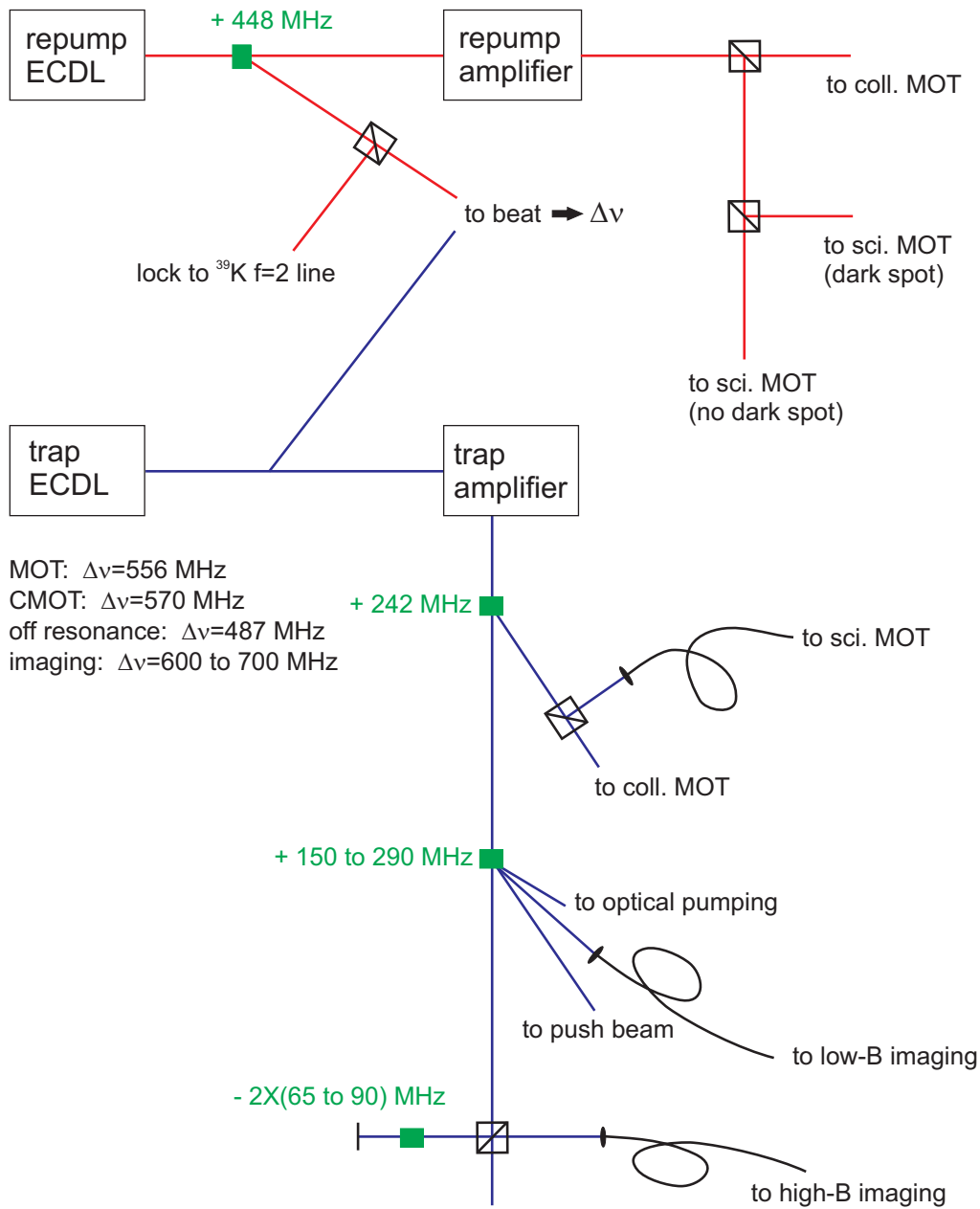

Figure 10.4: Laser frequencies used in the experiment for trapping, cooling, and probing of $^{40}$K . The green boxes are AOMs; the blue lines represent light near the $^{40}$K $f$=9/2 → $f'$=11/2 transition (trap light); the red lines represent light near the $^{40}$K $f$=7/2 → $f'$=9/2 transition (repump light). A + sign indicates a blue shift of the frequency, and a - sign indicates a red shift of the frequency. The loops depict optical fiber.

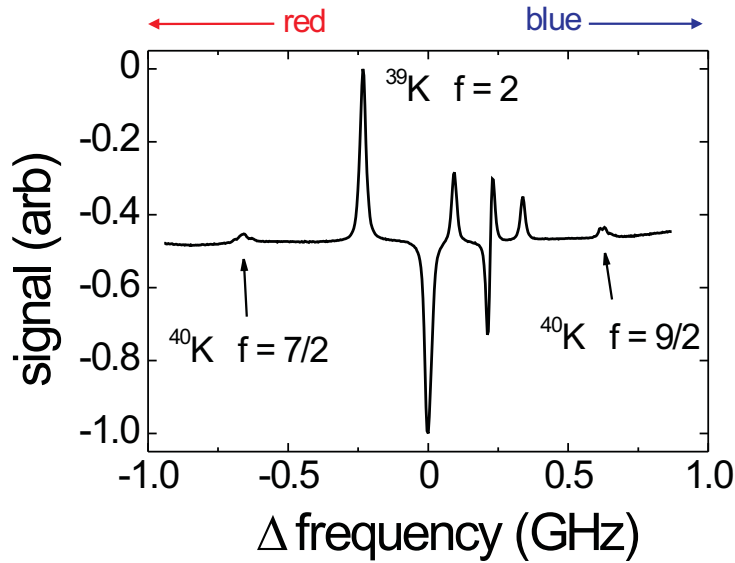

Figure 10.5: Saturated absorption spectroscopy of an enriched source of potassium atoms. Our lasers are locked to the $^{39}$K $f = 2$ line, which lies between the two $^{40}$K lines. Data from Ref. [182]. See, for instance, Ref. [183] for a discussion of diode laser stabilization using saturated absorption spectroscopy.

The ECDLs currently used in our apparatus each produce ~10 mW of laser power. Thus, for our experiments we amplify the power from the ECDLs to attain the power we require for the MOTs. For this purpose we use SDL Inc. power amplifiers (Fig. 10.4). In contrast to nearly everything else in the laser system, these amplifiers have not been modified or replaced at all in over 5 years. It is astounding that both of these sensitive devices have had such a long lifetime. Unfortunately, SDL no longer produces power amplifiers, but Eagleyard photonics now makes similar devices, yet with a somewhat lower gain and perhaps a shorter lifetime.

The laser setup of Fig. 10.4 has proven to be very robust in its ability to reliably operate day-to-day. Key features that contribute to this ability include: The design keeps the number of components to a minimum. There are only two lasers to maintain and only four AOMs to align and control. Further, for many of the AOMs and incidental beams in the bottom section of Fig. 10.4, there is always ample power available. This is a result of a design in which light for various purposes is siphoned off a main beam by AOMs.

## 10.3 A stable optical dipole trap

In Ch. 4 we discussed how an optical trap can be created using a gaussian laser beam far red-detuned of atomic transitions. In this section I outline our current system for creating a stable optical trap and provide tips for building and stabilizing such a trap.

The most important design parameters of an optical trap are the waist and the power. The power is an easily tunable parameter, while the waist is most often fixed during an experiment cycle. Together the waist and power determine the trap depth and frequencies (Eqns. 4.2 and 4.3). The depth of the trap is very important for evaporative cooling as it defines the temperature of the gas during evaporation (evaporative cut); the trap frequencies determine the collision rate for evaporation as well as the density of the final sample used for BCS-BEC crossover studies. A smaller waist will result in a higher mean trap frequency for a given trap depth (Eqn. 4.3); on the other hand, a larger waist will make a weaker trap for an equivalent depth. The waist size also affects the aspect ratio: A small waist results in a less extreme aspect ratio (a smaller $\nu_r/\nu_z$). If the waist is made too large the axial direction will become too weak at deep cuts to hold atoms against realistic magnetic-field gradients.

In Ch. 4 we discussed only the case of a trap formed by a single gaussian laser beam, and this is indeed the trap that was used for most of the experiments described in this thesis. However, a single beam trap has the disadvantage that the aspect ratio is severe and cannot be controlled independently of the trap strength. Thus, our current system contains a crossed dipole beam configuration in which a second red-detuned beam perpendicular to the first is overlapped with the atoms. In our apparatus this beam points along $\hat{y}$ and thus is referred to as the "vertical beam". To maintain a relatively low density atom gas the waist of this beam is made significantly larger than the first beam. For example, for the measurements in Ch. 9 the main beam had a 15 $\mu$m waist and the vertical beam a 200 $\mu$m waist. The technical challenges related to assembling a second beam are equivalent or simpler than those for the main beam. Thus, in the rest of this section I will focus solely on constructing our main optical trap beam.

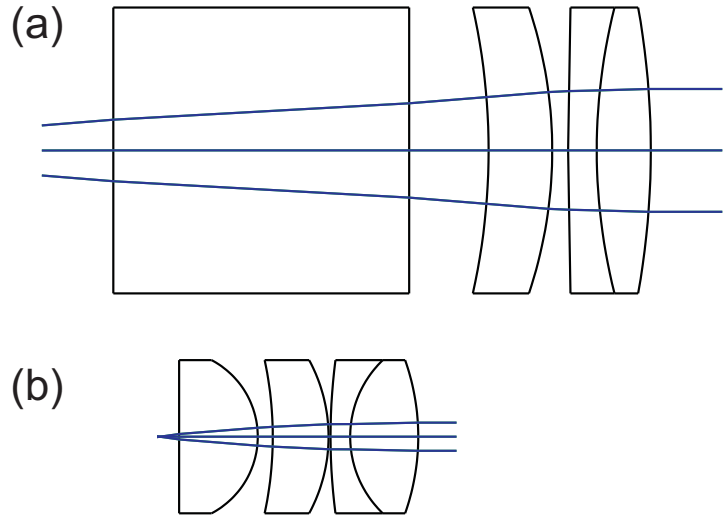

Figure 10.6: Useful options for collimating a 1064 nm beam out of a fiber. (a) A self-constructed multielement lens (design by M. Greiner). This set contains a piece of silica (n=1.45), a Melles Griot meniscus lens (LAM111/077), and a Melles Griot achromat (LAO111/077); it has an effective focal length of 48.2 mm and a diameter of 18 mm. (b) A pre-assembled multielement lens, the Melles Griot 06 GLC 001, which has an effective focal length of 6.56 mm and a 10 mm diameter.

To create our optical dipole trap we start with light from a CW laser at 1064 nm. We purchased a diode-pumped YAG laser made by Spectra Physics (Spectra Physics part number T20-BL-106C). This laser runs multimode (but single transverse mode) and can produce greater than 4 W of power. The light from the laser is coupled into a polarization maintaining fiber, which must be angle cut or AR coated to avoid intensity fluctuations due to cavity formation in the fiber. The light coming out of the fiber is then collimated to the appropriate size for creating the desired gaussian waist. For example, to create a 15 $\mu$m waist at the atoms using a 200 mm focusing lens the the beam must be collimated to w=4.5 mm.

Usually fiber output couplers are designed to collimate a beam to a much smaller size than 4.5 mm. Thus, if a typical fiber output coupler is used a telescope must be employed to expand the small beam. The optical design of Fig. 10.6(a) avoids these multiple steps and creates a much more compact setup by directly collimating the beam out of the fiber to a large size. Care must be taken in designing such a lens system to maintain low spherical aberration. It is important to optimize the thickness of the plate of glass that forms the first element of Fig. 10.6(a). This optimization can be done with a ray-tracing program that can calculate the geometrical spot size that results from propagation through a set of optics. A program we use for this purpose is a product of Lambda Research Corporation, OSLO Light. (OSLO stands for Optics Software for Layout and Optimization.) This program, in addition to carrying out computations based upon geometric optics, will perform propagation of a gaussian beam through a lens system using ABCD matrix analysis. Lens specification files can be imported directly into the program, and the software also creates drawings such as those in Fig. 10.6.

Fig. 10.6(b) shows a suggested coupler in the case where the beam must be collimated to a small size and later expanded with a telescope. Such a design might be useful if a small optic needs to be placed in the beam after the fiber, such as an AOM or a polarizer. Using the collimator also creates a slightly more flexible system because the size of the beam, and hence the waist, can be changed by altering just one lens in the telescope, instead of replacing and reoptimizing the collimating lens.

Following assembly of the optical components it is important to check for aberrations and astigmatism (different focal positions in $x$ and $y$) in the final focus. A useful tool for this is a beam-profiling camera; we have found that the WinCamD beam-profiling camera manufactured by DataRay Inc. works well. Fig.

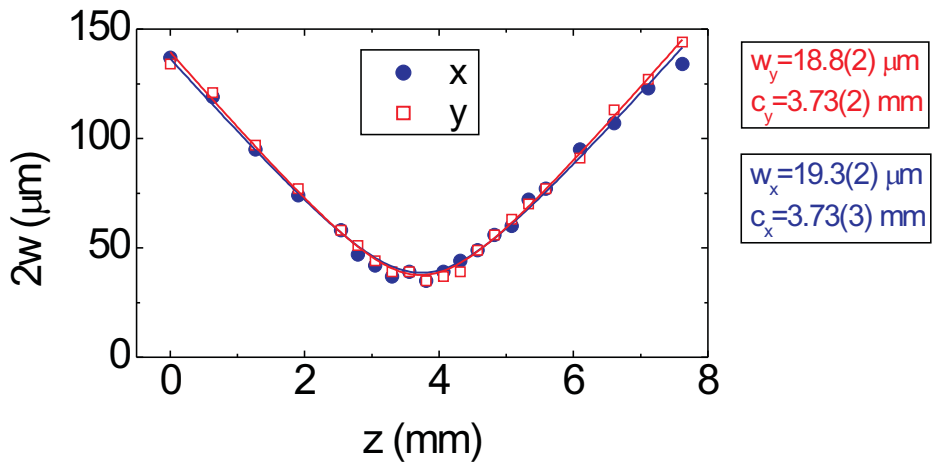

Figure 10.7: Measured optical trap axial profile. The beam sizes in the $x$ and $y$ planes are measured as a function of $z$ using a beam profiling camera. $\mathrm{w}_x$ and $\mathrm{w}_y$ are the fitted gaussian waist sizes and $c_x$ and $c_y$ the center positions. The fitting function is $2\mathrm{w}\sqrt{1 + (\frac{(z-c)\lambda}{\pi \mathrm{w}^2})^2}$.

10.7 shows the result of measuring the beam size on this camera as a function of the position along the axis of the beam. Good signs are that the profile is symmetric about the minimum of w, the points fit well to the expected form for a gaussian focus for 1064 nm light, and the two directions ($x$ and $y$) focus at the same position.

After creating a suitable gaussian focus, the next concern is the stability of the optical trap, as fluctuation in optical trap parameters can cause heating of the atom gas. Noise on the laser intensity leads to fluctuations in the trap spring constant, while noise in the pointing of the gaussian beam leads to fluctuations in the trap center. Thus, intensity noise leads to heating when the fluctuations are at twice the trap frequency, while pointing noise leads to heating when the fluctuations are at the trap frequency. A theoretical analysis of the effect of intensity and pointing noise on optically trapped atoms is presented in Ref. [184].

We minimize intensity fluctuations by servoing the optical trap power. Our servo uses feedback from a pickoff of the trapping light power after the optical fiber, and an AOM before the fiber controls the power. To achieve good pointing stability of the optical trap we do not actively stabilize the beam position, but rather rely on mounting the optical trap optics in a way that minimizes vibrations. When we first started building optical traps, we did not pay particular attention to the stability of the mounted optics. In this case we found that the heating rate of the atoms in the trap was severely limited by the stability of the optics. Thus, in subsequent configurations the optics have been mounted on thick posts, which we sometimes fill with lead shot to damp vibrations. Moving parts, such as mirror mounts or lens translators, are minimized. Following continual improvement to the stability of the optics the heating rate of our trap is now $\sim$5 nK/sec.

The last step is to align the optical trap focus on the atoms. In our apparatus we must send the beam through the science cell along an optical path already occupied by a MOT beam. Useful methods of combining the MOT light (767 nm) and the optical trapping light (1064 nm) include using motorized mirrors that can be triggered with a TTL signal during the course of the experiment and dichroic plates that reflect 767 nm light, yet transmit 1064 nm light. (CVI Laser manufactures dichroic plates in a variety of sizes with better than 99% reflectance at 767 nm and 95% transmission at 1064 nm.) The optical trap beam should be aligned at a few degrees with respect to the cell normal; this avoids interference patterns caused by back-reflections from the cell wall. Precisely aligning the optical trap beam onto the atoms is not a trivial task since the beam must be placed with $\sim$10 $\mu$m precision to trap the atoms. The best way to roughly align the beam is to image both the atoms and the optical trap beam onto the same position on a CCD camera. (One must be careful to account for any chromatic shifts, however.) This brings the beam close enough to trap a fraction of the atoms. We determined the final beam alignment by optimizing the number of atoms in the optical trap after evaporation.

To evaporate the atoms in the optical trap the power at the atoms is controlled with the AOM intensity servo described above. The power during a typical cycle of the experiment is shown in Fig. 10.8. The trap is ramped on in $\sim$100 ms while the magnetic trap is still on. Then the magnetic trap is switched off suddenly and the optical trap evaporation is immediately initiated.[2] At the end of the sequence the gas is expanded for time-of-flight imaging by turning off the optical trap power

---

[2]Switching off the magnetic trap suddenly will not result in a perfectly adiabatic transfer. However, for the hot gases we load into the trap the effect is negligible.

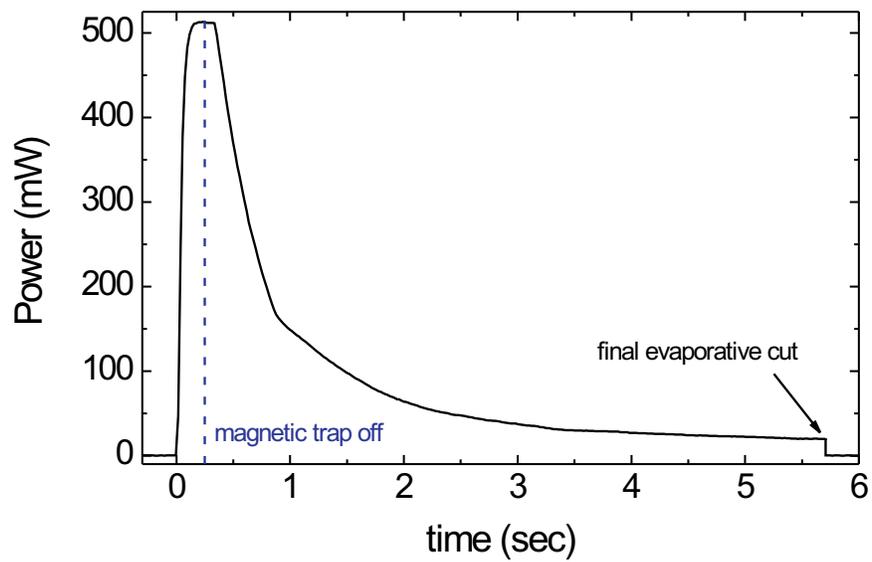

Figure 10.8: A typical optical trap evaporation trajectory to a modest final optical trap cut. The functional form of the decrease in intensity is a series of exponentials with changing time constants.

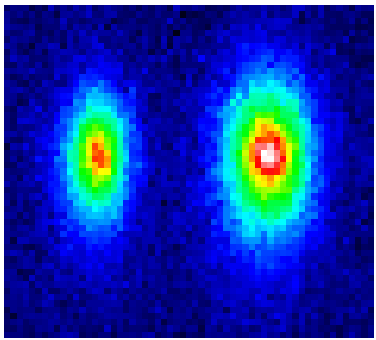

Figure 10.9: False color absorption image of two molecular condensates in a double-well potential created with a spatially-modulated optical trap. The gases are different sizes because there is an imbalance in the molecule number in the two wells.

very quickly ($< 1 \ \mu$s) using an AOM as a switch.

An additional feature of the current version of our optical trap is the placement of two crossed AOMs in the beam after the optical fiber. These AOMs can deflect the optical trap beam in $x$ and $y$. The position of the optical trap can thus be arbitrarily modulated at rates up to 1 MHz. Since the trap can be modulated on time scales much faster than the trap oscillator frequencies (at most a few kHz), a time-averaged potential can be created. This idea is similar to the time-orbiting-potential (TOP) magnetic trap [124] and was suggested by M. Greiner. With this design we succeeded in creating a double-well potential; Fig. 10.9 shows two molecular condensates that were creating in such a double-well potential. They are imaged by kicking the two condensates velocity in opposite directions just before time-of-flight expansion. A spatially-modulated optical trap could perhaps also be used to create a quartic potential or a rotating elongated trap for vortex creation.

## 10.4 Large magnetic-field control

For accessing Feshbach resonances we needed to a create a stable, homogeneous magnetic field at the atoms of up to ~250 G. To do this we modified the control circuitry for the cloverleaf Ioffe-Pritchard magnetic trap (Fig. 10.3) to allow the coils to be used not only as a trap, but also as a bias magnetic field. A schematic of our control over the coils is shown in Fig. 10.10. We can manipulate the current in the bias coils independently of the current through the petal and center coils. We start with a HP 6628A power supply operating in constant voltage mode. The current is monitored with F. W. Bell Hall effect current sensors ("H" in Fig. 10.10) placed in various locations. A low noise servo of these currents is built using high power FETs to control the current ("F" in Fig. 10.10). Each of the five FETs shown in Fig. 10.10 consists of a pair of water cooled, Advanced Power Technology MOSFETs in parallel. F1 and F3 are used only as switches for fast turn-on or turn-off of the current. F2, F4, and F5 are controlled via servos.

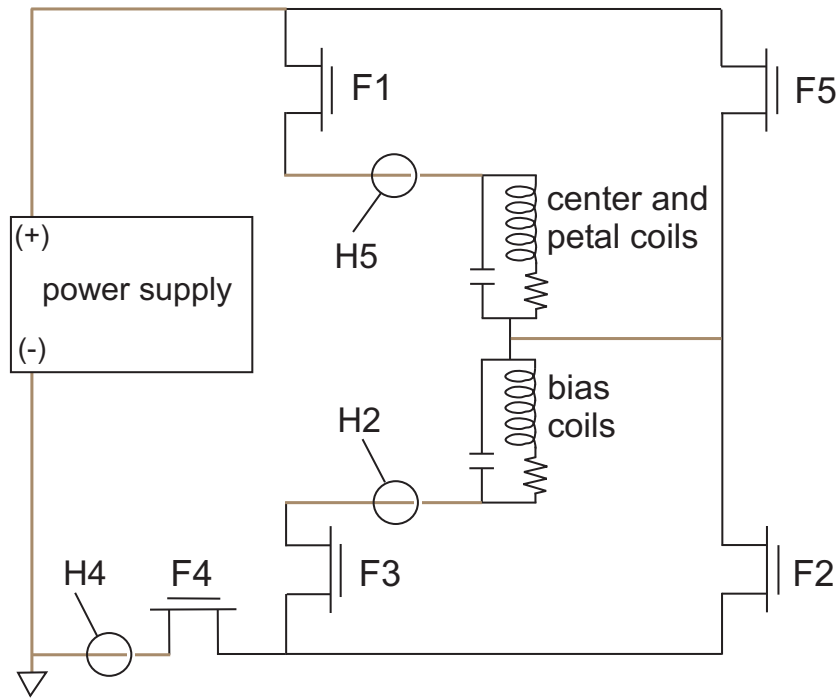

Figure 10.10: Schematic diagram of the circuit used to control the current through the coils of the Ioffe-Pritchard trap (Fig. 10.3). The coils can either be used to form a magnetic trap [83] or to provide a flat bias field for Feshbach resonance studies. The brown lines indicate long cables that contribute some resistance.

For operation of the magnetic trap F5 is switched off, F4 servos the full current (measured with H4), and F2 maintains the required current through the bias coil (measured with H2). For creating a bias field, F4 is switched on, F5 is allowed to pass current, and F2 again maintains the current through the bias coils. To make the magnetic field as homogeneous as possible across the axial direction of our trapped gas (see Fig. 10.15), we route a small amount of current through the center and petal coils to cancel the axial magnetic-field curvature created by the bias coils; this current is servoed with F5. Lastly, a small current through the science MOT quadrupole coils cancels out the residual axial magnetic field gradient (due to differences in the precise centers of the bias and center coil fields).

A key aspect of the magnetic-field control in our apparatus is its low noise and drift. To attain this we use PI servos with low-noise control voltages (see Ref. [83]). There is a designated ground for these magnetic-field servos to avoid 60 Hz noise from ground loops. The two most important current sensors, H2 and H4, are temperature stabilized and never used to monitor the performance of the servo (as monitoring the hall probe output on an oscilloscope can lead to ground loops and small shifts). At the position of H2 and H4 are additional "monitor" sensors, solely for monitoring the resulting current in the coils.

To test the magnetic-field stability that we achieve we probe the magnetic field with the atoms by driving rf transitions. Fig. 10.11 shows the result of such a measurement using optically trapped atoms at a bias field of $\sim$240 G. Atoms were transferred from the $m_f = -7/2$ spin state to the $m_f = -5/2$ spin state using a square rf pulse with a duration of $t_{pulse} = 240$ $\mu$s. A long rf pulse (and hence a narrow Fourier width) allowed us to sensitively measure the magnetic-field stability. We took multiple points at a frequency on the side of the line and measured the standard deviation. The points in Fig. 10.11 were taken over an hour and each error bar represents the standard deviation of over 10 points. This standard deviation corresponds to a 1 kHz fluctuation in the resonance frequency, which corresponds to a magnetic-field stability of 7 mG.

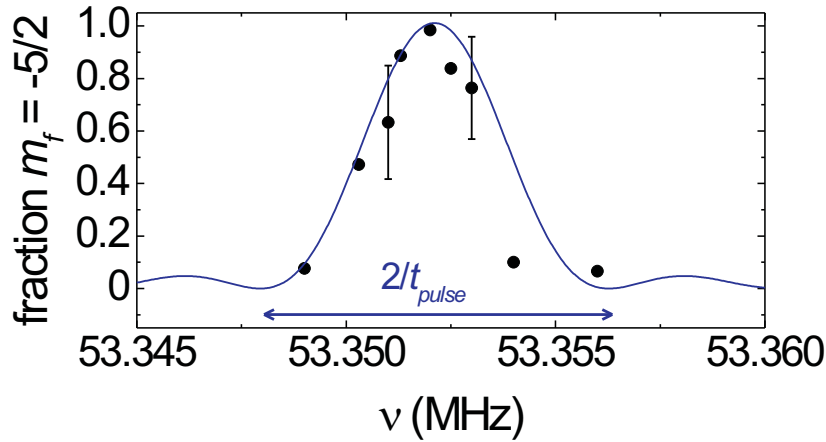

Figure 10.11: Transfer of atoms at 240 G between the $m_f = -7/2$ and $m_f = -5/2$ spin states as a function of rf frequency. The data are fit to the function for fourier transform of square pulse, $A\left[Sinc(\pi t_{pulse}(\nu - \nu_0))\right]^2$. $t_{pulse}$ is fixed at 240 $\mu$s, and $A$ and $\nu_0$ are allowed to vary.

## 10.5 Fast magnetic-field control

For many of the BCS-BEC crossover experiments we performed it was essential to be able to change the magnetic field very quickly. The large coils we use for the 200 G bias field are not well suited to this purpose since they have a large inductance and the servos that are used to control them were designed with a bandwidth of ~3 kHz. Thus, for magnetic-field ramps with inverse ramp rates of 1 $\mu$s/G to 100 $\mu$s/G we used low-inductance auxiliary coils and a high-bandwidth servo to ramp the magnetic field. A circuit diagram for this system is shown in Fig. 10.12. A Hall effect current sensor monitors the current and a Powerex IBGT controls the current through the circuit. The current is provided by discharging a capacitor that is charged to high voltage by a HP 6207B power supply [185]. Two coils, each mounted along the inner diameter of one of the center coils, produce the magnetic field at the atoms. The direction of this field can be quickly switched using the H-bridge configuration shown in Fig. 10.12. This allows us to jump the magnetic field either above or below the Feshbach resonance.

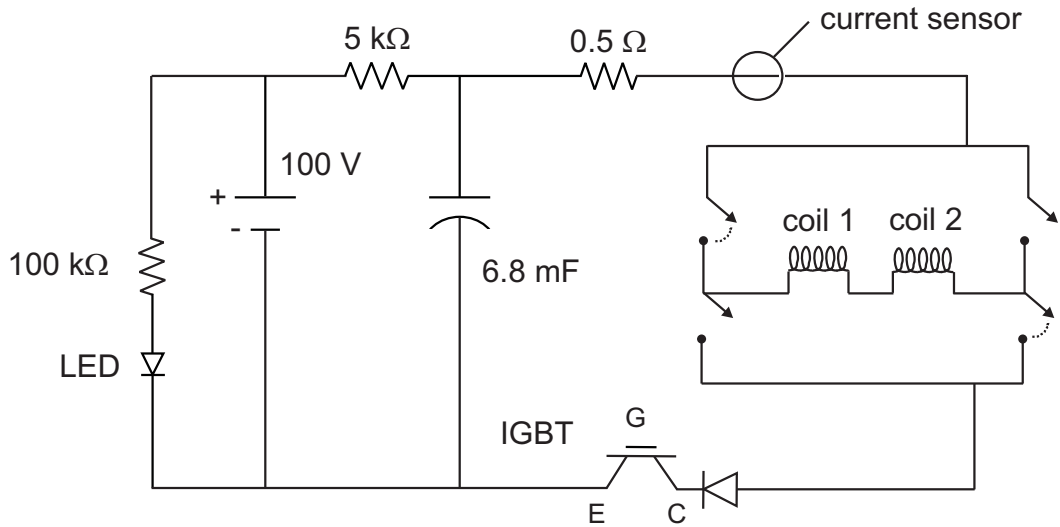

Figure 10.12: Circuit used to created fast magnetic-field pulses. The two coils shown are placed on opposite sides of the 1 inch wide cell holding the atoms, are 0.9 inches in diameter, have 5 turns each, and are wound from 24 AWG wire.

Fig. 10.13 shows the magnetic field produced with this circuit as determined by observing the current through the auxiliary coils as a function of time for various control voltage ramps. In the blue, red, and black curves the control voltage ramp occurs in 7 $\mu$s, 20 $\mu$s, and 40 $\mu$s, respectively. According to this plot, we find that we can ramp the magnetic field controllably with inverse rates less than 1 $\mu$s/G. Current induced in other coils in the system may reduce the true rate of the magnetic-field ramp.

## 10.6   Rf delivery

For the spin transfer to achieve a mixture of atoms in the Feshbach resonance states, we required a rf delivery system that could produce high power rf in two distinct frequency ranges. The transfer from the $m_f=+9/2, +7/2$ spin states to the $m_f=-9/2, -7/2$ at 20 G requires rf at $\sim$6 MHz, and the transfer amongst the $m_f=-5/2, -7/2$ and $-9/2$ states at magnetic fields near the Feshbach resonances requires rf in the 40 to 50 MHz range. The rf system we use consists of a 0-80 MHz function generator (Agilent 33250A), a high-power rf amplifier, and a 0.75 inch diameter coil consisting of 8 turns of wire located about 1.8 inches from the atoms.

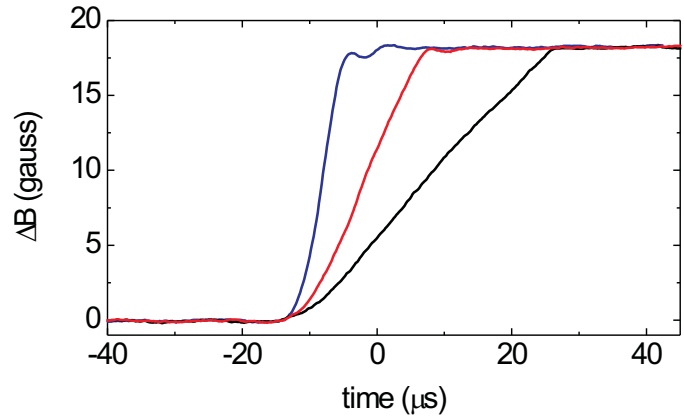

Figure 10.13: Magnetic field as determined through measuring the current through the current sensor in Fig. 10.12 as a function of time. For the three different curves the servo control voltage is changed at three different rates.

The key to delivering rf at the required frequencies was to properly impedance match the rf coil to the 50 $\Omega$ impedance of the transmission line. To deliver the most power to the atoms at a specific frequency it is best to purely reactively impedance match, i.e., not to add any resistance to the circuit. A system suitable for reactively impedance matching for our requirements is shown in Fig. 10.14 [186]. For the 40 to 50 MHz frequency range and our coil inductance, the required value of $L$ is a few $\mu$H and the value of C is a few pF. The advantage of this circuit is that since the capacitor is in parallel with the rf coil the dc response of the coil is maintained. This allows efficient delivery of rf at 6 MHz.

While the coil can deliver enough power throughout the 40 to 50 MHz range it is of course not perfectly impedance matched throughout this range. (Even at the optimum frequency only about 75% of the power is delivered to the coil.) Thus, invariably some power will be reflected back from the coil; this has the capability of damaging the amplifier. To avoid this problem we now use a class A amplifier made by ENI, the ENI525LA. The final result of our rf system is that we are able to apply a $\pi$ pulse between any of the $m_f=-5/2, -7/2, -9/2$ states in $\sim$15 $\mu$s.

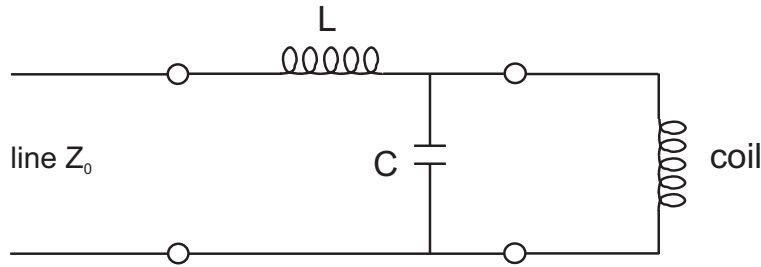

Figure 10.14: The circuit for the rf coil used to drive spin-changing transitions in the optical trap.

## 10.7 Absorption imaging

Nearly all of the information that we extract from our experiment comes from absorption imaging on the $f=9/2 \leftrightarrow f'=11/2$ cycling transition in $^{40}$K. The concept of absorption imaging was discussed in Ch. 4, and detailed derivations related to absorption imaging can be found in, for example, Refs. [124, 187]. In this section I will concentrate on the specific configurations that we use for absorption imaging $^{40}$K gases to extract the information needed for our BCS-BEC crossover experiments. Especially important for our crossover studies were the abilities to image spin selectively and at high magnetic field.

Fig. 10.15 shows the basic geometry for absorption imaging in our apparatus. The black oval represents the atom gas within the science cell and the two red arrows show the two possible resonant imaging beams. Each imaging configuration has a designated Princeton Instruments back-illuminated CCD camera, the CCD-512TKB along $\hat{x}$ and the CCD-512EBFT along $\hat{z}$.

In our first imaging option the resonant beam propagates along $\hat{x}$ and thus provides information about the axial and radial directions of the trapped gas. This imaging beam is perpendicular to the large bias magnetic field. Thus, imaging at high magnetic field with this beam is not optimum since the quantization is 90° off. Most imaging along this direction takes place at low magnetic field where a small ($\sim$4 G) magnetic field can be applied along $\hat{x}$ for the quantization. At low magnetic field the imaging light does not distinguish between different spin states. Thus, if we need to image the spin states separately a large magnetic field gradient ($\sim$80 G/cm [83]) is applied along $\hat{y}$ to separate the spin states spatially (Stern-Gerlach imaging). Since we have the capability to switch the direction of the small quantization field along this axis, this setup can image atoms in either the + or - spin states of the $f$=9/2 manifold.

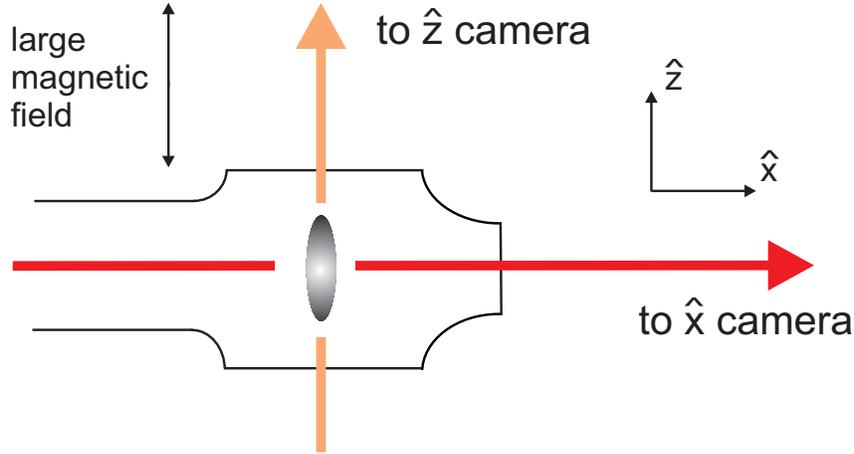

Figure 10.15: Imaging configurations. Gravity points into the page, and the large bias field and the axis of the optical trap both point along $\hat{z}$. The $^{40}$K gas (black oval) can be imaged along either $\hat{z}$ where there is a 1 inch window or $\hat{x}$ where there is a 1/2 inch window. Diagram not to scale.

In the second imaging option the resonant beam propagates along $\hat{z}$ where it is combined with the optical trap beam using dichroic plates. This setup images both radial directions of the trapped gas and is optimized for imaging the $m_f$=−5/2, −7/2, −9/2 spin states at high magnetic field near Feshbach resonances. The quantization is provided by the large bias magnetic field (Sec. 10.4), and the atoms can easily be imaged spin selectively since the transitions for the different spin states are separated by $\sim$30 MHz near 200 G due to the nonlinear Zeeman shift. We can collect an image for each of two spin states per experiment cycle using the kinetics mode and frame transfer capabilities of our CCD camera.

It is important that the optics used for imaging the atom gas are designed correctly to provide sufficient imaging resolution. The optics used to image the gas along the $\hat{z}$ direction are shown in Fig. 10.16. The blue lines roughly represent the image of the atom gas, which is collimated by the first set of lenses and focused onto the camera with the second. We expect the diffraction limit to be defined by an Airy disc pattern (see Fig. 10.17 red line) the diameter of which is $1.22\lambda/\text{NA}$, where NA$= d/2f$, $d$ is the lens diameter, and $f$ the lens focal length. Diffraction limited performance of the first lens set in Fig. 10.16 at 767 nm should result in an Airy disc diameter (spot size) of 4.5 $\mu$m.

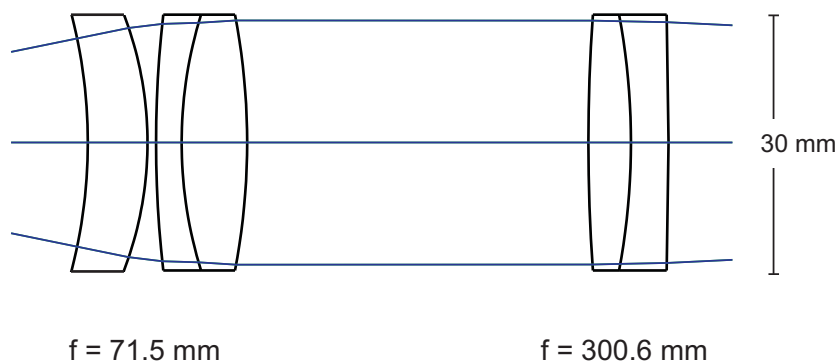

f = 71.5 mm                                        f = 300.6 mm

Figure 10.16: Optics used to image the atom gas along the $\hat{z}$ direction. The atoms are to the left and the CCD camera to the right. The meniscus and achromat lens pair provide diffraction limited performance.

To test whether this imaging system could provide diffraction limited performance, we imaged a 1 $\mu$m pinhole onto the WinCamD beam profiling camera using this set of lenses. Using a gaussian fit of the resulting spot on the camera, and accounting for the magnification and the difference in test wavelength and 767 nm, we found that the full-width-half-max (fwhm) of the spot was $2.1 \pm 0.3$ $\mu$m. Converting this result to an Airy disc diameter (Fig. 10.17), we found $5.1 \pm 0.6$ $\mu$m, which is consistent with diffraction limited performance. While we cannot create a gas small enough to fully test the resolution limit of this system in the actual experiment, measurements in Ref. [155] are consistent with a spot size of $\sim$5 $\mu$m.

The imaging optics for the path along $\hat{x}$ are similar to those in Fig. 10.16, with the differences being the first lens is an achromat (instead of an achromat/meniscus lens pair) with a 140 mm focal length and the second lens has a 500 mm focal length. The expected diffraction limited performance of this imaging system is a spot size diameter of 8.7 $\mu$m (fwhm of 3.7 $\mu$m). It is also possible to place two more lenses between the f=500 mm lens and the camera to make a low-magnification imaging system for probing the MOT or hot gases in the magnetic trap [83].

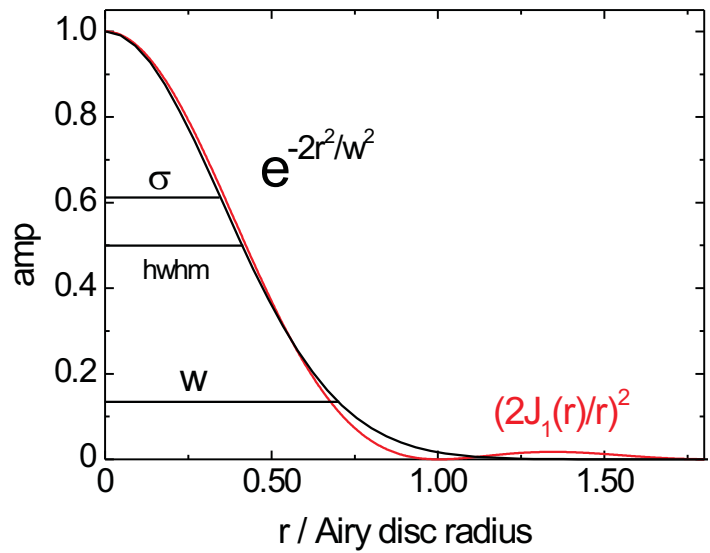

Figure 10.17: Gaussian distribution compared to an Airy disc pattern; a gaussian was fit to the Bessel function that describes the Airy disc pattern. Numbers used to characterize the radius of gaussian distributions include the rms size $\sigma$, the half-width-half-max (hwhm), and the waist (w). These are related through w=2$\sigma$=1.7 hwhm. The hwhm of the Airy disc and the hwhm of the gaussian occur at nearly the same position, 0.42 and 0.41 times the Airy disc radius, respectively.

# Chapter 11

# Conclusions and future directions

## 11.1 Conclusion

In this thesis I have presented the story of the realization of BCS-BEC crossover physics with a gas of $^{40}$K atoms. Experiments with $^{40}$K and $^6$Li have shown that a Fermi gas at a broad Feshbach resonance crosses a phase transition to a superfluid state and displays properties of the classic BCS-BEC crossover problem. With this system we can study the evolution from fermionic superfluidity of pairs nearly described by BCS theory to BEC of diatomic molecules. This new system provides a physical link between two descriptions of superfluid systems, BCS and BEC, that were historically thought to be distinct.

The fermionic superfluids created in these gases have extremely high transition temperatures, $T_c \approx 0.1\ T_F$. In these model systems the absolute value of $T_c \approx 100$ nK is very cold. However, for typical values of $T_F$ in metals the corresponding transition temperature would be above room temperature. Fully understanding this model system will perhaps contribute to the efforts to create higher transition temperature superconductors in real materials. These atomic gas systems also provide the opportunity to study aspects of quantum systems that are not typically accessible in solid state materials. For example, experiments such as those in Ch. 8 utilize dynamics for measurements. The ability to study the physics of the BCS-BEC crossover in real time may provide new insight into this many-body quantum system.

## 11.2 Future work

In the immediate future there are a plethora of experiments that can be done to study BCS-BEC crossover physics with this new system. This list includes both more precise versions of previous measurements as well as experiments designed to study entirely new phenomena. In our $^{40}$K experiments there are a number of possible direct extensions of measurements discussed in this thesis. For example, a better measurement of the Feshbach resonance position $B_0$ than that of Fig. 6.9 would produce more precise values of $1/k_F a$. To make a more precise measurement of $B_0$ one could use the magnetic field modulation technique for dissociating molecules to measure $E_b$ of a low density gas. The phase diagram of the BCS-BEC crossover measured through $N_0/N$ could be improved with better temperature measurements and better understanding of the systematics on $N_0/N$. With more precise measurements of experimental observables in the crossover, we will be able to test more sophisticated crossover theories.

Possible extensions of this study of BCS-BEC crossover physics are also numerous. Studies of higher partial wave pairing in an atomic system [57, 164, 188] would be an important contribution due to the relevance to cuprate superconductors, which have been shown to have $d$-wave symmetry. Predictions have been made for the behavior of an atomic Fermi gas in the BCS-BEC crossover with unequal particle number in each spin state of the two-component gas [189, 190]. Such a "magnetized" system has been the topic of a number of recent experimental pursuits [191, 192]. Further, quantum Fermi gases have now been studied in optical lattice potentials [193, 40]; studying crossover physics in the presence of such a lattice could more closely mimic conditions in real solids.

# Bibliography

[1] H. K. Onnes, On the sudden rate at which the resistance of mercury disappears, *Akad. van Wetenschappen* **14(113)**, 818 (1911).

[2] J. F. Allen and A. D. Misener, Flow of liquid helium II, *Nature* **141**, 75 (1938).

[3] P. Kapitza, Viscosity of liquid helium at temperatures below the lambda point, *Nature* **141**, 74 (1938).

[4] D. D. Osheroff, R. C. Richardson, and D. M. Lee, Evidence for a new phase of solid $^3$He, *Phys. Rev. Lett.* **28**, 885 (1972).

[5] J. G. Bednorz and K. Mueller, Possible high-$T_c$ superconductivity in the Ba-La-Cu-O system, *Z. Physik* **B64**, 189 (1986).

[6] D. W. Snoke and G. Baym, in *Bose-Einstein Condensation*, edited by A. Griffin, D. W. Snoke, and S. Stringari (Cambridge Univ., Cambridge, UK, 1995), pp. 1–11.

[7] M. Holland, S. J. J. M. F. Kokkelmans, M. L. Chiofalo, and R. Walser, Resonance superfluidity in a quantum degenerate Fermi gas, *Phys. Rev. Lett.* **87**, 120406 (2001).

[8] M. Randeria, in *Bose-Einstein Condensation*, edited by A. Griffin, D. W. Snoke, and S. Stringari (Cambridge Univ., Cambridge, UK, 1995), pp. 355–392.

[9] S. Bose, Plancks Gesetz und Lichtquantenhypothese, *Z. Phys.* **26**, 178 (1924).

[10] A. Einstein, Quantentheorie des einatomigen idealen Gases. Zweite Abhandlung, *Sitzungsber. Preuss. Akad. Wiss.* **1925**, 3 (1925).

[11] For a nice pedagogical description of BEC see Ch. 1 of J. F. Annett, *Superconductivity, Superfluids, and Condensates* (Oxford Univ. Press, Oxford, UK, 2004).

[12] F. London, On the Bose-Einstein condensation, *Phys. Rev.* **54**, 947 (1938).

[13] J. M. Blatt and S. T. Butler, Superfluidity of a boson gas, *Phys. Rev.* **96**, 1149 (1954).

[14] P. E. Sokol, in *Bose-Einstein Condensation*, edited by A. Griffin, D. W. Snoke, and S. Stringari (Cambridge Univ., Cambridge, UK, 1995), pp. 51–85.

[15] J. Bardeen, L. N. Cooper, and J. R. Schrieffer, Microscopic theory of super-conductivity, *Phys. Rev.* **106**, 162 (1957).

[16] J. Bardeen, L. N. Cooper, and J. R. Schrieffer, Theory of superconductivity, *Phys. Rev.* **108**, 1175 (1957).

[17] L. N. Cooper, Bound electron pairs in a degenerate Fermi gas, *Phys. Rev.* **104**, 1189 (1956).

[18] Q. Chen, J. Stajic, S. Tan, and K. Levin, BCS-BEC crossover: From high temperature superconductors to ultracold superfluids, *Phys. Rep.* **412**, 1 (2005).

[19] W. D. Philllips, Laser cooling and trapping of neutral atoms, *Rev. Mod. Phys.* **70**, 721 (1998).

[20] W. Ketterle, D. S. Durfee, and D. M. Stamper-Kurn, in *Proceedings of the International School of Physics - Enrico Fermi*, edited by M. Inguscio, S. Stringari, and C. E. Wieman (IOS Press, Amsterdam, 1999), p. 67.

[21] E. A. Cornell, J. R. Ensher, and C. E. Wieman, in *Proceedings of the International School of Physics - Enrico Fermi*, edited by M. Inguscio, S. Stringari, and C. E. Wieman (IOS Press, Amsterdam, 1999), p. 15.

[22] M. H. Anderson, J. R. Ensher, M. R. Matthews, C. E. Wieman, and E. A. Cornell, Observation of Bose-Einstein condensation in a dilute atomic vapor, *Science* **269**, 198 (1995).

[23] K. B. Davis, M.-O. Mewes, M. R. Andrews, N. J. van Druten, D. S. Durfee, D. M. Kurn, and W. Ketterle, Bose-Einstein condensation in a gas of sodium atoms, *Phys. Rev. Lett.* **75**, 3969 (1995).

[24] M. R. Andrews, C. G. Townsend, H.-J. Miesner, D. S. Durfee, D. M. Kurn, and W. Ketterle, Observation of interference between two Bose-Einstein condensates, *Science* **275**, 637 (1997).

[25] M. R. Matthews, B. P. Anderson, P. C. Haljan, D. S. Hall, C. E. Wieman, and E. A. Cornell, Vortices in a Bose-Einstein Condensate, *Phys. Rev. Lett.* **83**, 2498 (1999).

[26] C. Raman, M. Köhl, R. Onofrio, D. S. Durfee, C. E. Kuklewicz, Z. Hadzibabic, and W. Ketterle, Evidence for a critical velocity in a Bose-Einstein condensed gas, *Phys. Rev. Lett.* **83**, 2502 (1999).

[27] F. Dalfovo, S. Giorgini, L. P. Pitaevskii, and S. Stringari, Theory of Bose-Einstein condensation in trapped gases, *Rev. Mod. Phys.* **71**, 463 (1999).

[28] D. R. Tilley, *Superfluidity and Superconductivity* (Adam Hilger, Ltd., Accord, MA, 1986).

[29] S. Inouye, M. R. Andrews, J. Stenger, H.-J. Miesner, D. M. Stamper-Kurn, and W. Ketterle, Observation of Feshbach resonances in a Bose-Einstein condensate, *Nature* **392**, 151 (1998).

[30] J. L. Roberts, N. R. Claussen, J. P. Burke, Jr., C. H. Greene, E. A. Cornell, and C. E. Wieman, Resonant magnetic field control of elastic scattering of cold $^{85}$Rb, *Phys. Rev. Lett.* **81**, 5109 (1998).

[31] M. Greiner, O. Mandel, T. Esslinger, T. W. Hänsch, and I. Bloch, Quantum phase transition from a superfluid to a Mott insulator in a gas of ultracold atoms, *Nature* **415**, 39 (2002).

[32] B. DeMarco and D. S. Jin, Onset of Fermi degeneracy in a trapped atomic gas, *Science* **285**, 1703 (1999).

[33] A. G. Truscott, K. E. Strecker, W. I. McAlexander, G. B. Partridge, and R. G. Hulet, Observation of Fermi pressure in a gas of trapped atoms, *Science* **291**, 2570 (2001).

[34] F. Schreck, L. Khaykovich, K. L. Corwin, G. Ferrari, T. Bourdel, J. Cubizolles, and C. Salomon, Quasipure Bose-Einstein condensate immersed in a Fermi sea, *Phys. Rev. Lett.* **87**, 080403 (2001).

[35] G. Roati, F. Riboli, G. Modugno, and M. Inguscio, Fermi-Bose quantum degenerate $^{40}$K-$^{87}$Rb mixture with attractive interaction, *Phys. Rev. Lett.* **89**, 1804 (2002).

[36] S. R. Granade, M. E. Gehm, K. M. O'Hara, and J. E. Thomas, All-optical production of a degenerate Fermi gas, *Phys. Rev. Lett.* **88**, 120405 (2002).

[37] Z. Hadzibabic, C. A. Stan, K. Dieckmann, S. Gupta, M. W. Zwierlein, A. Görlitz, and W. Ketterle, Two species mixture of quantum degenerate Bose and Fermi gases, *Phys. Rev. Lett.* **88**, 160401 (2002).

[38] J. Goldwin, S. Inouye, M. Olsen, B. Newman, B. D. DePaola, and D. S. Jin, Measurement of the interaction strength in a Bose-Fermi mixture with $^{87}$Rb and $^{40}$K, *Phys. Rev. A* **70**, 021601 (2004).

[39] M. Bartenstein, A. Altmeyer, S. Riedl, S. Jochim, C. Chin, J. Hecker Denschlag, and R. Grimm, Crossover from a molecular Bose-Einstein condensate to a degenerate Fermi gas, *Phys. Rev. Lett.* **92**, 120401 (2004).

[40] M. Köhl, H. Moritz, T. Stöferle, K. Günter, and T. Esslinger, Fermionic atoms in a 3D optical lattice: Observing Fermi-surface, dynamics, and interactions, *Phys. Rev. Lett.* **94**, 080403 (2005).

[41] C. Ospelkaus, S. Ospelkaus, K. Sengstock, and K. Bongs, Interaction-driven dynamics of $^{40}$K / $^{87}$Rb Fermi-Bose gas mixtures in the large particle number limit, *cond-mat/0507219* (2005).

[42] H. T. C. Stoof, M. Houbiers, C. A. Sackett, and R. G. Hulet, Superfluidity of spin-polarized $^6$Li, *Phys. Rev. Lett.* **76**, 10 (1996).

[43] R. Combescot, Trapped $^6$Li: A high $T_c$ superfluid?, *Phys. Rev. Lett.* **83**, 3766 (1999).

[44] E. Timmermans, K. Furuya, P. W. Milonni, and A. K. Kerman, Prospect of creating a composite Fermi-Bose superfluid, *Phys. Lett. A* **285**, 228 (2001).

[45] Y. Ohashi and A. Griffin, BCS-BEC crossover in a gas of Fermi atoms with a Feshbach resonance, *Phys. Rev. Lett.* **89**, 130402 (2002).

[46] D. M. Eagles, Possible pairing without superconductivity at low carrier concentrations in bulk and thin-film superconducting semiconductors, *Phys. Rev.* **186**, 456 (1969).

[47] A. J. Leggett, Cooper pairing in spin-polarized Fermi systems, *J. Phys. C. (Paris)* **41**, 7 (1980).

[48] P. Nozieres and S. Schmitt-Rink, Bose condensation in an attractive fermion gas: From weak to strong coupling superconductivity, *J. Low-Temp. Phys.* **59**, 195 (1985).

[49] M. Randeria, J. M. Duan, and L. Y. Shieh, Superconductivity in two-dimensional Fermi gas: Evolution from Cooper pairing to Bose condensation, *Phys. Rev. B* **41**, 327 (1990).

[50] M. Dreschler and W. Zwerger, Crossover from BCS-superconductivity to Bose-condensation, *Ann. Physik* **1**, 15 (1992).

[51] R. Haussman, Properties of Fermi liquid at the superfluid transition in the crossover region between BCS superconductivity and Bose-Einstein condensation, *Phys. Rev. B* **49**, 12975 (1994).

[52] M. Houbiers, H. T. C. Stoof, W. I. McAlexander, and R. G. Hulet, Elastic and inelastic collisions of Li-6 atoms in magnetic and optical traps, *Phys. Rev. A* **57**, R1497 (1998).

[53] J. L. Bohn, Cooper pairing in ultracold $^{40}$K using Feshbach resonances, *Phys. Rev. A* **61**, 053409 (2000).

[54] T. Loftus, C. A. Regal, C. Ticknor, J. L. Bohn, and D. S. Jin, Resonant control of elastic collisions in an optically trapped Fermi gas of atoms, *Phys. Rev. Lett.* **88**, 173201 (2002).

[55] K. M. O'Hara, S. L. Hemmer, S. R. Granade, M. E. Gehm, J. E. Thomas, V. Venturi, E. Tiesinga, and C. J. Williams, Measurement of the zero crossing in a Feshbach resonance of fermionic $^{6}$Li, *Phys. Rev. A* **66**, 041401(R) (2002).

[56] K. Dieckmann, C. A. Stan, S. Gupta, Z. Hadzibabic, C. H. Schunck, and W. Ketterle, Decay of an ultracold fermionic lithium gas near a Feshbach resonance, *Phys. Rev. Lett.* **89**, 203201 (2002).

[57] C. A. Regal, C. Ticknor, J. L. Bohn, and D. S. Jin, Tuning $p$-wave interactions in an ultracold Fermi gas of atoms, *Phys. Rev. Lett.* **90**, 053201 (2003).

[58] K. M. O'Hara, S. L. Hemmer, M. E. Gehm, S. R. Granade, and J. E. Thomas, Observation of a strongly interacting degenerate Fermi gas of atoms, *Science* **298**, 2179 (2002).

[59] C. A. Regal and D. S. Jin, Measurement of positive and negative scattering lengths in a Fermi gas of atoms, *Phys. Rev. Lett.* **90**, 230404 (2003).

[60] S. Gupta, Z. Hadzibabic, M. W. Zwierlein, C. A. Stan, K. Dieckmann, C. H. Schunck, E. G. M. van Kempen, B. J. Verhaar, and W. Ketterle, Radio-frequency spectroscopy of ultracold fermions, *Science* **300**, 1723 (2003).

[61] M. E. Gehm, S. L. Hemmer, K. M. O'Hara, and J. E. Thomas, Unitarity-limited elastic collision rate in a harmonically trapped Fermi gas, *Phys. Rev. A* **68**, 011603(R) (2003).

[62] T. Bourdel, J. Cubizolles, L. Khaykovich, K. M. F. Magalhães, S. J. J. M. F. Kokkelmans, G. V. Shlyapnikov, and C. Salomon, Measurement of the interaction energy near a Feshbach resonance in a $^6$Li Fermi gas, *Phys. Rev. Lett.* **91**, 020402 (2003).

[63] C. A. Regal, C. Ticknor, J. L. Bohn, and D. S. Jin, Creation of ultracold molecules from a Fermi gas of atoms, *Nature* **424**, 47 (2003).

[64] K. E. Strecker, G. B. Partridge, and R. G. Hulet, Conversion of an atomic Fermi gas to a long-lived molecular Bose gas, *Phys. Rev. Lett.* **91**, 080406 (2003).

[65] J. Cubizolles, T. Bourdel, S. J. J. M. F. Kokkelmans, G. V. Shlyapnikov, and C. Salomon, Production of long-lived ultracold Li$_2$ molecules from a Fermi gas of atoms, *Phys. Rev. Lett* **91**, 240401 (2003).

[66] S. Jochim, M. Bartenstein, A. Altmeyer, G. Hendl, C. Chin, J. Hecker Denschlag, and R. Grimm, Pure gas of optically trapped molecules created from fermionic atoms, *Phys. Rev. Lett.* **91**, 240402 (2003).

[67] C. A. Regal, M. Greiner, and D. S. Jin, Lifetime of molecule-atom mixtures near a Feshbach resonance in $^{40}$K, *Phys. Rev. Lett.* **92**, 083201 (2004).

[68] M. Greiner, C. A. Regal, and D. S. Jin, Emergence of a molecular Bose-Einstein condensate from a Fermi gas, *Nature* **426**, 537 (2003).

[69] S. Jochim, M. Bartenstein, A. Altmeyer, G. Hendl, S. Riedl, C. Chin, J. Hecker Denschlag, and R. Grimm, Bose-Einstein condensation of molecules, *Science* **302**, 2101 (2003).

[70] M. W. Zwierlein, C. A. Stan, C. H. Schunck, S. M. F. Raupach, S. Gupta, Z. Hadzibabic, and W. Ketterle, Observation of Bose-Einstein condensation of molecules, *Phys. Rev. Lett.* **91**, 250401 (2003).

[71] T. Bourdel, L. Khaykovich, J. Cubizolles, J. Zhang, F. Chevy, M. Teichmann, L. Tarruell, S. Kokkelmans, and C. Salomon, Experimental study of the BEC-BCS crossover region in $^6$Li, *Phys. Rev. Lett.* **93**, 050401 (2004).

[72] G. B. Partridge, K. E. Strecker, R. I. Kamar, M. W. Jack, and R. G. Hulet, Molecular probe of pairing in the BEC-BCS crossover, *Phys. Rev. Lett.* **95**, 020404 (2005).

[73] C. A. Regal, M. Greiner, and D. S. Jin, Observation of resonance condensation of fermionic atom pairs, *Phys. Rev. Lett.* **92**, 040403 (2004).

[74] M. W. Zwierlein, C. A. Stan, C. H. Schunck, S. M. F. Raupach, A. J. Kerman, and W. Ketterle, Condensation of pairs of fermionic atoms near a Feshbach resonance, *Phys. Rev. Lett.* **92**, 120403 (2004).

[75] M. Bartenstein, A. Altmeyer, S. Riedl, S. Jochim, C. Chin, J. Hecker Denschlag, and R. Grimm, Collective excitations of a degenerate gas at the BEC-BCS crossover, *Phys. Rev. Lett.* **92**, 203201 (2004).

[76] J. Kinast, S. L. Hemmer, M. E. Gehm, A. Turlapov, and J. E. Thomas, Evidence for superfluidity in a resonantly interacting Fermi gas, *Phys. Rev. Lett.* **92**, 150402 (2004).

[77] J. Kinast, A. Turlapov, and J. E. Thomas, Breakdown of hydrodynamics in the radial breathing mode of a strongly interacting Fermi gas, *Phys. Rev. A* **70**, 051401 (2004).

[78] C. A. Regal, M. Greiner, S. Giorgini, M. Holland, and D. S. Jin, Momentum distribution of a Fermi gas of atoms in the BCS-BEC crossover, *Phys. Rev. Lett.* **95**, 250404 (2005).

[79] J. Kinast, A. Turlapov, J. E. Thomas, Q. Chen, J. Stajic, and K. Levin, Heat capacity of a strongly interacting Fermi gas, *Science* **307**, 1296 (2005).

[80] C. Chin, M. Bartenstein, A. Altmeyer, S. Riedl, S. Jochim, J. H. Denschlag, and R. Grimm, Observation of the pairing gap in a strongly interacting Fermi gas, *Science* **305**, 1128 (2004).

[81] M. Greiner, C. A. Regal, and D. S. Jin, Probing the excitation spectrum of a Fermi gas in the BCS-BEC crossover regime, *Phys. Rev. Lett.* **94**, 070403 (2005).

[82] M. Zwierlein, J. Abo-Shaeer, A. Schirotzek, C. Schunck, and W. Ketterle, Observation of high-$T_c$ superfluidity, *Nature* **435**, 1047 (2005).

[83] B. DeMarco, Quantum behavior of an atomic Fermi gas, Ph.D. thesis, University of Colorado - Boulder, 2001.

[84] J. Stenger, S. Inouye, M. R. Andrews, H.-J. Miesner, D. M. Stamper-Kurn, and W. Ketterle, Strongly enhanced inelastic collisions in a Bose-Einstein condensate near Feshbach resonances, *Phys. Rev. Lett.* **82**, 2422 (1999).

[85] J. L. Roberts, N. R. Claussen, S. L. Cornish, and C. E. Wieman, Magnetic field dependence of ultracold inelastic collisions near a Feshbach resonance, *Phys. Rev. Lett.* **85**, 728 (2000).

[86] P. O. Fedichev, M. W. Reynolds, and G. V. Shlyapnikov, Three-body recombination of ultracold atoms to a weakly bound s-level, *Phys. Rev. Lett.* **77**, 2921 (1996).

[87] E. Nielsen and J. H. Macek, Low-energy recombination of identical bosons by three-body collisions, *Phys. Rev. Lett.* **83**, 1566 (1999).

[88] B. D. Esry, C. H. Greene, and J. P. Burke, Recombination of three atoms in the ultracold limit, *Phys. Rev. Lett.* **83**, 1751 (1999).

[89] E. Braaten, H.-W. Hammer, and S. Hermans, Nonuniversal effects in the homogeneous Bose gas, *Phys. Rev. A* **63**, 063609 (2001).

[90] B. D. Esry, C. H. Greene, and H. Suno, Threshold laws for three-body rescombination, *Phys. Rev. A* **65**, 010705(R) (2001).

[91] D. S. Petrov, Three-body problem in Fermi gases with short-range interparticle interaction, *Phys. Rev. A* **67**, 010703(R) (2003).

[92] P. G. de Gennes, *Superconductivity of Metals and Alloys* (Addison-Wesley, California, 1966).

[93] M. Tinkham, *Introduction to Superconductivity* (Krieger, Malabar, Florida, 1980).

[94] J. J. Sakurai, *Modern quantum mechanics* (Addison-Wesley, Reading, MA, 1994).

[95] D. S. Petrov, C. Salomon, and G. V. Shlyapnikov, Weakly bound dimers of fermionic atoms, *Phys. Rev. Lett* **93**, 090404 (2004).

[96] G. A. Baker, Neutron matter model, *Phys. Rev. C* **60**, 054311 (1999).

[97] H. Heiselberg, Fermi systems with long scattering lengths, *Phys. Rev. A* **63**, 043606 (2001).

[98] J. Kinast, A. Turlapov, and J. E. Thomas, Damping of a unitary Fermi gas, *Phys. Rev. Lett.* **94**, 170404 (2005).

[99] J. E. Thomas, J. Kinast, and A. Turpalov, Virial theorem and universality in a unitary Fermi gas, *Phys. Rev. Lett.* **95**, 120402 (2005).

[100] M. Marini, F. Pistolesi, and G. C. Strinati, Evolution from BCS superconductivity to Bose condensation: Analytic results for the crossover in three dimensions, *Eur. Phys. J. B* **1**, 151 (1998).

[101] C. J. Pethick and H. Smith, *Bose-Einstein condensation in dilute gases* (Cambridge Univ. Press, Cambridge, UK, 2002).

[102] J. Carlson, S.-Y. Chang, V. R. Pandharipande, and K. E. Schmidt, Super-fluid Fermi gases with large scattering length, *Phys. Rev. Lett.* **91**, 050401 (2003).

[103] G. E. Astrakharchik, J. Boronat, J. Casulleras, and S. Giorgini, Equation of state of a Fermi gas in the BEC-BCS crossover: A quantum Monte Carlo study, *Phys. Rev. Lett.* **93**, 200404 (2004).

[104] M. J. Holland, C. Menotti, and L. Viverit, The role of boson-fermion corre-lations in the resonance theory of superfluids, *cond-mat/0404234* (2005).

[105] L. Vichi and S. Stringari, Collective oscillations of an interacting trapped Fermi gas, *Phys. Rev. A* **60**, 4734 (1999).

[106] U. Fano, Effects of configuration interaction on intensities and phase shifts, *Phys. Rev.* **124**, 1866 (1961).

[107] H. Feshbach, A unified theory of nuclear reactions II, *Ann. Phys.* **19**, 287 (1962).

[108] W. C. Stwalley, Stability of spin-aligned hydrogen at low temperatures and high magnetic fields: New field-dependent scattering resonance and predis-sociations, *Phys. Rev. Lett.* **37**, 1628 (1976).

[109] E. Tiesinga, B. J. Verhaar, and H. T. C. Stoof, Threshold and resonance phe-nomena in ultracold ground-state collisions, *Phys. Rev. A* **47**, 4114 (1993).

[110] J. P. Burke, Theoretical investigations of cold alkali atom collisions, Ph.D. thesis, University of Colorado - Boulder, 1999, and references therein.

[111] M. H. Szymańska, K. Góral, T. Köhler, and K. Burnett, Conventional char-acter of the BCS-BEC crossover in ultracold gases of $^{40}$K, *Phys. Rev. A* **72**, 013610 (2005).

[112] K. Góral, T. Köhler, S. A. Gardiner, E. Tiesinga, and P. S. Julienne, Adia-batic association of ultracold molecules via magnetic-field tunable interac-tions, *J. Phys. B* **37**, 3457 (2004).

[113] A. Derevianko, W. R. Johnson, M. S. Safronova, and J. F. Babb, High-precision calculations of dispersion coefficients, static dipole polarizabilities, and atom-wall interaction constants for alkali-metal atoms, *Phys. Rev. Lett.* **82**, 3589 (1999).

[114] M. H. Szymańska, Private communication; the full coupled channels calculation was done by P. Julienne.

[115] A. Corney, *Atomic and Laser Spectroscopy* (Clarendon Press, Oxford, UK, 1977).

[116] C. Chin, A simple model of Feshbach molecules, *cond-mat/0506313* (2005).

[117] G. F. Gribakin and V. V. Flambaum, Calculation of the scattering length in atomic collisions using the semiclassical approximation, *Phys. Rev. A* **48**, 546 (1993).

[118] C. Ticknor, private communication.

[119] R. B. Diener and T.-L. Ho, The condition for universality at resonance and direct measurement of pair wavefunctions using rf spectroscopy, *cond-mat/0405174* (2004).

[120] G. M. Bruun and C. J. Pethick, Effective theory of Feshbach resonances and many-body properties of Fermi gases, *Phys. Rev. Lett.* **92**, 140404 (2004).

[121] S. De Palo, M. L. Chiofalo, M. J. Holland, and S. J. J. M. F. Kokkelmans, Resonance effects on the crossover of bosonic to fermionic superfluidity, *Phys. Lett. A* **327**, 490 (2004).

[122] M. Bartenstein *et al.*, Precise determination of $^6$Li cold collision parameters by radio-frequency spectroscopy on weakly bound molecules, *Phys. Rev. Lett.* **94**, 103201 (2005).

[123] C. J. Myatt, Bose-Einstein condensation experiments in a dilute vapor of Rubidium, Ph.D. thesis, University of Colorado - Boulder, 1997.

[124] J. R. Ensher, The first experiments with Bose-Einstein condensation of $^{87}$Rb, Ph.D. thesis, University of Colorado - Boulder, 1998.

[125] B. DeMarco, J. L. Bohn, J. P. Burke, Jr., M. Holland, and D. S. Jin, Measurement of $p$-wave threshold law using evaporatively cooled fermionic atoms, *Phys. Rev. Lett.* **82**, 4208 (1999).

[126] B. DeMarco, S. B. Papp, and D. S. Jin, Pauli blocking of collisions in a quantum degenerate atomic Fermi gas, *Phys. Rev. Lett.* **86**, 5409 (2001).

[127] S. Chu, J. E. Bjorkholm, A. Ashkin, and A. Cable, Experimental observation of optically trapped atoms, *Phys. Rev. Lett.* **57**, 314 (1986).

[128] R. Grimm, M. Weidemüller, and Y. B. Ovchinnikov, in *Advances in Atomic, Molecular, and Optical Physics* (PUBLISHER, ADDRESS, 2000), Vol. 42, Chap. Optical dipole traps for neutral atoms, pp. 95–170.

[129] D. A. Butts and D. S. Rokhsar, Trapped Fermi gases, *Phys. Rev. A* **55**, 4346 (1997).

[130] L. Viverit, S. Giorgini, L. P. Pitaevskii, and S. Stringari, Adiabatic compression of a trapped Fermi gas, *Phys. Rev. A* **63**, 033603 (2001).

[131] Q. Chen, J. Stajic, and K. Levin, Thermodynamics of interacting fermions in atomic traps, *cond-mat/0411090* (2005).

[132] B. Jackson, P. Pedri, and S. Stringari, Collisions and expansion of an ultra-cold dilute Fermi gas, *Europhys. Lett.* **67(4)**, 524 (2004).

[133] C. R. Monroe, E. A. Cornell, C. A. Sackett, C. J. Myatt, and C. E. Wieman, Measurement of Cs-Cs elastic scattering at $T = 30 \ \mu K$, *Phys. Rev. Lett.* **70**, 414 (1993).

[134] J. L. Roberts, J. P. Burke, Jr., N. R. Claussen, S. L. Cornish, E. A. Donley, and C. E. Wieman, Improved characterization of elastic scattering near a Feshbach resonance in $^{85}$Rb, *Phys. Rev. A* **64**, 024702 (2001).

[135] J. P. Burke, Jr., C. H. Greene, J. L. Bohn, H. Wang, P. L. Gould, and W. C. Stwalley, Determination of $^{39}$K scattering lengths using photoassociation spectroscopy of the $O_g^-$ state, *Phys. Rev. A* **60**, 4417 (1999).

[136] C. Menotti, P. Pedri, and S. Stringari, Expansion of an interacting Fermi gas, *Phys. Rev. Lett.* **89**, 250402 (2002).

[137] Y. Kagan, E. L. Surkov, and G. V. Shlyapnikov, Evolution of a Bose gas in anisotropic time-dependent traps, *Phys. Rev. A* **55**, R18 (1997).

[138] S. Gupta, Z. Hadzibabic, J. R. Anglin, and W. Ketterle, Collisions in zero temperature Fermi gases, *Phys. Rev. Lett.* **92**, 100401 (2004).

[139] S. L. Cornish, N. R. Claussen, J. L. Roberts, E. A. Cornell, and C. E. Wieman, Stable $^{85}$Rb Bose-Einstein condensates with widely tunable interactions, *Phys. Rev. Lett.* **85**, 1795 (2000).

[140] J. M. Gerton, D. Strekalov, I. Prodan, and R. G. Hulet, Direct observation of growth and collapse of a Bose-Einstein condensate with attractive interactions, *Nature* **408**, 692 (2000).

[141] J. L. Roberts, N. R. Claussen, S. L. Cornish, E. A. Donley, E. A. Cornell, and C. E. Wieman, Controlled collapse of a Bose-Einstein condensate, *Phys. Rev. Lett.* **86**, 4211 (2001).

[142] D. M. Harber, H. J. Lewandowski, J. M. McGuirck, and E. A. Cornell, Effect of cold collisions in spin coherence and resonance shifts in a magnetically trapped ultracold gas, *Phys. Rev. A* **66**, 053616 (2002).

[143] C. Ticknor, C. A. Regal, D. S. Jin, and J. L. Bohn, Multiplet structure of Feshbach resonances in nonzero partial waves, *Phys. Rev. A* **69**, 042712 (2004).

[144] E. A. Donley, N. R. Claussen, S. T. Thompson, and C. E. Wieman, Atom-molecule coherence in a Bose-Einstein condensate, *Nature* **417**, 529 (2002).

[145] N. R. Claussen, E. A. Donley, S. T. Thompson, and C. E. Wieman, Microscopic dynamics in a strongly interacting Bose-Einstein condensate, *Phys. Rev. Lett.* **89**, 010401 (2002).

[146] E. Hodby, S. T. Thompson, C. A. Regal, M. Greiner, A. C. Wilson, D. S. Jin, E. A. Cornell, and C. E. Wieman, Production efficiency of ultracold Feshbach molecules in bosonic and fermionic systems, *Phys. Rev. Lett.* **94**, 120402 (2005).

[147] E. Timmermans, P. Tommasini, M. Hussein, and A. Kerman, Feshbach resonances in atomic Bose-Einstein condensates, *Phys. Rep.* **315**, 199 (1999).

[148] F. A. Abeelen and B. J. Verhaar, Time-dependent Feshbach resonance scattering and anomalous decay of a Na Bose-Einstein condensate, *Phys. Rev. Lett.* **83**, 1550 (1999).

[149] F. H. Mies, E. Tiesinga, and P. S. Julienne, Manipulation of Feshbach resonances in ultracold atomic collisions using time-dependent magnetic fields, *Phys. Rev. A* **61**, 022721 (2000).

[150] L. B. Ratcliff, J. L. Fish, and D. D. Konowalow, Electronic transition dipole moment functions for a transitions among the twenty-six lowest-lying states of $Li_2$, *J. Mol. Spectrosc.* **122**, 293 (1987).

[151] C. Chin and P. S. Julienne, Radio-frequency transitions on weakly-bound ultracold molecules, *Phys. Rev. A* **71**, 012713 (2005).

[152] K. G. Petrosyan, Fermionic atom laser, *JETP Lett.* **70**, 11 (1999).

[153] P. Törmä and P. Zoller, Laser probing of atomic Cooper pairs, *Phys. Rev. Lett.* **85**, 487 (2000).

[154] H. Moritz, T. Stöferle, K. Günter, M. Köhl, and T. Esslinger, Confinement induced molecules in a 1D Fermi gas, *Phys. Rev. Lett.* **94**, (2005).

[155] M. Greiner, C. A. Regal, J. T. Stewart, and D. S. Jin, Probing pair-correlated fermionic atoms through correlations in atom shot noise, *Phys. Rev. Lett.* **94**, 110401 (2005).

[156] S. T. Thompson, E. Hodby, and C. E. Wieman, Ultracold molecule production via a resonant oscillating magnetic field, *Phys. Rev. Lett.* **95**, 190404 (2005).

[157] J. P. D'Incao and B. D. Esry, Scattering length scaling laws for ultracold three-body collisions, *Phys. Rev. Lett.* **94**, 213201 (2005).

[158] H. Suno, B. D. Esry, and C. H. Greene, Recombination of three ultracold Fermionic atoms, *Phys. Rev. Lett.* **90**, 053202 (2003).

[159] S. T. Thompson, E. Hodby, and C. E. Wieman, Spontaneous dissociation of $^{85}$Rb Feshbach molecules, *Phys. Rev. Lett.* **94**, 020401 (2005).

[160] J. P. D'Incao and B. D. Esry, Enhancing observability of the Efimov effect in ultracold atomic gas mixtures, *cond-mat/0508474* (2005).

[161] N. Balakrishnan, R. C. Forrey, and A. Dalgarno, Quenching of $H_2$ vibrations in ultracold $^3$He and $^4$He collisions, *Phys. Rev. Lett.* **80**, 3224 (1998).

[162] R. C. Forrey, N. Balakrishnan, A. Dalgarno, M. R. Haggerty, and E. J. Heller, Quasiresonant energy transfer in ultracold atom-diatom collisions, *Phys. Rev. Lett.* **82**, 2657 (1999).

[163] P. Soldán, M. T. Cvitaš, J. M. Hutson, P. Honvault, and J.-M. Launay, Quantum dynamics of ultracold Na and $Na_2$ collisions, *Phys. Rev. Lett.* **89**, 153201 (2002).

[164] J. Zhang, E. G. M. van Kempen, T. Bourdel, L. Khaykovich, J. Cubizolles, F. Chevy, M. Teichmann, L. Tuarruell, S. J. J. M. F. Kokkelmans, and C. Salomon, Expansion of a lithium gas in the BEC-BCS crossover, *cond-mat/0410167* (2004).

[165] K. Xu, T. Mukaiyama, J. R. Abo-Shaeer, J. K. Chin, D. E. Miller, and W. Ketterle, Formation of quantum-degenerate sodium molecules, *Phys. Rev. Lett.* **91**, 210402 (2003).

[166] S. Dürr, T. Volz, A. Marte, and G. Rempe, Observation of molecules produced from a Bose-Einstein condensate, *Phys. Rev. Lett.* **92**, 020406 (2003).

[167] T. Weber, J. Herbig, M. Mark, H.-C. Nagerl, and R. Grimm, Three-body recombination at large scattering lengths in an ultracold atomic gas, *Phys. Rev. Lett.* **91**, 123201 (2003).

[168] C. Chin and R. Grimm, Thermal equilibrium and efficient evaporation of an ultracold atom-molecule mixture, *Phys. Rev. A* **69**, 033612 (2004).

[169] L. D. Carr, G. V. Shlyapnikov, and Y. Castin, Achieving a BCS transition in an atomic Fermi gas, *Phys. Rev. Lett.* **92**, 150404 (2004).

[170] S. Giorgini, L. P. Pitaevskii, and S. Stringari, Condensate fraction and critical temperature of a trapped interacting Bose gas, *Phys. Rev. A* **54**, R4633 (1996).

[171] M. W. Zwierlein, C. H. Schunck, C. A. Stan, S. M. F. Raupach, and W. Ketterle, Formation dynamics of fermion pair condensate, *Phys. Rev. Lett.* **94**, 180401 (2005).

[172] A. V. Avdeenkov and J. L. Bohn, Pair wave functions in atomic Fermi condensates, *Phys. Rev. A* **71**, 023609 (2005).

[173] E. Altman and A. Vishwanath, Dynamic projection on Feshbach molecules: a probe of pairing and phase flucutations, *Phys. Rev. Lett.* **95**, 110404 (2005).

[174] A. Perali, P. Pieri, and G. C. Strinati, Extracting the condensate density from projection experiments with Fermi gases, *Phys. Rev. Lett.* **95**, 010407 (2005).

[175] L. Viverit, S. Giorgini, L. Pitaevskii, and S. Stringari, Momentum distribution of a trapped Fermi gas with large scattering length, *Phys. Rev. A* **69**, 013607 (2004).

[176] M. L. Chiofalo, S. Giorgini, and M. Holland, Released momentum distribution of a Fermi gas in the BCS-BEC crossover, *in progress* (2005).

[177] E. Arimondo, M. Inguscio, and P. Violino, Experimental determinations of the hyperfine structure in the alkali atoms, *Rev. Mod. Phys.* **49**, 31 (1977).

[178] B. DeMarco, H. Rohner, and D. Jin, An enriched $^{40}$K source for fermionic atom studies, *Rev. Sci. Instrum.* **70**, 1967 (1999).

[179] D. Sesko, T. Walker, C. Monroe, A. Gallagher, and C. Wieman, Collisional losses from a light-force atom trap, *Phys. Rev. Lett.* **63**, 961 (1989).

[180] G. Modugno, C. Benko, P. Hannaford, G. Roati, and M. Inguscio, Sub-Doppler laser cooling of fermionic $^{40}$K atoms, *Phys. Rev. A* **60**, R3373 (1999).

[181] K. L. Corwin, Z. Lu, C. F. Hand, R. J. Epstein, and C. E. Wieman, Frequency-stabilized diode laser with Zeeman shift in an atomic vapor, *Applied Optics* **37**, 3295 (1998).

[182] J. Goldwin, Quantum degeneracy and interactions in the $^{87}$Rb-$^{40}$K Bose-Fermi mixture, Ph.D. thesis, University of Colorado - Boulder, 2005.

[183] K. B. MacAdam, A. Steinbach, and C. Wieman, A narrow band tunable diode laser system with grating feedback, and a saturated absorption spectrometer for Cs and Rb, *Am. J. Phys.* **60**, 1098 (1992).

[184] M. E. Gehm, K. M. O'Hara, T. A. Savard, and J. E. Thomas, Dynamics of noise-induced heating in atom traps, *Phys. Rev. A* **58**, 3914 (1998).

[185] N. R. Claussen, Dynamics of Bose-Einstein condensates near a Feshbach resonance in $^{85}$Rb, Ph.D. thesis, University of Colorado - Boulder, 2003.

[186] D. M. Pozar, *Microwave Engineering* (John Wiley and Sons, Inc., New York, 1998).

[187] J. L. Roberts, Bose-Einstein condensates with tunable atom-atom interactions: The first experiments with $^{85}$Rb BECs, Ph.D. thesis, University of Colorado - Boulder, 2001.

[188] C. H. Schunck, M. W. Zwierlein, C. A. Stan, S. M. F. Raupach, W. Ketterle, A. Simoni, E. Tiesinga, C. J. Williams, and P. S. Julienne, Feshbach resonances in $^{6}$Li, *Phys. Rev. A* **71**, 045601 (2005).

[189] T. Mizushima, K. Machida, and M. Ichioka, Direct imaging of spatially modulated superfluid phases in atomic fermion systems, *Phys. Rev. Lett.* **94**, 060404 (2005).

[190] D. E. Sheehy and L. Radzihovsky, BEC-BCS crossover in magnetized Feshbach-resonantly paired superfluids, *cond-mat/0508430* (2005).

[191] M. W. Zwierlein, A. Schirotzek, C. H. Schunck, and W. Ketterle, Fermionic superfluidity with imbalanced spin populations and quantum phase transitin to the normal state, *cond-mat/0511197* (2005).

[192] G. B. Partridge, W. Li, R. I. Kamar, Y. Liao, and R. G. Hulet, Pairing and phase separation in a polarized Fermi gas, *cond-mat/0511752* .

[193] H. Ott, E. de Mirandes, F. Ferlaino, G. Roati, G. Modugno, and M. Inguscio, Collisionally induced transport in periodic potentials, *Phys. Rev. Lett.* **92**, 140405 (2004).

[194] M. Greiner, C. A. Regal, C. Ticknor, J. L. Bohn, and D. S. Jin, Detection of spatial correlations in an ultracold gas of fermions, *Phys. Rev. Lett.* **92**, 150405 (2004).

# Appendix A

# Journal articles

In writing this thesis my goal was to discuss the main objective of my graduate work, which was to access BCS-BEC crossover physics with a Fermi gas of atoms. To adhere to this subject I necessarily omitted some of the work to which I contributed during my graduate career. For example, I devoted time to the study of $p$-wave Feshbach resonances in Fermi gases [57, 143] and to a feasibility study of atom shot noise as a probe of ultracold gases [155]. Thus, for completeness I list in Table A.1 journal articles I contributed to in varying extents in my graduate work.

Table A.1: Articles I co-authored as a graduate student.

| Paper | Title | Published | Citation |
|---|---|---|---|
| 1 | Resonant control of elastic collisions in an optically trapped Fermi gas of atoms | 2002 | [54] |
| 2 | Tuning p-wave interactions in an ultracold Fermi gas of atoms | 2003 | [57] |
| 3 | Measurement of positive and negative scattering lengths in a Fermi gas of atoms | 2003 | [59] |
| 4 | Creation of ultracold molecules from a Fermi gas of atoms | 2003 | [63] |
| 5 | Lifetime of molecule-atom mixtures near a Feshbach resonance in $^{40}$K | 2004 | [67] |
| 6 | Detection of spatial correlations in an ultracold gas of fermions | 2004 | [194] |
| 7 | Emergence of a molecular Bose-Einstein condensate from a Fermi gas | 2003 | [68] |
| 8 | Multiplet structure of Feshbach resonances in nonzero partial waves | 2004 | [143] |
| 9 | Observation of resonance condensation of fermionic atom pairs | 2004 | [73] |
| 10 | Probing the excitation spectrum of a Fermi gas in the BCS-BEC crossover regime | 2005 | [81] |
| 11 | Production efficiency of ultracold Feshbach molecules in bosonic and fermionic systems | 2005 | [146] |
| 12 | Probing pair-correlated fermionic atoms through correlations in atom shot noise | 2005 | [155] |
| 13 | Momentum distribution of a Fermi gas of atoms in the BCS-BEC crossover | 2005 | [78] |